\documentclass[a4paper,12pt]{article}

\usepackage{jheppub} 
\usepackage{hyperref}
\hypersetup{
    bookmarks=true,         
    unicode=false,          
    pdftoolbar=true,        
    pdfmenubar=true,        
    pdffitwindow=false,     
    pdfstartview={FitH},    
    pdftitle={My title},    
    pdfauthor={Author},     
    pdfsubject={Subject},   
    pdfcreator={Creator},   
    pdfproducer={Producer}, 
    pdfkeywords={keyword1} {key2} {key3}, 
    pdfnewwindow=true,      
    colorlinks=true,       
    linkcolor=red,          
    citecolor=violet,        
    filecolor=magenta,      
    urlcolor=blue,           
    linktocpage=true
}
\usepackage{graphics}
\usepackage{rotating}
\usepackage{subfigure}
\usepackage{pstricks}
\usepackage{color}
\usepackage{amsfonts}
\usepackage{mathrsfs}
\usepackage{epsfig}
\usepackage{amsmath,amssymb,amsthm,graphicx,latexsym}
\usepackage{ulem}

\newcommand{\be}{\begin{equation}}
\newcommand{\ee}{\end{equation}}
\newcommand{\bea}{\begin{eqnarray}}
\newcommand{\eea}{\end{eqnarray}}
\newcommand{\bml}{\begin{subequations}}
\newcommand{\eml}{\end{subequations}}
\newcommand{\bfig}{\begin{figure}}
\newcommand{\efig}{\end{figure}}

\newcommand{\slashed}{\hspace{-1.1ex}/}

\begin{document} 
$~~~~~~~~~~~~~~~~~~~~~~~~~~~~~~~~~~~~~~~~~~~~~~~~~~~~~~~~~~~~~~~~~~~~~~~~~~~~~~~~~~~~$\textcolor{red}{\bf TIFR/TH/15-13}
\title{\textsc{\fontsize{22.8}{40}\selectfont \sffamily \bfseries Constraining brane inflationary magnetic field from cosmoparticle physics   
after Planck}}

\author[a]{Sayantan Choudhury}

\affiliation[a]{Department of Theoretical Physics, Tata Institute of Fundamental Research, Colaba, Mumbai - 400005, India
\footnote{\textcolor{purple}{\bf Presently working as a Visiting (Post-Doctoral) fellow at DTP, TIFR, Mumbai, \\$~~~~~$Alternative
 E-mail: sayanphysicsisi@gmail.com}. ${}^{}$}}

\emailAdd{sayantan@theory.tifr.res.in}

\abstract{In this article, I have studied the cosmological and particle physics constraints on a generic class of large
field ($|\Delta\phi|>M_{p}$) and small field ($|\Delta\phi|<M_{p}$) models of brane inflationary
magnetic field from: (1) tensor-to-scalar ratio ($r$), (2) reheating, (3) leptogenesis
and (4) baryogenesis in case of Randall-Sundrum single braneworld gravity (RSII) framework. I also establish
a direct connection between the magnetic field at the present epoch ($B_{0}$) and primordial gravity waves
($r$), which give a precise estimate of non-vanishing CP asymmetry ($\epsilon_{CP}$) in leptogenesis and baryon
asymmetry ($\eta_{B}$) in baryogenesis scenario respectively. Further assuming the conformal invariance to be restored
after inflation in the framework of RSII, I have explicitly shown that the requirement of the sub-dominant feature
of large scale coherent magnetic field after inflation gives two fold non-trivial characteristic
constraints- on equation of state parameter ($w$) and the corresponding energy scale during
reheating ($\rho^{1/4}_{rh}$) epoch. Hence giving the proposal for avoiding the contribution
of back-reaction from the magnetic field I have established a bound on the generic reheating
characteristic parameter ($R_{rh}$) and its rescaled version ($R_{sc}$), to achieve large scale
magnetic field within the prescribed setup and further apply the CMB constraints as obtained from
recently observed Planck 2015 data and Planck+BICEP2+Keck Array joint constraints. Using all these
derived results I have shown that it is possible to put further stringent constraints on various classes
of large and small field inflationary models to break the degeneracy between various cosmological
parameters within the framework of RSII. Finally, I have studied the consequences from two specific models of brane inflation- monomial and hilltop,
after applying the constraints obtained from inflation and primordial magnetic field. 
}

\maketitle
\flushbottom

\section{Introduction}
Large scale magnetic fields are ubiquitously present across the entire universe. They are a major component of the
interstellar medium e.g. stars, galaxies and galactic clusters of galaxies~\footnote{Magnetic fields in galaxies have a strength
, ${\cal O}(5\times 10^{-6}-10^{-4})$ Gauss \cite{Chyzy:2004hr} and the detected strength within clusters of galaxies is,
${\cal O}(10^{-6}-10^{-5})$ Gauss \cite{Vogt:2003su}. }. It has been verified by different astronomical
observations, but their true origin is a big mystery of cosmology and astro-particle physics \cite{Grasso:2000wj,Giovannini:2003yn,Kandus:2010nw,Durrer:2013pga}. 
The proper origin and the limits of the magnetic fields within the range ${\cal O}(5\times 10^{-17}-10^{-14})$ Gauss \cite{Essey:2010nd} in the intergalactic medium
have been recently studied using combined constraints from the Atmospheric Cherenkov Telescopes and
the Fermi Gamma-Ray Space Telescope on the spectra of distant blazars.
The upper bound on primordial magnetic fields could be also obtained from the Cosmic
Microwave Background (CMB) and the Large Scale Structure (LSS) observations, and the current
upper bound is given by ${\cal O}(10^{-9})$ Gauss from Faraday rotations \cite{Kronberg:1993vk,Kahniashvili:2010wm} and the lower bound is fixed at ${\cal O}(10^{-15})$ Gauss by HESS and Fermi/LAT observations \cite{Tavecchio:2010mk,Neronov:1900zz,Dolag:2010ni}.
If the magnetic fields are originated in the early universe, then they mimics the role of seed for the observed
galactic and cluster magnetic field, as well as directly explain the origin of the magnetic fields present at the interstellar medium.
Among various possibilities, inflationary (primordial) magnetic field is one of the plausible candidates, through which the origin of cosmic magnetic field 
at the early universe can widely be explained. 
Within this prescribed setup, large
scale coherent magnetic fields and the primordial curvature perturbations are generated
from the quantum fluctuations. However explaining the origin of cosmic magnetic field via inflationary paradigm is
not possible in a elementary fashion, as in the context
of standard electromagnetic theory the action~\footnote{ In Eq~(\ref{q1r}), $F_{\mu\nu}$ is the electromagnetic field strength tensor, which is defined as, 
\be F_{\mu\nu}=\partial_{[\mu}A_{\nu]},\ee where $A_{\mu}$ is the $U(1)$ gauge field. }: \bea\label{q1r} S_{EM}&=&-\frac{1}{4}\int d^{4}x~\sqrt{-g}~g^{\alpha\mu}g^{\beta\nu}F_{\mu\nu}F_{\alpha\beta} \eea is conformally invariant. Consequently
in FLRW cosmological background for a comoving observer $u_{\nu}$
the magnetic field: 
\bea B^{\mu}=-\frac{1}{2}\epsilon^{\mu\nu\alpha\beta}u_{\nu}F_{\alpha\beta}=-{{}^*}~F^{\mu\nu}u_{\nu} \eea always decrease with the scale factor in a inverse square manner and implies the rapid decay of magnetic field during inflation.
 In a flat universe,
this issue can be resolved by breaking the conformal invariance of the electromagnetic theory
during inflationary epoch~\footnote{One of the simplest, gauge invariant 
model of inflationary magnetogenesis is described by the following effective action \cite{Ratra:1991bn}:\bea\label{q2r} S_{EM}&=&-\frac{1}{4}\int d^{4}x~\sqrt{-g}f^{2}(\phi)~g^{\alpha\mu}g^{\beta\nu}F_{\mu\nu}F_{\alpha\beta} \eea  where 
the conformal invariance of the $U(1)$ gauge field $A_{\mu}$ is broken by a time dependent function $f(\phi)(\propto a^{\alpha})$ of
inflaton $\phi$ and at the end of inflation \be f(\phi_{end})\rightarrow 1.\ee}. See 
refs.~\cite{Turner:1987bw,Ratra:1991bn,Dolgov:1993mu,Dolgov:1993vg,Ratra:1994vw,Tornkvist:2000js,Subramanian:2015lua,Atmjeet:2013yta,Atmjeet:2014cxa,Subramanian:2009fu,Byrnes:2011aa,Ferreira:2013sqa,Ferreira:2014hma,Jain:2012ga} for the further details of this issue. Due to the breaking of conformal invariance of the electromagnetic theory the magnetic field gets amplified. On the other hand, during inflation
the back-reaction effect of the electromagnetic field spoil the underlying picture. Also the theoretical origin and the specific technical details of the conformal invariance breaking mechanism makes the
back-reaction effect model dependent. However, in this paper, during the analysis it is assumed that after the end of inflation conformal invariance is restored in absence of source and the magnetic field
decrease with the scale factor in a inverse square fashion. Also by suppressing the effect of back-reaction after inflation, in this work, I derive various useful constraints on- reheating, leptogenesis and
baryogenesis in a model independent way~\footnote{Additionally it is important to mention here that the back-reaction problem is true for some class of inflationary models. But on the contrary there exist also 
many inflationary models in cosmology 
literature in which back-reaction is not at all a problem \cite{Byrnes:2011aa,Ferreira:2013sqa,Kanno:2009ei}. For completeness it is also
mention here that, in the original model proposed as in \cite{Ratra:1991bn}, back-reaction is not an big issue in the relevant part of the parameter space.}.

The prime objective of this paper is
to establish
a theoretical constraint for a generic class of large field ($|\Delta\phi|>M_{p}$~\footnote{Field excursion of the inflation filed is defined as: \be \Delta \phi=\phi_{cmb}-\phi_{end},\ee where $\phi_{cmb}$ represent the field value of the inflaton 
at the momentum scale $k$ which satisfies the equality, \be k=aH=-\eta^{-1}\approx k_{*},\ee where $(a,~H,~\eta)$ represent the scale factor 
, Hubble parameter, the conformal time and pivot momentum scale respectively. Also $\phi_{end}$ is the field value of the inflaton defined at the end of inflation.}) and small field ($|\Delta\phi|<M_{p}$) model of inflation to explain the origin of
primordial magnetic field in the framework of Randall-Sundrun braneworld gravity (RSII) 
\cite{Randall:1999ee,Randall:1999vf,Choudhury:2011sq,Choudhury:2011rz,Choudhury:2012ib,Choudhury:2014sua,Choudhury:2013yg,Choudhury:2013aqa,Choudhury:2013eoa,Himemoto:2000nd,Mukhopadhyaya:2002jn,Maartens:2010ar,Brax:2004xh}
from various probes: \begin{enumerate}
                      \item Tensor-to-scalar ratio ($r$),
                      \item Reheating,
                      \item Leptogenesis \cite{Fong:2013wr,Davidson:2008bu,Buchmuller:2005eh} and 
                      \item Baryogenesis \cite{Morrissey:2012db,Trodden:1998ym,Cline:2006ts,Allahverdi:2012ju}.
                     \end{enumerate}
Throughout the analysis of the paper
 I assume:
 \begin{enumerate}
  \item Inflaton 
 field $\phi$ is localized in the membrane of RSII set up and also minimally coupled to the gravity sector at the membrane in the absence of any electromagnetic interaction. 
 In this situation the representative action in RSII membrane set up can be expressed as:
 \bea \label{wasq1} S&=&\int d^{5}x \sqrt{-G}\left[\frac{M^3_5}{2}R_5 -2\Lambda_5 -\left\{\left(\frac{1}{2}\left(\partial\phi\right)^2+V(\phi)\right)+\sigma\right\}\delta(y)\right],~~~~~~~~~~~
 \eea
 where the extra dimension ``y'' is non-compact for which the covariant formalism is applicable. Here $M_5$ represents the 5D
quantum gravity cut-off scale, $\Lambda_5$ represents the 5D bulk cosmological constant, $\phi$ is the scalar inflaton localized 
at the brane and $\sqrt{-G}$ is the determinant of the 5D metric. It is important to mention that, the scalar
inflaton degrees of freedom is embedded on the 3 brane which has a positive brane tension $\sigma$ and it is localized at the
position of orbifold point $y=0$. The exact connecting relationship between $M_5$, $\Lambda_5$ and $\sigma$ is explicitly mentioned in the later section 
of this article. Also for the sake of simplicity, in the RSII membrane set-up, during cosmological analysis one can choose the following sets of parameters to be free:
\begin{itemize}
 \item 5D bulk cosmological constant $\Lambda_5$ is the most important parameter of RSII set up. Only the upper bound of $\Lambda_5$ is fixed 
 to validate the Effective Field Theory framework within the prescribed set up. Once I choose the value of $\Lambda_5$ below its upper bound 
 value, the other two parameters- 5D quantum gravity cut-off scale $M_5$ and the brane tension $\sigma$ is fixed from their
 connecting relationship as discussed later. In this paper, I fix the values of all of these RSII braneworld gravity model parameters by using 
 Planck 2015 data and Planck+BICEP2/Keck Array joint constraints.
 \item The rest of the free parameters are explicitly appearing through the structural form of the inflationary potential $V(\phi)$. For example 
 in this article I have studied the cosmological features from monomial and hilltop potential. For both the cases the characteristic index 
 $\beta$, which controls the structural form of the brane inflationary potential are usually considered to be the free parameter in the present context.
 Additionally, for both the potentials the tunable energy scale $V_0$ is also treated as the free parameter within RSII set up. Finally, the mass parameter
 $\mu$ can also be treated as the free parameter of hilltop potential. Most importantly, all of these parameters can be constrained by applying the 
 observational constraints obtained from Planck 2015 and Planck+BICEP2/Keck Array joint data.
\end{itemize}

 \item Once the contribution from the electromagnetic interaction is switched on at the RSII membrane, the inflaton field $\phi$ gets non-minimally coupled with gravity as well as $U(1)$ gauge fields as depicted in Eq~(\ref{q2r}).
 But for the clarity it is important to note that, in this paper I have not explicitly discussed the exact generation mechanism of inflationary magnetic field
 within the framework of RSII membrane paradigm. Most precisely, here I explicitly assume a preexisting magnetic field parametrized by an amplitude, spectral index and running of the magnetic power spectrum. Consequently the exact structural form 
 of the non-minimal coupling is not exactly known in terms of the RSII model parameters. Additionally, it is important to mention here that in the rest of the paper I assume that the initial magnetic field is originated through some 
 background mechanism during inflation in RSII membrane set up. Here the representative action in RSII membrane set up can be modified as:
 \bea \label{wasq2} S&=&\int d^{5}x \sqrt{-G}\left[\frac{M^3_5}{2}R_5 -2\Lambda_5 -\left\{\left(\frac{1}{2}\left(\partial\phi\right)^2+V(\phi)\right)\right.\right.\nonumber\\ &&\left.\left.~~~
 ~~~~~~~~~~~~~~~~~~~~~~~~~~~~~~~~~~~~+\frac{1}{4}f^{2}(\phi)~g^{\alpha\mu}g^{\beta\nu}F_{\mu\nu}F_{\alpha\beta}+\sigma\right\}\delta(y)\right],~~~~~~~~~~~
 \eea
 where $f(\phi)$ plays the role of inflaton field dependent non-minimal coupling in the present context.
 \item The conformal symmetry of the quantized version of the $U(1)$ gauge fields breaks down in curved space-time through which it is possible to generate sizable amount of magnetic field during the phase of single field 
 inflation. Conformal invariance is restored at the end of inflation such that the magnetic field decays as inverse square of the scale factor.
 \item Slow-roll prescription perfectly holds good for the RSII braneworld version of the inflationary paradigm.
 \item I also assume the instantaneous transitions between inflation, reheating, radiation and matter dominated
 epoch which involves entropy injection. In the prescribed framework specifically reheating phenomena is characterized by the following sets of parameters:
 \begin{itemize}
  \item Instantaneous equation of state 
 parameter: \be w({\cal N}_{b})=P({\cal N}_{b})/\rho({\cal N}_{b}),\ee where ${\cal N}_{b}$ is the number of e-foldings  and 
 $P({\cal N}_{b})$ and $\rho({\cal N}_{b})$ characterize the instantaneous pressure and energy density in RSII membrane set up.
 \item Mean equation of state parameter: 
 \be \bar{w}_{reh}=\frac{\int^{{\cal N}_{reh;b}}_{{\cal N}_{end;b}}w({\cal N}_{b})d{\cal N}_{b}}
 {\int^{{\cal N}_{reh;b}}_{{\cal N}_{end;b}}d{\cal N}_{b}},\ee
 where ${\cal N}_{reh;b}$ and ${\cal N}_{end;b}$ represent the number of e-foldings during reheating epoch and at the end of inflation respectively.
 \item Reheating energy density $\rho_{reh}$.
 \item Reheating temperature $T_{reh}$.
 \item Reheating parameter and its rescaled version:
 \bea R_{rad}&=&\left(\frac{\rho_{reh}}{\rho_{end}}\right)^{\frac{1-3\bar{w}_{reh}}{12\left(1+\bar{w}_{reh}\right)}},\\
      R_{sc}&=& R_{rad}\times \frac{\rho^{1/4}_{end}}{M_p},\eea
  where $\rho_{end}$ and $M_p$ represent the energy density at the end of inflation and 4D effective Planck mass. 
  \item Change of relativistic degrees of freedom between reheating and present epoch is characterized by a parameter ${\cal A}_{reh}$, which is explicitly defined 
  in the later section of this paper.  
 \end{itemize}
 \item Contribution from the correction coming from the non-relativistic neutrinos are negligibly small.
\item Initial condition for inflation is guided via the Bunch-Davies vacuum. 
\item The effective sound speed during inflation is fixed at $c_{S}=1$.
 \end{enumerate}
                     The plan of the paper is as follows.
                     \begin{itemize}
                      \item In the section \ref{a1}, I will explicitly mention the various parametrization of magnetic power spectrum and its cosmological implications.
                      \item In the section \ref{a2}, I will explicitly show that for all of these generic class of inflationary models it is possible to predict the
amount of magnetic field at the present epoch ($B_{0}$), by measuring
non-vanishing CP asymmetry ($\epsilon_{CP}$) in leptogenesis and baryon asymmetry ($\eta_{B}$) in baryogenesis or the tensor-to-scalar
ratio.
\item In this paper I
use various constraints arising from Planck 2015 data on the 
amplitude of scalar power spectrum, scalar spectral tilt, the upper bound on tenor to scalar ratio, lower bound on rescaled characteristic reheating parameter
and the bound on the reheating energy density
within $1.5\sigma-2\sigma$ statistical CL.

\item I also mention that the GR limiting result ($\rho<<\sigma$) and the difference between the high energy
  limit result ($\rho>>\sigma$) of RSII.
  
  \item  Further assuming the conformal invariance to be restored after inflation in the framework of Randall-Sundrum single
braneworld gravity (RSII), I will show that the requirement of the 
sub-dominant feature of large scale magnetic field after inflation gives two fold
non-trivial characteristic constraints- on equation of state parameter ($w$) and the corresponding 
energy scale during reheating ($\rho^{1/4}_{rh}$) epoch in section \ref{a2}.

\item Hence in section \ref{a3} and \ref{a4}, avoiding the contribution of back-reaction from the magnetic field, I have established
a bound on the reheating characteristic
parameter ($R_{rh}$) and its rescaled version ($R_{sc}$), to achieve large scale magnetic field within the prescribed setup
and apply the Cosmic Microwave Background (CMB) constraints as obtained from recent Planck 2015 data \cite{Ade:2015cva,Planck:2015xua,Ade:2015lrj} and the joint constraint obtained from Planck+BICEP2+Keck Array \cite{Ade:2015tva}. 

\item Finally in section \ref{a5}, I will explicitly 
study the cosmological consequences from two specific models of 
brane inflation- monomial (large field) and hilltop (small field), after applying all the constraints obtained in this paper. 

\item Moreover, by doing parameter estimation from both of these simple class of models, I will explicitly 
show the magneto-reheating constraints can be treated as one of the probes through which one can distinguish between the prediction from both of these inflationary models.
                     \end{itemize}

\section{Parametrization of magnetic power spectrum}
\label{a1}
A Gaussian random magnetic field for a statistically
homogeneous and isotropic system is described by the equal time two-point correlation function 
in momentum space as \cite{Byrnes:2011aa}:
\be\begin{array}{lll}\label{eq1}
    \displaystyle \langle B^{*}_{i}({\bf k},\eta)B_{j}({\bf k^{'}},\eta)\rangle=(2\pi)^{3}\delta^{(3)}({\bf k}-{\bf k^{'}})
{\cal P}_{ij}(\hat{{\bf k}})\frac{2\pi^2}{k^3}
P_{\bf B}(k),
   \end{array}\ee
where $P_{\bf B}(k)$ represents the magnetic power spectrum~\footnote{It is important to note that here for magnetic power spectrum equivalently one can use the following definition of two-point correlation function \cite{Shiraishi:2010kd,Shiraishi:2012rm}:
\be\begin{array}{lll}\label{eqqw1}
    \displaystyle \langle B^{*}_{i}({\bf k},\eta)B_{j}({\bf k^{'}},\eta)\rangle=(2\pi)^{3}\delta^{(3)}({\bf k}-{\bf k^{'}})
{\cal P}_{ij}(\hat{{\bf k}})
\bar{P}_{\bf B}(k),
   \end{array}\ee
   where $\bar{P}_{\bf B}(k)$ is a magnetic power spectrum.} and ${\cal P}_{ij}(\hat{{\bf k}})$ characterize
the dimensionless plane projector onto the transverse
plane is defined as \cite{Shiraishi:2010kd,Shiraishi:2012rm}:
\be\begin{array}{lllll}\label{eq2}
   \displaystyle {\cal P}_{ij}(\hat{{\bf k}})=\sum_{\lambda=\pm 1}e^{\lambda}_{i}(\hat{{\bf k}})
e^{-\lambda}_{j}(\hat{{\bf k}})=(\delta_{ij}-\hat{\bf k}_{i}\hat{\bf k}_{j})
   \end{array}\ee
in which the divergence-free nature of the magnetic field is imposed via the orthogonality condition, \be \hat{\bf k}^{i}\epsilon^{\pm 1}_{i}=0.\ee
Here $\hat{\bf k}_{i}$ signifies the unit vector which can be expanded in terms of spin spherical harmonics. See ref.~\cite{Shiraishi:2010kd} for the details of the useful properties of the projection tensors of magnetic modes.
Additionally, it is worthwhile to mention that in the present context,
$P_{\bf B}(k)$ be the   
part of the power spectrum
for the primordial magnetic
field which will only contribute to the cosmological
perturbations for the scalar modes and the Faraday Rotation at the phase of decoupling~\footnote{It is important to mention here that, the exact form of the magnetic power power spectrum strongly 
depends on the production mechanism of primordial magnetic field within RSII membrane setup, which I have not studied in this paper.}.

The non-helical part of the primordial magnetic power spectrum is parameterized within the upper and lower cut-off momentum scale ($k_{L}\leq k\leq k_{\Lambda}$) as~\footnote{It is important to note
that here if I start with Eq~(\ref{eqqw1}), then equivalently one can use the following parametrization of magnetic power spectrum $\bar{P}_{\bf B}(k)$:
\be\begin{array}{llll}\label{eq3aa}
\displaystyle \bar{P}_{\bf B}(k) =\frac{k^{3}}{2\pi^2}
P_{\bf B}(k)=\left\{\begin{array}{ll}
                    \displaystyle \bar{A}_{\bf B}~\left(\frac{k}{k_{*}}\right)^{3} &
 \mbox{\small {\bf for \underline{Case A}}}  
\\ 
         \displaystyle  \bar{A}_{\bf B}~\left(\frac{k}{k_{*}}\right)^{n_{\bf B}+3}~~~~~~~~~~ & \mbox{\small {\bf for \underline{Case B}}}
         \\ 
         \displaystyle  \bar{A}_{\bf B}~\left(\frac{k}{k_{*}}\right)^{n_{\bf B}+3
+\frac{\alpha_{\bf B}}{2}\ln \left(\frac{k}{k_{*}}\right)}~~~~~~~~~~ & \mbox{\small {\bf for \underline{Case C}}}
\\ 
         \displaystyle  \bar{A}_{\bf B}~\left(\frac{k}{k_{*}}\right)^{n_{\bf B}+3
+\frac{\alpha_{\bf B}}{2}\ln \left(\frac{k}{k_{*}}\right) +\frac{\kappa_{\bf B}}{6}\ln^{2} \left(\frac{k}{k_{*}}\right)}~~~~~~~~~~ & \mbox{\small {\bf for \underline{Case D}}}
          \end{array}
\right.
\end{array}         
\ee
   where $\bar{A}_{\bf B}$ represents the amplitude of the magnetic power spectrum defined as:
   \be\label{vbvb} \bar{A}_{\bf B}=\frac{{A}_{\bf B}}{2\pi^2}k^{3}_{*}.\ee
   Here ${A}_{\bf B}$ characterizes the amplitude of the magnetic power spectrum as defined in Eq~(\ref{eq3}) and $k_{*}$ be the pivot scale of momentum. But instead of using the above structure of 
   magnetic power spectrum in the rest of the paper I use the parametrization of the magnetic power spectrum mentioned in Eq~(\ref{eq3}).} \cite{Choudhury:2014hua}:
\be\begin{array}{llll}\label{eq3}
\displaystyle P_{\bf B}(k) =\left\{\begin{array}{ll}
                    \displaystyle A_{\bf B} &
 \mbox{\small {\bf for \underline{Case I}}}  
\\ 
         \displaystyle  A_{\bf B}~\left(\frac{k}{k_{*}}\right)^{n_{\bf B}}~~~~~~~~~~ & \mbox{\small {\bf for \underline{Case II}}}
         \\ 
         \displaystyle  A_{\bf B}~\left(\frac{k}{k_{*}}\right)^{n_{\bf B}
+\frac{\alpha_{\bf B}}{2}\ln \left(\frac{k}{k_{*}}\right)}~~~~~~~~~~ & \mbox{\small {\bf for \underline{Case III}}}
\\ 
         \displaystyle  A_{\bf B}~\left(\frac{k}{k_{*}}\right)^{n_{\bf B}
+\frac{\alpha_{\bf B}}{2}\ln \left(\frac{k}{k_{*}}\right) +\frac{\kappa_{\bf B}}{6}\ln^{2} \left(\frac{k}{k_{*}}\right)}~~~~~~~~~~ & \mbox{\small {\bf for \underline{Case IV}}}
          \end{array}
\right.
\end{array}         
\ee
where $A_{\bf B}$ represents the amplitude of the magnetic power spectrum,
 $n_{\bf B}$ is the magnetic spectral tilt, $\alpha_{\bf B}$ is the running and $\kappa_{\bf B}$ be the running of the magnetic spectral tilt. Here the upper
 cut-off momentum scale ($k_{\Lambda}$) corresponds
to the Alfv$\acute{e}$n wave damping length-scale, representing the dissipation of magnetic energy due to
the generation of magneto-hydrodynamic (MHD) waves. Additionally, $k_{*}$ being the pivot or normalization scale of momentum. Now let me briefly discuss the physical significance of the 
above mentioned four possibilities~\footnote{ If one follows the convention as stated in Eq~(\ref{eq3aa}), the physical interpretation of the magnetic power spectrum parametrization 
for the four possibilities are changed as:
\begin{itemize}
                        \item \underline{\bf Case A} stands for a physical situation where the magnetic power spectrum is scale dependent and follows the cubic power law,
\item  \underline{\bf Case B} stands for a physical situation where the magnetic power spectrum follows power law $(n_{\bf B}+3)$ feature 
in presence of magnetic spectral tilt $n_{\bf B}$. In this case the scale invariant power spectrum can be achieved when we take $n_{\bf B}=-3$.
\item \underline{\bf Case C} signifies a physical situation where the magnetic power spectrum shows deviation from power law behaviour in presence of running of the
 magnetic spectral tilt $\alpha_{\bf B}$ and
\item \underline{\bf Case D} characterizes
 a physical situation in which the magnetic power spectrum is further modified 
by allowing running of the running of the magnetic spectral tilt $\kappa_{\bf B}$.
                       \end{itemize}
        However for all the cases the amplitude $\bar{A}_{\bf B}$
is pivot scale dependent by following the relation stated in Eq~(\ref{vbvb}).                }:
\begin{itemize}
                        \item \underline{\bf Case I} stands for a physical situation where the magnetic power spectrum is exactly scale invariant and it is characterized by $n_{\bf B}=0$,
\item  \underline{\bf Case II} stands for a physical situation where the magnetic power spectrum follows power law feature 
in presence of magnetic spectral tilt $n_{\bf B}$,
\item \underline{\bf Case III} signifies a physical situation where the magnetic power spectrum shows deviation from power law behaviour in presence of running of the
 magnetic spectral tilt $\alpha_{\bf B}$ along with logarithmic correction in the momentum scale (as appearing in the exponent) and
\item \underline{\bf Case IV} characterizes
 a physical situation in which the magnetic power spectrum is further modified compared to the \underline{\bf Case III},
by allowing running of the running of the magnetic spectral tilt $\kappa_{\bf B}$ along with square of the momentum dependent logarithmic correction.
                       \end{itemize}
\begin{figure*}[htb]
\centering
\subfigure[$P_{S}$ vs $n_{S}$.]{
    \includegraphics[width=7.2cm,height=8.3cm] {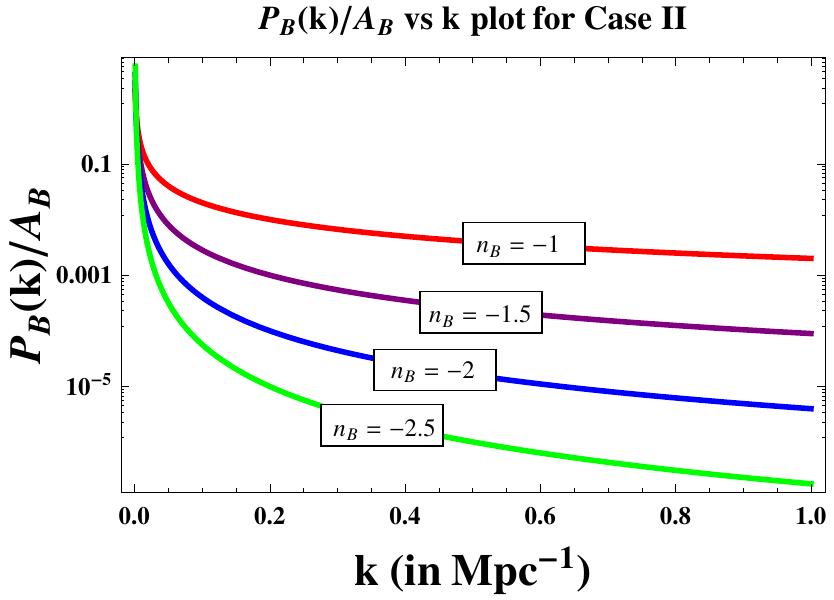}
    \label{f1}
}
\subfigure[$P_{S}$ vs $\beta$.]{
    \includegraphics[width=7.2cm,height=8.3cm] {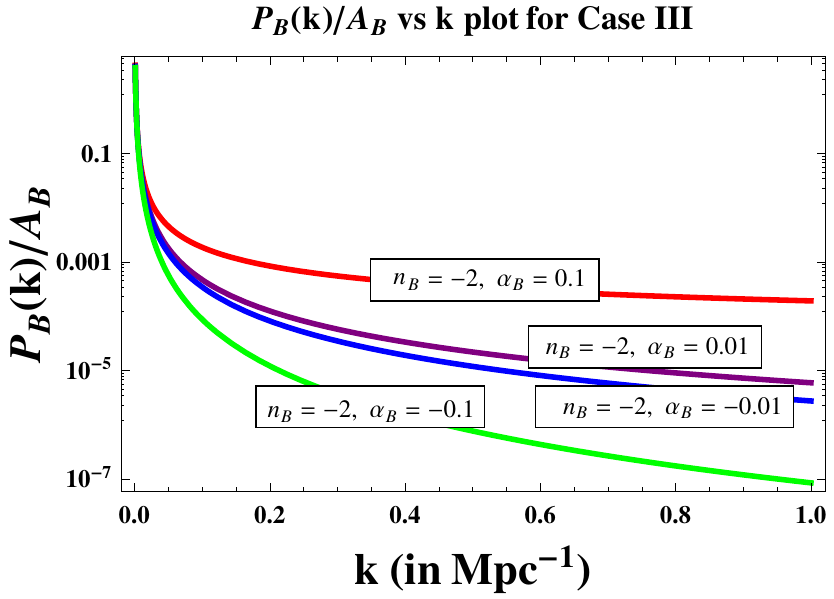}
    \label{f2}
}
\subfigure[$n$ vs $\beta$.]{
    \includegraphics[width=12.8cm,height=10cm] {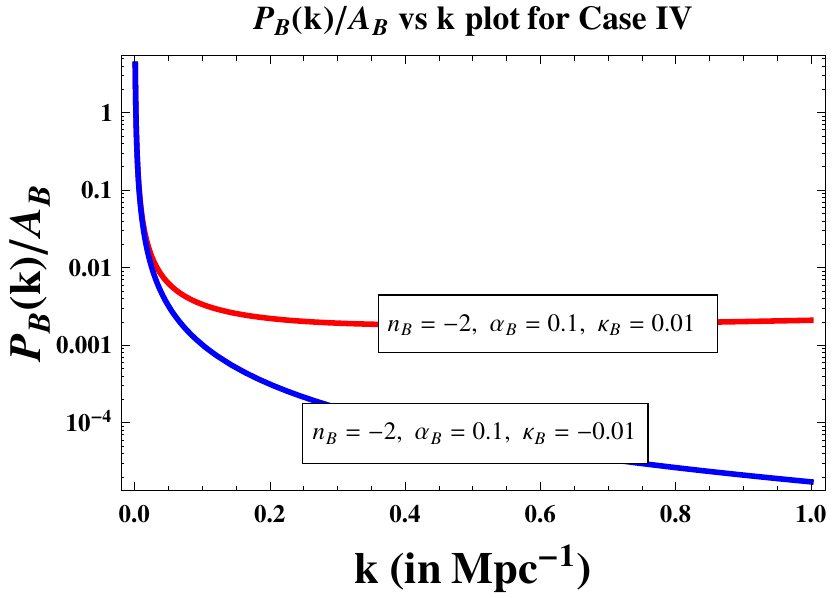}
    \label{f3}
}
\caption[Optional caption for list of figures]{Variation of the magnetic power spectrum with respect to momentum scale $k$ for 
\ref{f1} $n_{\bf B}<0, \alpha_{\bf B}=0, \kappa_{\bf B}=0$, \ref{f2}$ n_{\bf B}<0, \alpha_{\bf B}\neq 0, \kappa_{\bf B}=0$
and \ref{f3} $n_{\bf B}<0, \alpha_{\bf B}\neq 0, \kappa_{\bf B}\neq 0$.} 
\label{f1vv1}
\end{figure*}
In fig.~\ref{f1}-fig.~\ref{f3} by following the convention stated in Eq~(\ref{eq1}), I have explicitly shown the variation of the magnetic power spectrum with respect to momentum scale $k$ for 
 \begin{enumerate}
  \item $n_{\bf B}<0, \alpha_{\bf B}=0, \kappa_{\bf B}=0$,
  \item $n_{\bf B}<0, \alpha_{\bf B}\neq 0, \kappa_{\bf B}=0$ and 
  \item $n_{\bf B}<0, \alpha_{\bf B}\neq 0, \kappa_{\bf B}\neq 0$ respectively.
 \end{enumerate}
It is important to note that the most recent observational constraint
from CMB temperature anisotropies on the amplitude
and the spectral index of a primordial magnetic field has
been predicted by using Planck 2015 data as~\footnote{Here $B_{1~ Mpc}$ represents the comoving field amplitude at a scale
of $1~{\rm  Mpc}$.} \cite{Ade:2015cva,Planck:2015xua}
\be B_{1~ {\rm Mpc}} < 4.4 {\rm nG}\ee with magnetic spectral tilt \be n_{\bf B} < 0\ee at $2\sigma$ CL.
If, in near future, Planck or any other observational probes can predict the signatures for $\alpha_{\bf B}$ and $\kappa_{\bf B}$ in the primordial
magnetic power spectrum (as already predicted in case of primordial scalar power spectrum within $1.5-2\sigma$ CL \cite{Ade:2015lrj}), then it is possible to put 
further stringent constraint on
the various models of inflation. 
\section{Constraint on inflationary magnetic field from leptogenesis and baryogenesis}
\label{a2}
In the present section, I am interested in the mean square amplitude of the primordial magnetic field on a given characteristic scale $\xi$, on which I smooth the magnetic power
spectrum using a Gaussian filter~\footnote{In standard prescriptions, Gaussian filter is characterized by a Gaussian window function $\exp\left(-k^{2}\xi^{2}\right)$, defined in a characteristic scale $\xi$. } as given by \cite{Shiraishi:2012rm}:
\be\begin{array}{lll}\label{eq4}
    \displaystyle B^{2}_{\xi}=\langle B_{i}({\bf x})B_{i}({\bf x})\rangle_{\xi}
=\frac{1}{2\pi^{2}}\int^{\infty}_{0}\frac{dk}{k_{*}}~\left(\frac{k}{k_{*}}\right)^{2}P_{\bf B}(k)\exp\left(-k^{2}\xi^{2}\right).
   \end{array}\ee
Here in \underline{\bf Case III} and \underline{\bf Case IV} of Eq~(\ref{eq3}) describes a more generic picture where 
the magnetic power spectrum deviates from its exact power law form in presence of logarithmic correction. Consequently, the resulting 
mean square primordial magnetic field is logarithmically divergent in both the limits of the integral as presented in Eq~(\ref{eq4}).
But in \underline{\bf Case I} and \underline{\bf Case II} of Eq~(\ref{eq3}) no such divergence is appearing.
To remove the divergent contribution from the mean square amplitude of the primordial magnetic field as appearing in \underline{\bf Case III} and \underline{\bf Case IV} of Eq~(\ref{eq4}), 
I introduce here cut-off regularization technique in which I have re-parameterized the integral in terms of
regulated UV (high) and IR (low) momentum scales. Most importantly, for the sake of completeness in all four cases, here I introduce the high and low 
cut-offs $k_{\Lambda}$ and $k_{L}$ are momentum regulators to collect
only the finite contributions from Eq~(\ref{eq4}). Finally I get the following expression for the regularized magnetic field:
\be\begin{array}{lll}\label{eq5}
  \displaystyle B^{2}_{\xi}(k_{L};k_{\Lambda})
 =\frac{I_{\xi}(k_{L};k_{\Lambda})}{2\pi^{2}}A_{\bf B}
 \end{array}\ee
 where  
 \be\begin{array}{lll}\label{eq5a}
 I_{\xi}(k_{L};k_{\Lambda})=
\left\{\begin{array}{ll}
                    \displaystyle \int^{k_{\Lambda}}_{k_{L}}\frac{dk}{k_{*}}~\exp\left(-k^{2}\xi^{2}\right)\left(\frac{k}{k_{*}}\right)^{2}&
 \mbox{\small {\bf for \underline{Case I}}}  
\\ 
         \displaystyle \int^{k_{\Lambda}}_{k_{L}}\frac{dk}{k_{*}}~\exp\left(-k^{2}\xi^{2}\right)
\left(\frac{k}{k_{*}}\right)^{n_{\bf B}+2} & \mbox{\small {\bf for \underline{Case II}}}
         \\ 
         \displaystyle  \int^{k_{\Lambda}}_{k_{L}}\frac{dk}{k_{*}}~\exp\left(-k^{2}\xi^{2}\right)
\left(\frac{k}{k_{*}}\right)^{n_{\bf B}+2
+\frac{\alpha_{\bf B}}{2}\ln \left(\frac{k}{k_{*}}\right)} & \mbox{\small {\bf for \underline{Case III}}}
\\ 
         \displaystyle  \int^{k_{\Lambda}}_{k_{L}}\frac{dk}{k_{*}}~\exp\left(-k^{2}\xi^{2}\right)
\left(\frac{k}{k_{*}}\right)^{n_{\bf B}+2
+\frac{\alpha_{\bf B}}{2}\ln \left(\frac{k}{k_{*}}\right) +\frac{\kappa_{\bf B}}{6}\ln^{2} \left(\frac{k}{k_{*}}\right)}& \mbox{\small {\bf for \underline{Case IV}}}.
          \end{array}
\right.
   \end{array}\ee
The exact expression for the regularized integral function $I_{\xi}(k_{L};k_{\Lambda})$ are explicitly mentioned in the appendix \ref{a7} for all four cases.
It is important to mention here that, for \underline{\bf Case I} and \underline{\bf Case II}, $I_{\xi}(k_{L}\rightarrow 0;k_{\Lambda}\rightarrow \infty)$ is finite. But for rest of the two cases, $I_{\xi}(k_{L}\rightarrow 0;k_{\Lambda}\rightarrow \infty)\rightarrow \infty$.
On the other hand, in absence of any Gaussian filter, the magnetic energy density can be expressed in terms of the
 mean square primordial magnetic field as \cite{Shiraishi:2012rm}:
\be\begin{array}{lll}\label{eq6}
    \displaystyle \rho_{\bf B}=\frac{1}{8\pi}\langle B_{i}({\bf x})B_{i}({\bf x})\rangle
=\frac{1}{8\pi^{2}}\int^{\infty}_{0}\frac{dk}{k_{*}}~\left(\frac{k}{k_{*}}\right)^{2}P_{\bf B}(k)
   \end{array}\ee
which is logarithmically divergent in UV and IR end for \underline{\bf Case III} and \underline{\bf Case IV}. For rest of the two cases also the contribution become divergent, but the behaviour of the divergences are different 
compared to the \underline{\bf Case III} and \underline{\bf Case IV}. After introducing the momentum cut-offs as mentioned earlier, I get the  following expression for the 
regularized magnetic energy density as:
\be\begin{array}{lll}\label{eq7}
    \displaystyle \rho_{\bf B}(k_{L};k_{\Lambda})=\frac{J(k_{L};k_{\Lambda})}{8\pi^2}A_{\bf B}=\frac{J(k_{L};k_{\Lambda})B^{2}_{\xi}(k_{L};k_{\Lambda})}{4I_{\xi}(k_{L};k_{\Lambda})}
    \end{array}\ee 
    where 
    \be\begin{array}{lll}\label{eq7q}
    \displaystyle J(k_{L};k_{\Lambda})=
\left\{\begin{array}{ll}
                    \displaystyle \int^{k_{\Lambda}}_{k_{L}}\frac{dk}{k_{*}}~\left(\frac{k}{k_{*}}\right)^{2}~~~&
 \mbox{\small {\bf for \underline{Case I}}}  
\\ 
         \displaystyle \int^{k_{\Lambda}}_{k_{L}}\frac{dk}{k_{*}}~\left(\frac{k}{k_{*}}\right)^{n_{\bf B}+2}~~~ & \mbox{\small {\bf for \underline{Case II}}}
         \\ 
         \displaystyle  \int^{k_{\Lambda}}_{k_{L}}\frac{dk}{k_{*}}~\left(\frac{k}{k_{*}}\right)^{n_{\bf B}+2
+\frac{\alpha_{\bf B}}{2}\ln \left(\frac{k}{k_{*}}\right)}~~~ & \mbox{\small {\bf for \underline{Case III}}}
\\ 
         \displaystyle  \int^{k_{\Lambda}}_{k_{L}}\frac{dk}{k_{*}}~\left(\frac{k}{k_{*}}\right)^{n_{\bf B}+2
+\frac{\alpha_{\bf B}}{2}\ln \left(\frac{k}{k_{*}}\right) +\frac{\kappa_{\bf B}}{6}\ln^{2} \left(\frac{k}{k_{*}}\right)}~~~ & \mbox{\small {\bf for \underline{Case IV}}}
          \end{array}
\right.
   \end{array}\ee
where I use Eq~(\ref{eq5}). Here the regularized integral function $J(k_{L};k_{\Lambda})$ are explicitly written in the appendix \ref{a8} for all four possibilities.

Now to derive a phenomenological constraint here I further assume the fact that
 the primordial magnetic field is made up of relativistic degrees of freedom.
In this physical prescription, the regularized magnetic energy
density can be expressed as \cite{Long:2013tha}:
\be\begin{array}{llll}\label{eq8}
    \displaystyle \rho_{\bf B}(k_{L};k_{\Lambda})\sim \frac{\pi^{2}}{30}g_{*}T^{4}\sim{\cal O}(10^{-13})
\times \frac{T^{4}}{\epsilon_{\bf CP}}
   \end{array}\ee
where the CP asymmetry parameter $\epsilon_{\bf CP}$ is defined as:
 \be\begin{array}{lll}\label{eq9cvx}
     \displaystyle \epsilon_{\bf CP}=\frac{\Gamma_{L}(N_{R}\rightarrow L_{i}\Phi)-\Gamma_{L^c}(N_{R}\rightarrow L^{c}_{i}\Phi^{c})
}{\Gamma_{L}(N_{R}\rightarrow L_{i}\Phi)+\Gamma_{L^c}(N_{R}\rightarrow L^{c}_{i}\Phi^{c})}\approx {\cal O}(|\lambda|^2)\sin\theta_{\bf CP}
    \end{array}\ee
for the standard leptogenesis scenario \cite{Fukugita:1986hr,Buchmuller:2005eh} where the Majorana neutrino ($N_{R}$) decays through Yukawa matrix interaction ($\lambda$) with the 
Higgs ($\Phi$) and lepton ($L$) doublets. Here $\theta_{\bf CP}$ is the CP-violating phase and for heavy majorana neutrino ($N_{R}$) mass \be M_{N_{R}}\sim 10^{10}~{GeV}\ee the Yukawa coupling is given by, \be |\lambda|^2= {\cal O}(10^{-16}).\ee
Now combining Eq~(\ref{eq7}) and Eq~(\ref{eq8}), I derive the following simplified expression for the root mean square
value of the primordial magnetic field at the present epoch in terms of the CP asymmetry parameter ($\epsilon_{\bf CP}$) as:
\be\begin{array}{lll}\label{eq11}
 \displaystyle B_{0}\sim
{\cal O}(10^{-14})\times\sqrt{\frac{I_{\xi}(k_{L}=k_{0};k_{\Lambda})}{J(k_{L}=k_{0};k_{\Lambda})\epsilon_{\bf CP}}}~{\rm Gauss}
\end{array}\ee
where I use the temperature at the present epoch \be T_{0}\sim 2\times 10^{-4}~{\rm eV}\ee and \be 1~{\rm Gauss}=7\times 10^{-20}~{\rm GeV^{2}}.\ee
In addition, here in this paper, I fix the IR cut-off scale of the momentum at the present epoch i.e. $k_{L}=k_{0}$. Consequently the momentum integrals satisfy the following constraint:
\be \label{cont}\sqrt{\frac{I_{\xi}(k_{L}=k_{0};k_{\Lambda})}{J(k_{L}=k_{0};k_{\Lambda})}}\sim 10^{-8}.\ee
Further using Eq~(\ref{eqintk1}) and Eq~(\ref{eqintk11}) in Eq~(\ref{cont}) one can write the following constarints for all four cases of the parametrization of magnetic power spectrum as:
\bea\label{cxz1}
\underline{\bf Case~ I}: ~~~~~ k_{*}&\sim & {\cal O}(8.17\times 10^{-9})\times \sqrt{\frac{\sqrt{\xi}\left[k^{3}_{\Lambda}-k^{3}_{L}\right]}{\sqrt{\pi} \left[{\rm erf(\xi k_{\Lambda})}-{\rm erf(\xi k_{L})}\right]}},\eea 
\bea\label{cxz2}
\underline{\bf Case~ II}: ~ \frac{1}{\xi^{n_B+3}} &\sim & {\cal O}(2\times 10^{-16})\times\frac{\left[k^{n_{B}+3}_{\Lambda}-k^{n_{B}+3}_{L}\right]}{(n_B+3)\left[\Gamma \left(\frac{\left(n_B+3\right)}{2} ,\xi ^2 k^2_{L}\right)
         -\Gamma \left(\frac{\left(n_B+3\right)}{2} ,\xi ^2 k^2_{\Lambda}\right)\right]},~~~~~~~~~~~
\eea
\be\begin{array}{lll}\label{cxz3}
\underline{\bf Case~ III}: ~ \displaystyle \left[\frac{\sqrt{\pi}~{\rm erf}\left(\xi k\right)}{2\xi}\left\{1+{\cal Q}\ln\left(\frac{k}{k_{*}}\right)
+{\cal P}\ln^{2}\left(\frac{k}{k_{*}}\right)\right\}\right.\\ \left. 
\displaystyle~~~~~~~~~~~~~~~~+k
\left\{2{\cal P}~_PF_Q\left[\left\{\frac{1}{2},\frac{1}{2},\frac{1}{2}\right\}~;\left\{\frac{3}{2},\frac{3}{2},\frac{3}{2}\right\}~;
-\xi^{2}k^{2}\right]
\right.\right.\\ \left.\left.\displaystyle 
~~~~~~~~~~~~~~~~~~
\displaystyle -\left({\cal Q}+2{\cal P}\ln\left(\frac{k}{k_{*}}\right)\right)
~_PF_Q\left[\left\{\frac{1}{2},\frac{1}{2}\right\}~;\left\{\frac{3}{2},\frac{3}{2}\right\}~;
-\xi^{2}k^{2}\right]\right\}\right]^{k=k_{\Lambda}}_{k=k_{L}}\\
\displaystyle \sim {\cal O}(10^{-16})\times \left\{k\left[(1+2{\cal P}-{\cal Q})+({\cal Q}-2{\cal P})\ln\left(\frac{k}{k_{*}}\right)
+{\cal P}\ln^{2}\left(\frac{k}{k_{*}}\right)\right]\right\}^{k=k_{\Lambda}}_{k=k_{L}} ,~~~~~~~~~~~
\end{array}\ee
\be\begin{array}{lll}\label{cxz4}
\underline{\bf Case~ IV}: ~ \displaystyle \left[\frac{\sqrt{\pi}~{\rm erf}\left(\xi k\right)}{2\xi k_{*}}\left\{1+{\cal Q}\ln\left(\frac{k}{k_{*}}\right)
+{\cal P}\ln^{2}\left(\frac{k}{k_{*}}\right)+{\cal F}\ln^{3}\left(\frac{k}{k_{*}}\right)\right\}\right.\\ \left. 
\displaystyle~~~~~~~~~~~~+\left(\frac{k}{k_{*}}\right)
\left\{-6{\cal F}~_PF_Q\left[\left\{\frac{1}{2},\frac{1}{2},\frac{1}{2},\frac{1}{2}\right\}~;\left\{\frac{3}{2},\frac{3}{2},\frac{3}{2},\frac{3}{2}\right\}~;
-\xi^{2}k^{2}\right]\right.\right.\\ \left.\left.\displaystyle 
~~~~~~~~~~~+2\left({\cal P}+3{\cal F}\ln\left(\frac{k}{k_{*}}\right)\right)
~_PF_Q\left[\left\{\frac{1}{2},\frac{1}{2},\frac{1}{2}\right\}~;\left\{\frac{3}{2},\frac{3}{2},\frac{3}{2}\right\}~;
-\xi^{2}k^{2}\right]\right.\right.\\ \left.\left.\displaystyle 
~~~~~~~~~~~~
\displaystyle -\left({\cal Q}+2{\cal P}\ln\left(\frac{k}{k_{*}}\right)+6{\cal F}\ln^{2}\left(\frac{k}{k_{*}}\right)\right)\right.\right.\\ \left.\left.\displaystyle
~~~~~~~~~~~~~~~~~~~\times~_PF_Q\left[\left\{\frac{1}{2},\frac{1}{2}\right\}~;\left\{\frac{3}{2},\frac{3}{2}\right\}~;
-\xi^{2}k^{2}\right]\right\}\right]^{k=k_{\Lambda}}_{k=k_{L}}\\
\displaystyle ~~~~~~~~\sim {\cal O}(10^{-16})\times \left\{\left(\frac{k}{k_{*}}\right)\left[(1-6{\cal F}+2{\cal P}-{\cal Q})+(6{\cal F}-2{\cal P}+{\cal Q})\ln\left(\frac{k}{k_{*}}\right)
\right.\right.\\ \left.\left.\displaystyle~~~~~~~~~~~~~~~~~~~~~~~~~~~~-(3{\cal F}-{\cal P})\ln^{2}\left(\frac{k}{k_{*}}\right)+{\cal F}\ln^{3}\left(\frac{k}{k_{*}}\right)\right]\right\}^{k=k_{\Lambda}}_{k=k_{L}} ,~~~~~~~~~~~
\end{array}\ee
where \bea {\cal Q}&=&n_{ B}+2,\\ {\cal P}&=&\alpha_{ B}/2,\\ {\cal F}&=&\kappa_{ B}/6.\eea
The conformal symmetry of the quantized electromagnetic field breaks down in curved space-time which 
is able to generate a sizable amount of magnetic field during a phase
of slow-roll inflation. Such primordial magnetism is characterized by the renormalized mean square amplitude 
 of the primordial magnetic field at leading order in slow-roll approximation for comoving observers as \cite{Agullo:2013tba}:
\be\begin{array}{llll}\label{eq12}
    \displaystyle  \rho_{\bf B}(k_{L};k_{\Lambda})=\frac{1}{8\pi}\langle B_{i}({\bf x})B_{i}({\bf x})\rangle \approx\frac{V^{4}(\phi)\epsilon_{b}(\phi)}{8640\pi^{3}M^{4}_{p}\sigma^2}
   \end{array}\ee
where $V(\phi)$ represents the inflationary potential, $\sigma$ represents the brane tension of RSII setup and  $M_{p}\sim 2.43 \times 10^{18}~{\rm GeV}$ be the four dimensional reduced Planck mass. 
Within RSII setup the visible brane tension $\sigma$ can be expressed as \cite{Choudhury:2014sua}:
\begin{eqnarray}\label{lam}
 \sigma&=&\sqrt{-\frac{3}{4\pi}M^{3}_{5}\Lambda_{5}}=
\sqrt{-24M^{3}_{5}\tilde{\Lambda}_{5}}>0
\end{eqnarray}
where $\tilde{\Lambda}_{5}$ be the scaled 5D bulk cosmological constant defined as \cite{Choudhury:2014sua}:
\be\label{lamc}
\tilde{\Lambda}_{5}= \frac{\Lambda_{5}}{32\pi}<0.
\ee
Also the 5D quantum gravity cut-off scale can be expressed in terms of 5D cosmological constant and the 4D effective Planck scale as:
\begin{eqnarray}\label{mass1}
M^{3}_{5}&=&\sqrt[3]{-\frac{4\pi\Lambda_{5}}{3}}M^{4/3}_{p}=\sqrt[3]{-\frac{128\pi^2\tilde{\Lambda}_{5}}{3}}M^{4/3}_{p}.
\end{eqnarray}
In the high energy regime the energy density $\rho>>\sigma$ the slow-roll parameter $\epsilon_{b}(\phi)$ in the visible brane can be expressed as \cite{Choudhury:2014sua}:
\begin{eqnarray}
 \epsilon_{b}(\phi)&\approx& \frac{2M^{2}_{p}\sigma (V^{'}(\phi))^{2}}{V^{3}(\phi)}.
\end{eqnarray}
It is important to note that Eq~(\ref{eq12}) is insensitive to
the intrinsic ambiguities of renormalization in curved space-times. See the appendix where I have mentioned the inflationary consistency conditions within RSII setup.
Around the pivot scale $k=k_{*}$ I can write:
 \be \epsilon_{b}(k_{*})\approx \frac{r(k_{*})}{24}+\cdots,\ee where $\cdots$ includes the all the higher
 order slow-roll contributions. Here $r=P_T/P_S$ represents the tensor-to-scalar ratio.
The recent observations from Planck (2013 and
2015) and Planck+BICEP2+Keck Array puts an upper bound on  the amplitude of {\it primordial gravitational waves}
via tensor-to-scalar ratio. This bounds the potential energy stored in the inflationary potential within RSII setup as \cite{Choudhury:2014sua}:
\be\begin{array}{lll}\label{iopk}
\displaystyle \sqrt[4]{{ V}_{inf}}\approx\sqrt[12]{2\pi^2 P_{S}(k_{*})r(k_{*})}M^{1/3}_{p}\sigma^{1/6}\lesssim \sqrt[4]{\frac{3}{2}P_{S}(k_{*})r(k_{*})\pi^{2}}~M_p\\
~~~~~~~~~~~~~~~~~~~~~~~~~~~~~~~~~~~~~~~~~~~~~~~~\displaystyle =(1.96\times 10^{16}{\rm GeV})\times\left(\frac{r(k_{*})}{0.12}\right)^{1/4}
\end{array}\ee 
where $P_{S}(k_*)$ represents the amplitude of the scalar power spectrum. More precisely Eq~(\ref{iopk}) can be recast as a stringent constraint on the upper bound on the brane tension in RSII setup during inflation as:
\be\label{newconstraint}
\sigma < \frac{3\sqrt{3}}{4}\pi^2 P_{S}(k_*)r(k_*)M^4_p.
\ee
It is important to note that, to validate the effective field theory prescription within the framework of small field models of inflation, the model independent bound on the brane tension, the 5D cut-off scale and 5D bulk cosmological 
constant can be written as \cite{Choudhury:2014sua}:
\be\begin{array}{llll}\label{efqzx}
\large 
\displaystyle \sigma\leq {\cal O}(10^{-9})~M^4_p,~~~M_5\leq {\cal O}(0.04-0.05)~M_p,~~~\tilde{\Lambda}_{5}\geq -{\cal O}(
10^{-15})~M^5_p.
\end{array}\ee
If I go beyond the above mentioned bound on the characteristic parameters of RSII then one can describe the inflationary paradigm in large field regime. Please see ref.~\cite{Choudhury:2014sua} for further details.

 Finally using this constraint along with Eq~(\ref{eq7}) in Eq~(\ref{eq12}) I get the following simplified expression
for the root mean square value of
the primordial magnetic field in terms of the
tensor-to-scalar ratio $r$ in RSII setup as~\footnote{In case of the low energy limit of RSII setup i.e. when the energy density of the matter content $(\rho_{m})$ is much higher compared to the 
RSII brane tension $\sigma$ then the actual version of the Friedmann equations in RSII setup are mapped into the Friedmann equations known for General Relativistic setup. Technically this statement can be expressed as:
\bea H^{2}&=&\frac{\rho_{m}}{3M^2_p}\left(1+\underbrace{\frac{\rho_{m}}{2\sigma}}_{<<1}\right)\approx\frac{\rho_{m}}{3M^2_p}.\eea
In the low
energy regime of RSII the lower bound of the CP asymmetry parameter can be written as \cite{Choudhury:2014hua}:
\be\begin{array}{llll}\label{eq15c}
     \displaystyle B_{\xi}(k_{L};k_{\Lambda})\lesssim
{\cal O}(10^{44})\times\left(\frac{r(k_{*})}{0.12}\right)^{3/2}\underbrace{\Sigma^{3/2}(k_{L},k_{*})}_{\bf Regulator~in~ GR}\times\sqrt{\frac{I_{\xi}(k_{L};k_{\Lambda})}{J(k_{L};k_{\Lambda})}}~{\rm Gauss}.
    \end{array}\ee
    where $\Sigma(k_{L}=k_0,k_{*})$ plays the GR analogue of the regulator and satisfies the following stringent constraint:
    \be {\cal O}(10^{-2/3})\leq \Sigma(k_{L}=k_{0},k_{*})\leq {\cal O}(10^{-30}).\ee}:
\be\begin{array}{llll}\label{eq15}
     \displaystyle B_{\xi}(k_{L};k_{\Lambda})\lesssim
{\cal O}(10^{35})\times\left(\frac{r(k_{*})}{0.12}\right)^{5/2}\underbrace{\Sigma^{5/2}_{b}(k_{L},k_{*})\times\left(\frac{M^4_p}{\sigma}\right)}_{\bf Regulator~in~ RSII}\times\sqrt{\frac{I_{\xi}(k_{L};k_{\Lambda})}{J(k_{L};k_{\Lambda})}}~{\rm Gauss}.
    \end{array}\ee
At the present epoch the regulating factor $\Sigma_{b}(k_{L}=k_{0},k_{*})$ appearing in Eq~(\ref{eq15}) is lying within the window,
 \be {\cal O}(4.77\times10^{13})\leq \Sigma_{b}(k_{L}=k_{0},k_{*})\times\left(\frac{M^4_p}{\sigma}\right)^{2/5}\leq {\cal O}(10^{-17.6}),\ee for the tensor-to-scalar ratio, \be 10^{-29}\leq r_*\leq 0.12\ee
 at the momentum pivot scale, 
$k_*\sim 0.002~{\rm Mpc}^{-1}$. Here the ``b'' subscript is used to specify the fact that the analysis is done within RSII setup. Now by setting $k_{L}=k_{0}$ at the present epoch, the estimated numerical value of the primordial magnetic field from RSII setup turns out to be:
\be B_{0}=B_{\xi}(k_{L}=k_{0};k_{\Lambda})\sim {\cal O}(10^{-9})~{\rm Gauss}.\ee
Further using Eq~(\ref{eq11}) I get following expression for the lower bound of the CP asymmetry parameter
within RSII setup as~\footnote{In case of low energy regime of RSII or equivalently for GR prescribed setup the lower bound of the CP asymmetry parameter can be written as \cite{Choudhury:2014hua}:
\be\begin{array}{llll}\label{eq16c}
     \displaystyle \epsilon_{\bf CP}\gtrsim
{\cal O}(10^{-116})\times\left(\frac{0.12}{r(k_{*})}\right)^{3}\Sigma^{-3}(k_{L}=k_0,k_{*}),
    \end{array}\ee
    where $\Sigma(k_{L}=k_0,k_{*})$ plays the GR analogue of the regulator.}:
\be\begin{array}{llll}\label{eq16}
     \displaystyle \epsilon_{\bf CP}\gtrsim
{\cal O}(10^{-98})\times\left(\frac{0.12}{r(k_{*})}\right)^{5}\Sigma^{-5}_{b}(k_{L}=k_0,k_{*})\times\left(\frac{\sigma}{M^4_p}\right)^2,
    \end{array}\ee
which is pointing towards the following possibilities within RSII setup:
\begin{enumerate}
 \item For the large tensor-to-scalar ratio the significant features of CP asymmetry can be possible to detect 
in future collider experiments. For an example we consider a situation where the tensor-to-scalar ratio is, \be r(k_*)\sim 0.12\ee and in such a case the lower bound of 
CP asymmetry is given by \be \epsilon_{\bf CP}\gtrsim 10^{-10}\ee in RSII braneworld. For GR one can also compute the lower bound of CP asymmetry parameter and it turns out to be \be \epsilon_{\bf CP}\gtrsim 10^{-16}\ee
for GR limit \cite{Choudhury:2014hua}.
 \item For very small tensor-to-scalar ratio the CP asymmetry is largely suppressed and can't be possible to detect 
in the particle colliders. For an example if tensor-to-scalar ratio, \be r(k_*)\sim 10^{-29}\ee then the lower bound of 
CP asymmetry is given by \be \epsilon_{\bf CP}\gtrsim 10^{-26}\ee in RSII braneworld. Similarly the lower bound of CP asymmetry parameter in GR prescribed setup can be computed as \cite{Choudhury:2014hua},
\be \epsilon_{\bf CP}\gtrsim 10^{-30}.\ee 
\end{enumerate}
If, in near future, any direct/indirect observational probe detects the signatures of primordial gravitational waves by measuring large detectable amount of 
tensor-to-scalar ratio then it will follow the first possibility. For a rough estimate for CP asymmetry in terms of neutrino masses one can write:
\bea \label{zaw1}
\epsilon_{\bf CP}\sim \frac{3}{16\pi}\frac{M_{1}m_{3}}{v^{2}}\sim 0.1 \frac{M_{1}}{M_{3}}.
\eea
This implies that in the first case it is highly
possible to achieve the upper bound of CP asymmetry parameter \cite{Choudhury:2014hua}, \be \epsilon_{\bf CP}\lesssim 10^{-6}\ee  for \be M_{1}/M_{3}\sim m_{u}/m_{t} \sim 10^{-5},\ee
by tuning the regulating factor as well the brane tension of RSII setup at the pivot scale $k_{*}\sim 0.002~{\rm Mpc}^{-1}$ to the following value~\footnote{In case of low energy
regime of RSII or equivalently for GR prescribed setup
the upper bound on the tuning in the regulator can be expressed as:
\bea \label{tune1}
  \Sigma(k_{L}=k_{0},k_{*})\lesssim {\cal O}(2.1\times 10^{-37}).
\eea
}:
\bea \label{tune}
  \Sigma_{b}(k_{L}=k_{0},k_{*})\times\left(\frac{M^4_p}{\sigma}\right)^{2/5}\lesssim {\cal O}(4\times 10^{-19}),
\eea
which is required to accommodate mass hierarchy of the heavy Majorana neutrino at the scale of $10^{10}~{\rm GeV}$. Additionally it is important mention here that 
the heavy Majorana neutrino $N_{R}$ is the ideal candidate for baryogenesis as 
decays to lepton-Higgs pairs yield lepton asymmetry \be \langle L\rangle_{T} \neq 0,\ee partially
converted to baryon asymmetry \be \langle B\rangle_{T} \neq 0.\ee Also the baryon asymmetry $\eta_{B}$ for given CP asymmetry $\epsilon_{\bf CP}$ can be expressed as:
\bea \label{zaw2}
\eta_{B}=\frac{n_{B}-n_{\bar{B}}}{n_{\gamma}}=\frac{\kappa}{f}c_{\Delta}\epsilon_{\bf CP}
\eea
where $f\sim 10^{2}$ is the dilution factor which accounts for the increase of the number
of photons in a comoving volume element between baryogenesis and today, $c_{\Delta}$ represents the fraction which is responsible for the conversion of lepton asymmetry to baryon asymmetry and exactly quantified by the 
following expression:
\bea 
c_{\Delta}=\frac{\langle B\rangle_{T}}{\langle B-L\rangle_{T}}=\frac{1}{1-\frac{\langle L\rangle_{T}}{\langle B\rangle_{T}}}. 
\eea
Usually the conversion factor \be c_{\Delta}\sim {\cal O}(1)\ee and in the context of Standard Model \be c_{\Delta}=28/79.\ee
Also the determination of the washout factor $\kappa$ requires the details of modified Boltzmann equations within RSII setup. But for realistic estimate one can fix \be \kappa\sim {\cal O}(10^{-2}-10^{-1}).\ee
The baryon asymmetry is generated around a temperature \be T_{B}\sim 10^{10}~{\rm GeV},\ee which is exactly same as mass scale of the heavy Majorana neutrino and
this has possibly interesting implications for the nature of dark matter. The observed value of the baryon asymmetry \cite{Planck:2015xua}, \be \eta_{B}\sim 10^{-9}\ee  is obtained as
consequence of a large hierarchy of the heavy neutrino masses, leading to a
small CP asymmetry, and the kinematical factors $f$ and $\kappa$. In case of RSII setup using Eq~(\ref{zaw2}) the lower bound on baryon asymmetry parameter can be expressed as
~\footnote{In case of GR prescribes setup the lower bound of the CP asymmetry parameter can be written as:
\bea \label{zaw3c}
\eta_{B}\gtrsim {\cal O}(10^{-119})\times\left(\frac{0.12}{r(k_{*})}\right)^{3}\Sigma^{-3}(k_{L}=k_0,k_{*}).
\eea}:
\bea \label{zaw3}
\eta_{B}\gtrsim {\cal O}(10^{-101})\times\left(\frac{0.12}{r(k_{*})}\right)^{5}\Sigma^{-5}_{b}(k_{L}=k_0,k_{*})\times\left(\frac{\sigma}{M^4_p}\right)^2.
\eea
which implies the following possibilities within RSII setup:
\begin{enumerate}
 \item For the large tensor-to-scalar ratio the significant features of baryon asymmetry can be possible to detect 
in future. For an example we consider a situation where the tensor-to-scalar ratio is, \be r(k_*)\sim 0.12\ee and in such a case the lower bound of 
baryon asymmetry is given by \be \eta_{B}\gtrsim 10^{-14}\ee in RSII braneworld. This also implies that in this case it is highly
possible to achieve the observed baryon asymmetry parameter, \be \eta_{B}\sim 10^{-9}\ee by adjusting the regulating factor as well the brane tension of RSII setup at the pivot scale $k_{*}\sim 0.002~{\rm Mpc}^{-1}$
by following the upper bound as stated in Eq~(\ref{tune}). In case of low energy regime of RSII or equivalently for GR prescribed setup the lower bound of 
baryon asymmetry is given by \be \eta_{B}\gtrsim 10^{-26}\ee with tensor to scalar ratio \be r(k_*)\sim 0.12.\ee
 \item For very small tensor-to-scalar ratio the baryon asymmetry is largely suppressed and can't be possible to detect via future experiments.
 For an example if tensor-to-scalar ratio, \be r(k_*)\sim 10^{-29}\ee then the lower bound of 
baryon asymmetry parameter is given by \be \eta_{B}\gtrsim 10^{-30}\ee in RSII braneworld. Similarly in the low energy regime of RSII or in GR limit 
the lower bound of 
baryon asymmetry is given by \be \eta_{B}\gtrsim 10^{-33}\ee with $r(k_*)\sim 0.12$.
\end{enumerate}

\section{Brane inflationary magnetic field via reheating}
\label{a3}
\subsection{Basic assumptions}
Before going to the critical details of the computation, let me first briefly mention the underlying assumptions and basics of the present setup:
\begin{itemize}
 \item  The primordial magnetic field is created via quantum vacuum fluctuation and amplified during the epoch of inflation.
 
 \item  Conformal invariance is restored at the end of inflation such that the magnetic field subsequently decays as $a^{-2}$, where $a$ is the cosmological scale factor. 
 Consequently the physical strength of the magnetic field today on the large scale is given by:
 \be \label{m1}
 B_{0}=\frac{B_{end}}{\left(1+z_{end}\right)^2}.
 \ee
 where $B_{0}$ and $B_{end}$ are the magnetic field today and at the end of inflation respectively. Also $z_{end}$ signifies the redshift at the end of inflation and in terms of scale factor it is defined as:
 \be\label{m2}
 z_{end}=\frac{a_{0}}{a_{end}}-1.
 \ee
 In this work I will explicitly show that for all classes of the models of originating brane inflationary magnetic field, the redshift $z_{end}$ depends on the properties of reheating. During the epoch of inflation the corresponding wave
 number can be expressed as:
 \be\label{m3}
 \frac{k_*}{a}=\frac{k_*}{a_0}\left(1+z_{end}\right)~e^{{\cal N}_{end;b}-{\cal N}_{b}}
 \ee
 where the subscript ``b'' is used to specify the braneworld gravity setup and exactly consistent with Eq~(\ref{m2}).
 \item Further I assume the instantaneous transitions between inflation, reheating, radiation and matter dominated epoch one can write:
 \be\label{m4}
 \left(1+z_{end}\right)=\left(1+z_{eq}\right)\left(\frac{\rho_{reh}}{\rho_{eq}}\right)^{1/4}\left(\frac{a_{reh}}{a_{end}}\right)
 \ee
 where the subscript ``reh'' and ``eq'' stand for end of reheating and the matter radiation equality.
 \item I also assume that at the present epoch the contribution from the correction coming from the non-relativistic neutrinos are negligibly small and so that I neglected the contribution from the computation.
\end{itemize}
\subsection{Reheating parameter}
Let us first start with the reheating parameter $R_{rad}$ defined by~\footnote{If I fix $R_{rad}=1$ then from Eq~(\ref{m5}) it implies that $\rho\propto a^{-4}$, which exactly mimics the role of the energy density 
during radiation dominated era.} \cite{Demozzi:2012wh}:
\be\label{m5}
R_{rad}\equiv \frac{a_{end}}{a_{reh}}\left(\frac{\rho_{end}}{\rho_{reh}}\right)^{1/4},
\ee
where the subscript ``reh'' can be interpreted as the end of reheating era and also the beginning of radiation dominated era. More 
precisely $R_{rad}$ measures the deviation between reheating and radiation dominated era.
Now using Eq~(\ref{m5}) in Eq~(\ref{m4}) one can write:
\be\label{m6}
\left(1+z_{end}\right)=\left(1+z_{eq}\right)\times\left(\frac{a_{eq}}{a_{reh}}\right)\times\left(\frac{a_{reh}}{a_{end}}\right)
=\frac{1}{R_{rad}}\left(\frac{\rho_{end}}{{\cal A}_{reh}\rho_{\gamma}}\right)^{1/4}
\ee
where in the high energy regime of RSII braneworld the radiation energy density can be expressed as: \be \rho_{\gamma}=\sqrt{6\sigma\Omega_{rad}}H_0 M_{p}\ee represents the energy density of radiation at present epoch and 
\be {\cal A}_{reh}\equiv \frac{g_{reh}}{g_{0}}\left(\frac{q_0}{q_{reh}}\right)^{4/3}\ee is the measure of the change of relativistic degrees of freedom between the reheating epoch and present epoch. Also 
$q$ and $g$ denotes the number of entropy and relativistic degrees of freedom at the epoch of interest respectively. Here $H_0$ represents the Hubble parameter at the present epoch and $\Omega_{rad}$ signifies the dimensionless
density parameter during radiation dominated era. To proceed further here I start with the expression for the number of e-foldings at any arbitrary momentum scale]
as \cite{Choudhury:2014kma,Choudhury:2014wsa,Choudhury:2013iaa,Choudhury:2013jya,Choudhury:2013woa}:
\bea\label{m7} 
{\cal N}_{b}(k)= 71.21-\ln\left(\frac{k}{k_{0}}\right)+\frac{1}{4}\ln\frac{V_*}{M^4_p}+ \frac{1}{4}\ln\frac{V_*}{\rho_{end}}+\frac{1-3\bar{w}_{reh}}{12\left(1+\bar{w}_{reh}\right)}\ln\left(\frac{\rho_{reh}}{\rho_{end}}\right)
\eea
where $\rho_{end}$ is the energy density at the end of inflation, $\rho_{reh}$ is an energy scale during
reheating, $k_{0} = a_{0}H_{0}$ is the present Hubble scale, $V_*$ corresponds to the potential
energy when the relevant modes left the Hubble patch during inflation corresponding
to the momentum scale $k_*\approx k_{cmb}$, and $\bar{w}_{reh}$ characterizes the effective equation of state
parameter between the end of inflation and the energy scale during reheating. Further using Eq~(\ref{m7}) one can write:
\bea\label{m8}
{\Delta\bar{\cal N}_{b}}={\cal N}_{reh;b}-{\cal N}_{end;b}=\ln\left(\frac{a_{reh}}{a_{end}}\right)=\ln\left(\frac{k_{end}}{k_{reh}}\right).                                                                        
\eea
Now using only the energy conservation one can derive the following expression for the reheating energy density:
\bea\label{m9} 
\rho_{reh}=\rho_{end}\exp\left[-3\int^{{\cal N}_{reh;b}}_{{\cal N}_{end;b}}\left(1+w({\cal N}_{b})\right)d{\cal N}_{b}\right]\approx \rho_{end}\exp\left[-3{\Delta\bar{\cal N}_{b}}\left(1+\bar{w}_{reh}\right)\right]~~~~~
\eea
where ${\Delta\bar{\cal N}_{b}}$ is defined in Eq~(\ref{m8}) and the mean equation of state parameter $\bar{w}_{reh}$ is defined as:
\bea\label{m10}
\bar{w}_{reh}\equiv \frac{\int^{{\cal N}_{reh;b}}_{{\cal N}_{end;b}}w({\cal N}_{b})d{\cal N}_{b}}{{\Delta\bar{\cal N}_{b}}}.
\eea
where \be w({\cal N}_{b})=P({\cal N}_b)/\rho({\cal N}_b)\ee represents the instantaneous equation of state parameter. Further using Eq~(\ref{m9}) in Eq~(\ref{m5}) one can derive the following expression for reheating parameter:
\bea \label{m11}
R_{rad}=\left(\frac{\rho_{reh}}{\rho_{end}}\right)^{\frac{1-3\bar{w}_{reh}}{12(1+\bar{w}_{reh})}}.
\eea
Here Eq~(\ref{m11}) also implies that for \be \bar{w}_{reh}=1/3\ee the reheating parameter \be R_{rad}=1.\ee 
\subsection{Evading magnetic back-reaction}
To evade magnetic back-reaction on the cosmological background in this paper I consider the following two physical situations:
\begin{enumerate}
 \item  {\bf In the first situation} the reheating epoch characterizes by the lower bound on the equation of state parameter at, \be \bar{w}_{reh}\geq 1/3\ee and the corresponding energy density during reheating decays very faster compared
 to the energy density during radiation dominated era. In this case, the magnetic back-reaction on the length scales of interest is evaded for the following constraint on the ratio of the energy densities \cite{Demozzi:2012wh}:
 \be \label{m12}\frac{\rho_{\bf B}(z_{reh})}{\rho_{reh}}=\frac{\rho_{\bf B_0}}{\rho_{\gamma}}<1 \ee
 Now further using the Planckian unit system one can write, $1~{\rm Gauss}\simeq 3.3 \times 10^{-57}~M^2_{p}$ and using this unit conversion the photon energy density can be written in terms of the magnetic unit as \cite{Demozzi:2012wh},
 \be\rho_{\gamma}\simeq 5.7\times 10^{-125}~M^4_p=5.2\times 10^{-12}~{\rm Gauss}^{2}.\ee  Using Eq~(\ref{m11}) one can further show that for $\bar{w}_{reh}\geq 1/3$ the reheating parameter \be R_{rad}\geq 1.\ee This clearly implies that
 magnetic back-reaction effect can evaded using this constraint.
 \item {\bf In the second situation} the reheating epoch characterizes by \be \bar{w}_{reh}< 1/3\ee and the corresponding energy density of the magnetic field dominate over the energy density during reheating epoch. Within 
 this prescription the effect of magnetic back-reaction 
 can be neglected, provided the magnetic energy density remains smaller compared to the background total energy density at any epoch i.e.
 \be\label{m13} 
\frac{\rho_{B_{end}}}{\rho_{end}}<1. 
 \ee
 where the magnetic energy density $\rho_{B_{end}}$ can be written in terms of the energy density at the end of inflationary epoch as:
 \be\label{m14}
 \rho_{B_{end}}=\frac{B^2_0}{2R^4_{rad}\rho_{\gamma}}\rho_{end}.\ee
 Further substituting Eq~(\ref{m14}) in Eq~(\ref{m13}) one can compute the lower bound on the reheating parameter as \cite{Demozzi:2012wh}:
 \be \label{m15}
 R_{rad}>\frac{\sqrt{B_0}}{(2\rho_{\gamma})^{1/4}}.
 \ee
 The physical interpretation of the bound on reheating parameter is as follows:
 \begin{itemize}
  \item \underline{\bf Firstly} it is important to note that the lower bound on reheating parameter is true for any models of inflation and 
  completely independent on any prior knowledge of inflationary models.  
  \item \underline{\bf Secondly} to hold this bound it necessarily requires that the conformal invariance has to be satisfied during the decelerating phase of the Universe. 
 \end{itemize}
\end{enumerate}
Further using Eq~(\ref{m9}), eq~(\ref{m11}) and Eq~(\ref{m15}) I get the following simplified expression for the reheating constraint: 
\be \label{m16}
\frac{\sqrt{B_0}}{(2\rho_{\gamma})^{1/4}}\exp\left[\frac{{\Delta\bar{\cal N}_{b}}}{4}\left(1-3\bar{w}_{reh}\right)\right]<1
 \ee
 from which one can compute the following analytical constraint on the mean equation of state parameter $\bar{w}_{reh}$ as:
 \be \label{m16v2}
 \bar{w}_{reh}< \frac{1}{3}\left(1+\frac{4}{\Delta\bar{\cal N}_{b}}\ln\left(\frac{\sqrt{B_0}}{(2\rho_{\gamma})^{1/4}}\right)\right)
 \ee
 For an example if I fix the magnetic field at the present epoch within \be B_{0}\sim {\cal O}(10^{-15}~{\rm Gauss}-10^{-9}~{\rm Gauss})\ee then the lower bound
 of the reheating parameter is constrained within \be R_{rad}>{\cal O}(1.76\times 10^{-5}-10^{-2}).\ee Consequently the bound on the mean equation of state parameter $\bar{w}_{reh}$ can be computed as:
 \be\label{m17}
 \bar{w}_{reh}<\frac{1}{3}\left(1+\frac{{\cal C}}{{\Delta\bar{\cal N}_{b}}}\right)
 \ee
 where the numerical factor ${\cal C}\sim {\cal O}(18.42 -43.81)>0$ for $B_{0}\sim {\cal O}(10^{-15}~{\rm Gauss}-10^{-9}~{\rm Gauss})$.
 \section{Reheating constraints on brane inflationary magnetic field}
\label{a4}
To derive the expression for the scale of reheating and also its connection with the inflationary magnetic field within RSII I start with Eq~(\ref{m11}) and using this input one can write:
\bea \label{m18}
\rho_{reh}=\rho_{end} R^{\frac{12(1+\bar{w}_{reh})}{1-3\bar{w}_{reh}}}_{rad}.
\eea
Further using the lower limit of the reheating parameter as stated in Eq~(\ref{m15}), one can derive the lower bound of the reheating energy density as:
\bea \label{m19}
\rho_{reh}>\rho_{end} \left(\frac{B_0}{\sqrt{2\rho_{\gamma}}}\right)^{\frac{6(1+\bar{w}_{reh})}{1-3\bar{w}_{reh}}}=\rho_{end} \left(\frac{B_0}{\sqrt{2\rho_{\gamma}}}
\right)^{-2\left(1+\frac{{\Delta\bar{\cal N}_{b}}}{\ln\left(\frac{\sqrt{B_0}}{(2\rho_{\gamma})^{1/4}}\right)}\right)}.
\eea
Now in the high density or high energy regime of RSII, $\rho>>\sigma$ and using the Friedmann equation one can write \cite{Maartens:2010ar,Brax:2004xh}:
\be \label{m20}
H\approx\frac{\rho}{\sqrt{6\sigma}M_p}.
\ee
where $\sigma$ is the brane tension in RSII setup. Hence using Eq~(\ref{m20}) the lower bound of the reheating energy density can be recast within RSII setup as
~\footnote{In the low density regime of RSII braneworld or equivalently in GR limit the lower bound on the reheating energy density can be expressed as: \bea \label{m21xcx}
\rho_{reh}>\sqrt{3}M_p H_{end} \left(\frac{B_0}{\sqrt{2\rho_{\gamma}}}\right)^{\frac{6(1+\bar{w}_{reh})}{1-3\bar{w}_{reh}}}=\sqrt{3}M_p H_{end} \left(\frac{B_0}{\sqrt{2\rho_{\gamma}}}
\right)^{-2\left(1+\frac{{\Delta\bar{\cal N}_{b}}}{\ln\left(\frac{\sqrt{B_0}}{(2\rho_{\gamma})^{1/4}}\right)}\right)}~~~~~~~
\eea}:
\bea \label{m21}
\rho_{reh}>\sqrt{6\sigma}M_p H_{end} \left(\frac{B_0}{\sqrt{2\rho_{\gamma}}}\right)^{\frac{6(1+\bar{w}_{reh})}{1-3\bar{w}_{reh}}}=\sqrt{6\sigma}M_p H_{end} \left(\frac{B_0}{\sqrt{2\rho_{\gamma}}}
\right)^{-2\left(1+\frac{{\Delta\bar{\cal N}_{b}}}{\ln\left(\frac{\sqrt{B_0}}{(2\rho_{\gamma})^{1/4}}\right)}\right)}~~~~~~~
\eea
where $H_{end}$ represents the Hubble parameter at the end of reheating and additionally Eq~(\ref{m17}) has to satisfied to avoid magnetic back-reaction. Here Eq~(\ref{m21}) implies that if the magnetic field is generated via inflation in braneworld then by knowing the Hubble scale at the 
end of inflation as well as the constraint on the brane tension $\sigma$ it is possible to constraint the lower bound of the scale of reheating. It is important to note that if \be \bar{w}_{reh}\rightarrow 1/3\ee the equality in 
Eq~(\ref{m21}) will not hold at all and also in such a situation the exponent diverges i.e. \be \frac{6(1+\bar{w}_{reh})}{1-3\bar{w}_{reh}}\rightarrow \infty.\ee This clearly implies that the lower bound of the reheating energy density is 
zero and compatible with the understandings of the physics of originating inflationary magnetic field. In the present context the field value at the end of inflation is determined by the violation of the slow-roll conditions. See Appendix \ref{a7} 
for the details. Consequently one can derive the following sets of constraints on the generic form of inflationary potential
and its derivatives at the end of inflation as:
\bea \label{m22}
V(\phi_{end})=\left(2M^2_p\sigma\right)^{1/3}\left(V^{'}(\phi_{end})\right)^{2/3},\\
\label{m23} V(\phi_{end})=\left(2M^2_p\sigma\right)^{1/2}\left(V^{''}(\phi_{end})\right)^{1/2},\\
\label{m24} V(\phi_{end})=\left(4M^4_p\sigma^2\right)^{1/4}\left(V^{'}(\phi_{end})V^{'''}(\phi_{end})\right)^{1/2},\\
\label{m25} V(\phi_{end})=\left(8M^6_p\sigma^3\right)^{1/6}\left(V^{'}(\phi_{end})\right)^{1/3}\left(V^{''''}(\phi_{end})\right)^{1/6}.
\eea
For more stringent constraint the system need to satisfy all of the equations as mentioned in Eq~(\ref{m22}-\ref{m25}) to fix the scale of inflationary potential
at the end of inflation. In this case the derivatives or more precisely the Taylor expansion co-efficients of the inflationary potential at the 
end of inflation are not independent at all. But if the system relaxes any three of the previously mentioned constraints, then also it possible
to constrain the scale of potential at the end epoch of inflation. Consequently Eq~(\ref{m21}) can be recast in terms of the generic form of the 
inflationary potential as:
\bea \label{m26}
\rho_{reh}>V(\phi_{end}) \left(\frac{B_0}{\sqrt{2\rho_{\gamma}}}\right)^{\frac{6(1+\bar{w}_{reh})}{1-3\bar{w}_{reh}}}\approx V(\phi_{end})  \left(\frac{B_0}{\sqrt{2\rho_{\gamma}}}
\right)^{-2\left(1+\frac{{\Delta\bar{\cal N}_{b}}}{\ln\left(\frac{\sqrt{B_0}}{(2\rho_{\gamma})^{1/4}}\right)}\right)}~~~~~~~
\eea
Here it is important to note that during reheating both kinetic and potential contribution play crucial role in the energy density. Later I will explicitly show
the estimation algorithm of $V(\phi_{end})$ from a generic as well as for specified form of inflationary potential for the determination of the lower bound of
the energy density during reheating.

In the high energy regime of RSII setup during reheating one can write the total decay width for the decay of heavy Majorana neutrinos as \cite{Choudhury:2011rz,Choudhury:2012ib}:
\bea\label{m27}
\Gamma_{total}=\Gamma_{L}(N_{R}\rightarrow L_{i}\Phi)+\Gamma_{L^c}(N_{R}\rightarrow L^{c}_{i}\Phi^{c})=3H(T_{reh})\approx\sqrt{\frac{3}{2\sigma}}\frac{\rho_{reh}}{M_p}
\eea
where $H(T_{reh})$ be the Hubble parameter during reheating and $\rho_{reh}$ represents the energy density during reheating. In the context of statistical theormodynamics one can express the reheating energy 
density as:
\be\label{m28}
\rho_{reh}=\frac{\pi^2}{30}g_{*}T^{4}_{reh}
\ee
where $g_*$ signifies the effective number of relativistic degrees of freedom. In a more generalized prescription $g_*$ can be expressed as:
\be\label{m29}
g_*= g_{B*}+\frac{7}{8}g_{F*}
\ee
where $g_{B*}$ and $g_{F*}$ are the number of bosonic and fermionic degrees of freedom respectively. It is worth mentioning that the reheating temperature within RSII does not depend on the initial value of the inflaton field from where inflation starts and is solely 
determined by the elementary particle theory of the early universe. 
Further using Eq~(\ref{m27}) and Eq~(\ref{m28}) the reheating temperature within the high energy regime of RSII
setup can be expressed as~\footnote{In the low energy regime of RSII or equivalently in the GR limit the reheating temperature can be expressed as \cite{Choudhury:2013jya,Choudhury:2011jt,Mazumdar:2010sa}:
\be\label{m30a}
T_{reh}=\left(\frac{30}{\pi^2 g_*}\right)^{1/4}\times\left(\frac{\Gamma_{total} M_p}{\sqrt{3}}\right)^{1/2}.
\ee}  \cite{Choudhury:2011rz,Choudhury:2012ib}:
\be\label{m30}
T_{reh}=\left(\frac{30}{\pi^2 g_*}\right)^{1/4}\times\left(\Gamma_{total} M_p\right)^{1/4}\times\left(\frac{2\sigma}{3}\right)^{1/8}.
\ee
On the other hand the reheating temperature can be expressed in terms of the tensor-to-scalar ratio as:
\be 
\label{m31}
T_{reh}\approx \left(\frac{30}{\pi^2 g_*}\right)^{1/4}\times(1.96\times 10^{16}{\rm GeV})\times\left(\frac{r(k_{*})}{0.12}\right)^{1/4}.
\ee
Now eliminating reheating temperature from Eq~(\ref{m30}) and Eq~(\ref{m31}) one can express the total decay width in terms of inflationary
tensor-to-scalar ratio as~\footnote{In the low energy regime of RSII or equivalently in the GR limit total decay width of the heavy Majorana neutrino can be written as:
\be\label{m32a}
\Gamma_{total}=1.13\times 10^{-4} M_{p}\times\left(\frac{r(k_{*})}{0.12}\right)^{1/2}.
\ee}:
\be\label{m32}
\Gamma_{total}=4.23\times 10^{-9} M^{3}_{p}\times \sqrt{\frac{3}{2\sigma}}\times\left(\frac{r(k_{*})}{0.12}\right).
\ee
Further combining Eq~(\ref{m9}) and Eq~(\ref{m28}) the energy density of inflaton at the end of inflation can be expressed in terms of tensor-to-scalar ratio as:
\be\label{m33}
\rho_{end}\approx V(\phi_{end})=(1.96\times 10^{16}{\rm GeV})^{4}\times\left(\frac{r(k_{*})}{0.12}\right)\times\exp\left[3\left(1+\bar{w}_{reh}\right)\Delta\bar{\cal N}_{b}\right].
\ee
Similarly using Eq~(\ref{m33}) in Eq~(\ref{m26}) the reheating energy density or more precisely the scale of reheating can be expressed in terms of the tensor-to-scalar ratio, mean equation of reheating $\bar{w}_{reh}$ and 
magnetic field at the present epoch as:
\bea \label{m34}
\rho_{reh}>(1.96\times 10^{16}{\rm GeV})^{4}\times\left(\frac{r(k_{*})}{0.12}\right)\times\exp\left[3\left(1+\bar{w}_{reh}\right)\Delta\bar{\cal N}_{b}\right]
\times\left(\frac{B_0}{\sqrt{2\rho_{\gamma}}}\right)^{\frac{6(1+\bar{w}_{reh})}{1-3\bar{w}_{reh}}}.~~~~~~~
\eea
Further applying the constraint in the mean equation of reheating parameter as stated in Eq~(\ref{m16v2}) the lower bound of the scale of reheating energy density can be recast as:
\bea \label{m35}
\rho_{reh}>(1.96\times 10^{16}{\rm GeV})^{4}\times\left(\frac{r(k_{*})}{0.12}\right)\times\exp\left[4\Delta\bar{\cal N}_{b}\right]\times\left(\frac{B_0}{\sqrt{2\rho_{\gamma}}}
\right)^{-\frac{4{\Delta\bar{\cal N}_{b}}}{\ln\left(\frac{B_0}{\sqrt{2\rho_{\gamma}}}\right)}}.~~~~~~~
\eea
 Next I will explicitly derive the expression for the density parameter during radiation dominated epoch ($\Omega_{rad}$) and further I will connect this to the density parameter of the magnetic field ($\Omega_{B_{end}}$).
 To serve this purpose I start with the analysis in the high energy regime of the RSII braneworld in which the dimensionless density parameter can be expressed as:
 \be \label{m36}
 \Omega=\frac{\rho^{2}}{\rho_{c}\rho_{0}}
 \ee
 where the critical energy density in RSII braneworld can be written as:
 \be\label{m37}
 \rho_{c}=2\sigma
 \ee
 and the energy density at the present epoch can be written as:
 \be \label{m38}
 \rho_{0}=3M^{2}_{p}H^{2}_{0}.
 \ee
 Now using Eq~(\ref{m6}) in Eq~(\ref{m14}) one can write the magnetic energy density in terms of redshift as:
 \be\label{m41}
 \rho_{B_{end}}=\frac{B^{2}_{0}}{2}\left(1+z_{eq}\right)^{4}\exp\left[\Delta\bar{\cal N}_{b}\left(1-3\bar{w}_{reh}\right)\right]
 \ee
 and using Eq~(\ref{m41}) the dimensionless density parameter for magnetic field can be written as:
 \be\label{m42}
 \Omega_{B_{end}}=\frac{B^{4}_{0}}{24\sigma H^{2}_{0}M^{2}_{p}}\left(1+z_{eq}\right)^{8}\exp\left[2\Delta\bar{\cal N}_{b}\left(1-3\bar{w}_{reh}\right)\right].
 \ee
 In the high energy regime of RSII braneworld one can write the density parameter at the end of inflation in terms of the density parameter at the radiation dominated era and redshift as:
 \be\label{m43}
 \Omega_{end}=\left(1+z_{eq}\right)^{8}\Omega_{rad}.
 \ee
 Further substituting Eq~(\ref{m43}) in Eq~(\ref{m42}) I get the following constraint relationship:
 \be\label{m42}
 \Omega_{B_{end}}=\frac{B^{4}_{0}\Omega_{end}}{24\sigma H^{2}_{0}M^{2}_{p}\Omega_{rad}}\exp\left[2\Delta\bar{\cal N}_{b}\left(1-3\bar{w}_{reh}\right)\right].
 \ee
 Next using Eq~(\ref{m33}) one can write down the expression for the dimensionless parameter at the end of inflationary epoch as:
 \be\label{m43}
 \Omega_{end}=\frac{M^{6}_{p}}{6\sigma H^{2}_{0}}\times(1.79\times 10^{-17})\times\left(\frac{r(k_{*})}{0.12}\right)^2\times\exp\left[6\left(1+\bar{w}_{reh}\right)\Delta\bar{\cal N}_{b}\right].
 \ee
Further applying the constraint in the mean equation of reheating parameter as stated in Eq~(\ref{m16v2}) the dimensionless density parameter can be expressed in terms of the magnetic field at the 
 present epoch as:
 \be\label{m44}
 \Omega_{end}=\frac{M^{6}_{p}}{6\sigma H^{2}_{0}}\times(1.79\times 10^{-17})\times\left(\frac{r(k_{*})}{0.12}\right)^2\times
 \exp\left[8\left(\Delta\bar{\cal N}_{b}+\ln\left(\frac{\sqrt{B_0}}{(2\rho_{\gamma})^{1/4}}\right)\right)\right].
 \ee
 Finally substituting Eq~(\ref{m44}) in Eq~(\ref{m42}) I get~\footnote{In the low energy regime of RSII dimensionless density parameter for magnetic field can be expressed as:
 \be\label{m45a}
 \Omega_{B_{end}}=\frac{B^{2}_{0}}{6M^2_{p}H^{2}_{0}\Omega_{rad}}\times(2.17\times 10^{-5})\times\left(\frac{r(k_{*})}{0.12}\right) \times\exp\left[4\Delta\bar{\cal N}_{b}\right].
 \ee}:
 \be\label{m45}
 \Omega_{B_{end}}=\frac{B^{4}_{0}M^{4}_{p}}{144\sigma^2 H^{4}_{0}\Omega_{rad}}\times(1.79\times 10^{-17})\times\left(\frac{r(k_{*})}{0.12}\right)^2 \times\exp\left[8\Delta\bar{\cal N}_{b}\right].
 \ee
 Here the dimensionless density parameter during the epoch of radiation domination is given by the following expression:
 \be\label{m46}
 \Omega_{rad}=\frac{\rho^2_{\gamma}}{\rho_{c}\rho_{0}}=\frac{\rho^2_{\gamma}}{6\sigma H^{2}_{0}M^{2}_{p}}
 \ee
 where $\rho_{\gamma}\simeq 5.7\times 10^{-125}~M^4_p=5.2\times 10^{-12}~{\rm Gauss}^{2}$. Further using Eq~(\ref{m46}) in 
 Eq~(\ref{m45}) I get:
 \be\label{m47}
 \Omega_{B_{end}}=\frac{B^{4}_{0}M^{6}_{p}}{24\sigma H^{2}_{0}\rho^2_{\gamma}}\times(1.79\times 10^{-17})\times\left(\frac{r(k_{*})}{0.12}\right)^2 \times\exp\left[8\Delta\bar{\cal N}_{b}\right].
 \ee
 Next using Eq~(\ref{eq11}) in Eq~(\ref{m47}) finally I get the following relationship between the density parameter of the magnetic field and the CP asymmetry parameter within the high energy regime of RSII 
 braneworld as~\footnote{In the low energy regime of RSII or equivalently in Gr limit the dimensionless density parameter for magnetic field can be expressed as:
 \be\label{m45a}
 \Omega_{B_{end}}=\frac{1}{2\epsilon_{\bf CP}}\times(4.17\times 10^{-38})\times\left(\frac{r(k_{*})}{0.12}\right) \times\exp\left[4\Delta\bar{\cal N}_{b}\right].
 \ee}:
 \be\label{m48}
 \Omega_{B_{end}}=\frac{M^{6}_{p}}{24\sigma H^{2}_{0}\epsilon^{2}_{\bf CP}}\times(6.63\times 10^{-83})\times\left(\frac{r(k_{*})}{0.12}\right)^{2} \times\exp\left[8\Delta\bar{\cal N}_{b}\right].
 \ee
 \section{Constraining brane inflationary magnetic field from CMB}
 \label{a5}
 Before going to the details of the constraints on the various models of describing the origin of brane inflationary magnetic field from CMB, let me introduce a rescaled reheating parameter $R_{sc}$ defined as \cite{Demozzi:2012wh}:
 \be\label{m49}
 R_{sc}\equiv R_{rad} \times \frac{\rho^{1/4}_{end}}{M_p}=\left(\frac{\rho_{reh}}{\rho_{end}}\right)^{\frac{1-3\bar{w}_{reh}}{12(1+\bar{w}_{reh})}}\times \frac{\rho^{1/4}_{end}}{M_p}
 =\frac{a_{end}}{a_{reh}}\times\left(\frac{\rho^{1/2}_{end}}{\rho^{1/4}_{reh}M_p}\right)
 \ee
 which is relevant for further analysis. Further using Eq~(\ref{m33}) the lower bound of the rescaled reheating parameter can be expressed in terms of the tensor-to-scalar ratio and the magnetic field at the present epoch as: 
 \be\label{m49}
 R_{sc}>8.07\times 10^{-3}\times\left(\frac{r(k_{*})}{0.12}\right)^{1/4}\times\exp\left[\Delta\bar{\cal N}_{b}
 \right]
 \times\left(\frac{B_0}{\sqrt{2\rho_{\gamma}}}\right).~~~~~
 \ee
 In the following subsections I will explicitly discuss about the CMB constraints on two types of models of describing the origin of brane inflationary magnetic field. But in principle one can carry forward the prescribed methodology for rest of the 
 brane inflationary models also.
\subsection{Monomial Models}
\begin{figure}[htb]
	\centering
	\includegraphics[width=12cm,height=9cm]{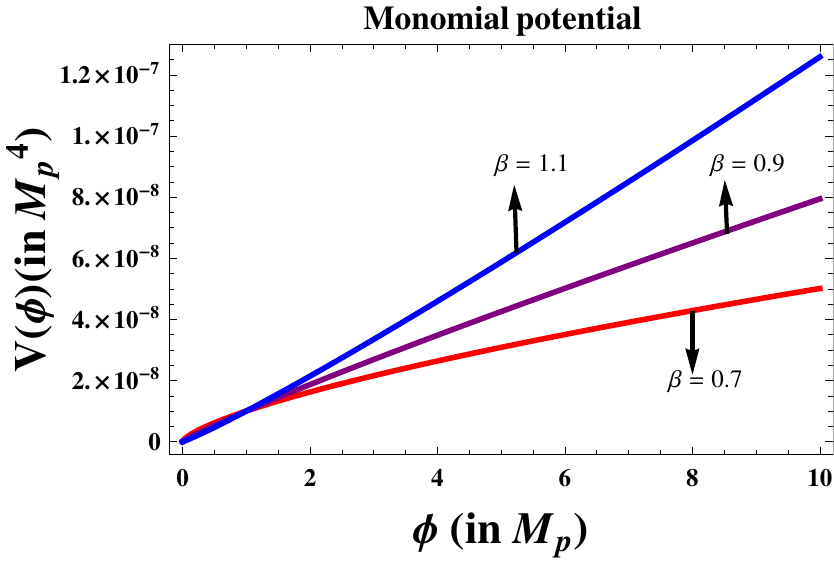}
	\caption{Variation of the monomial potential for the index $\beta=0.7, 0.9, 1.1$. Here I fix the tunable scale at $\sqrt[4]{V_{0}}=4.12\times 10^{-3}~M_{p}=10^{16}~GeV$.
	}
	\label{fig1}
\end{figure}

In case of monomial models the inflationary potential can be represented by the following functional form:
\be\label{m50}
V(\phi)=V_{0}\left(\frac{\phi}{M_p}\right)^{\beta}
\ee
where $V_{0}=M^4$ is the tunable energy scale, which is necessarily required to fix the amplitude of the CMB anisotropies and $\beta$ is the monomial index which characterizes the feature of the potential. 
The variation of the monomial potential for the index $\beta=0.7, 0.9, 1.1$ and the tunable scale \be \sqrt[4]{V_{0}}=4.12\times 10^{-3}~M_{p}=10^{16}~{\rm GeV}\ee is shown in fig.~\ref{fig1}.

\begin{figure*}[htb]
	\centering
	\subfigure[$r$ vs $n_{S}$.]{
		\includegraphics[width=10.2cm,height=6.5cm] {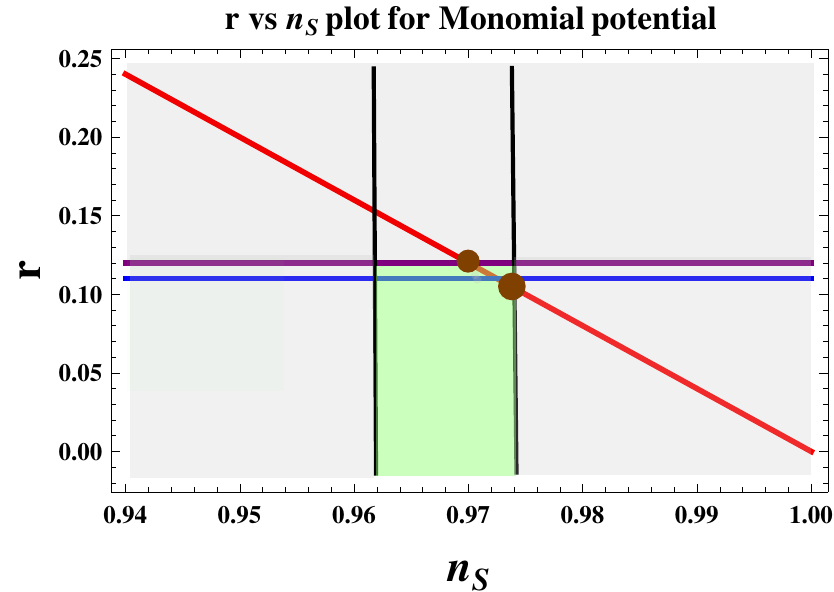}
		\label{fig3a}
	}
	\subfigure[$r$ vs $\beta$.]{
		\includegraphics[width=10.2cm,height=6.5cm] {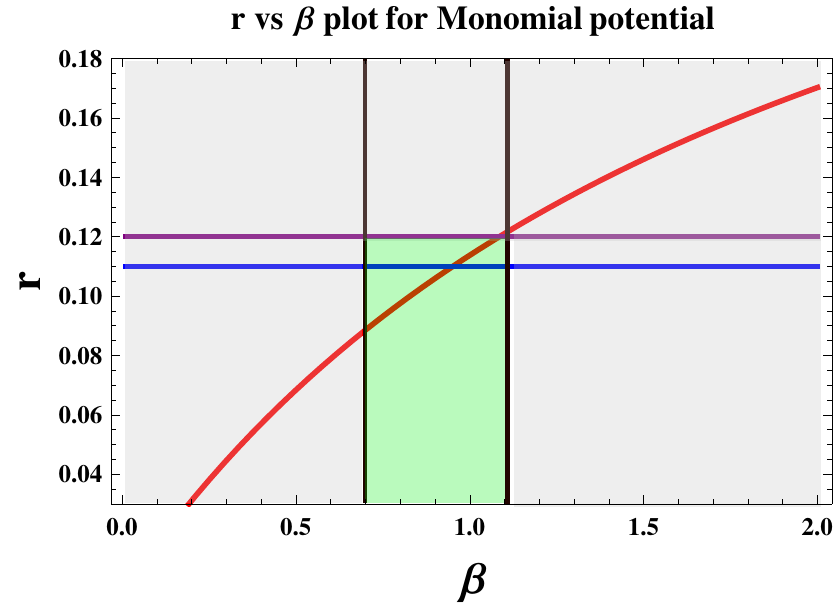}
		\label{fig3b}
	}
	\caption[Optional caption for list of figures]{Behaviour of the tensor-to-scalar ratio $r$ with respect to \ref{fig3a} the scalar spectral index $n_{S}$ and \ref{fig3b} the characteristic parameter of the monomial potential $\beta$.
		The purple and blue coloured line represent the upper bound of tenor-to-scalar ratio allowed by Planck+ BICEP2+Keck Array joint constraint and only Planck 2015 data respectively. The small and the big bubbles
		represent two consecutive points in $r-n_{S}$ plane, where for the small bubble $\Delta{\cal N}_{b}=50, r=0.124, n_{S}=0.969$ and for the big bubble $\Delta{\cal N}_{b}=70, r=0.121, n_{S}=0.970$ respectively,
		The green shaded region bounded by two vertical black coloured lines represent the Planck $2\sigma$ allowed region and the rest of the light grey shaded region is disfavoured by the Planck data and Planck+ BICEP2+Keck Array joint constraint. From \ref{fig3a} it 
		is observed that, within $50<\Delta{\cal N}_{b}<70$ the monomial potential is favoured only for the characteristic index $0.7<\beta<1.1$, by Planck 2015 data and Planck+ BICEP2+Keck Array joint analysis.
		In \ref{fig3b} I have explicitly shown that the in $r-\beta$ plane the observationally favoured window for the monomial index is $0.7<\beta<1.1$.} 
	\label{fig3}
\end{figure*}

In the present context both the rescaled reheating parameter $R_{sc}$ and the energy density at the end of inflation $\rho_{end}$ are constrained~\footnote{After marginalization over the monomial index of the potential 
within $0.2<\beta<5$ and over the cosmological parameters the following CMB constraints are obtained within $2\sigma$ CL \cite{Martin:2010kz}:
\bea \label{o1}R_{sc}>2.81\times 10^{-13},\\ \label{o2} 4\times 10^{15}~{\rm GeV}<\rho^{1/4}_{end}<1.2\times 10^{16}~{\rm GeV}. \eea}. To analyze the features of the potential in detail here I start with the definition of number of e-foldings
$\Delta{\cal N}_{b}(\phi)$ in the high energy regime of RSII setup (see Appendix \ref{a7} for details), using which I get: 
\be\label{m51}
\Delta{\cal N}_{b}(\phi)=\frac{V_{0}}{2\sigma \beta\left(\beta+2\right) M^{\beta+2}_{p}}\left(\phi^{\beta+2}-\phi^{\beta+2}_{end}\right).
\ee
Further setting $\phi=\phi_{cmb}$ in Eq~(\ref{m51}), the field value at the horizon crossing can be computed as:
\be\label{m52}
\phi_{cmb}=\phi_{end}\left[1+\frac{2\sigma\beta\left(\beta+2\right)M^{\beta+2}_{p}\Delta{\cal N}_{b}}{\phi^{\beta+2}_{end}V_0}\right]^{\frac{1}{\beta+2}}
\ee
where $\phi_{end}$ represents the field value of inflaton at the end of inflation. Within RSII setup from the violation of the slow-roll conditions one can compute:
\be\label{m53}
\phi_{end}\approx\left(\frac{2\sigma \beta^{2}}{V_{0}}\right)^{\frac{1}{\beta+2}}M_p.
\ee
From monomial models of inflation the scale of the potential at the horizon crossing and at the end of inflation can be computed as:
\bea\label{m54}
\rho_{cmb}&\approx& V(\phi_{cmb})=V_{0}\left(\frac{\phi_{cmb}}{M_p}\right)^{\beta}=V^{\frac{2}{\beta+2}}_{0}\left(2\sigma \beta^{2}\right)^{\frac{\beta}{\beta+2}}
\left[1+\left(1+\frac{2}{\beta}\right)\Delta{\cal N}_{b}\right]^{\frac{\beta}{\beta+2}},~~~~~~~~~~~~~\\
\rho_{end}&\approx& V(\phi_{end})=V_{0}\left(\frac{\phi_{end}}{M_p}\right)^{\beta}=V^{\frac{2}{\beta+2}}_{0}\left(2\sigma \beta^{2}\right)^{\frac{\beta}{\beta+2}}.
\eea
Further using the consistency condition in the high energy regime of RSII braneworld, as stated in Eq~(\ref{cv1}) of the Appendix C,
one can derive the following expressions for the amplitude of the scalar power spectrum, tensor to scalar ratio and scalar spectral tilt as:
\bea\label{m55cv} P_{S}(k_*)&=&
\frac{V^{\frac{2}{\beta+2}}_{0}\left(2\sigma \beta^{2}\right)^{\frac{\beta}{\beta+2}}}{36\pi^2}
\left[1+\left(1+\frac{2}{\beta}\right)\Delta{\cal N}_{b}\right]^{\frac{2(\beta+1)}{\beta+2}}
,\\
\label{m55} r(k_*)&=&\frac{24}{
\left[1+\left(1+\frac{2}{\beta}\right)\Delta{\cal N}_{b}\right]},\\
\label{m55v1} n_{S}(k_*)-1&\approx&-\frac{6}{
\left[1+\left(1+\frac{2}{\beta}\right)\Delta{\cal N}_{b}\right]}.\eea
and to satisfy the joint constraint on the scalar spectral tilt and upper bound of tensor-to-scalar ratio as observed by Planck (2013 and
2015) and Planck+BICEP2+Keck Array, one need the following constraint on the monomial index $\beta$ of the inflationary potential~\footnote{For a realistic estimate, if we fix $\Delta{\cal N}_{b}\approx {\cal O}(50-70)$, 
then the monomial index $\beta$ is constrained as, $0.7<\beta<1.1$.}:
\be\label{m55v2}
\beta<\frac{2}{\frac{199}{\Delta{\cal N}_{b}}-1}.
\ee 
The behaviour of the tensor-to-scalar ratio $r$ with respect to the scalar spectral index $n_{S}$ and the characteristic parameter of the monomial potential $\beta$ are plotted in fig.~\ref{fig3a} and fig.~\ref{fig3b} respectively. 
\begin{figure*}[htb]
	\centering
	\subfigure[$P_{S}$ vs $n_{S}$.]{
		\includegraphics[width=7.2cm,height=7cm] {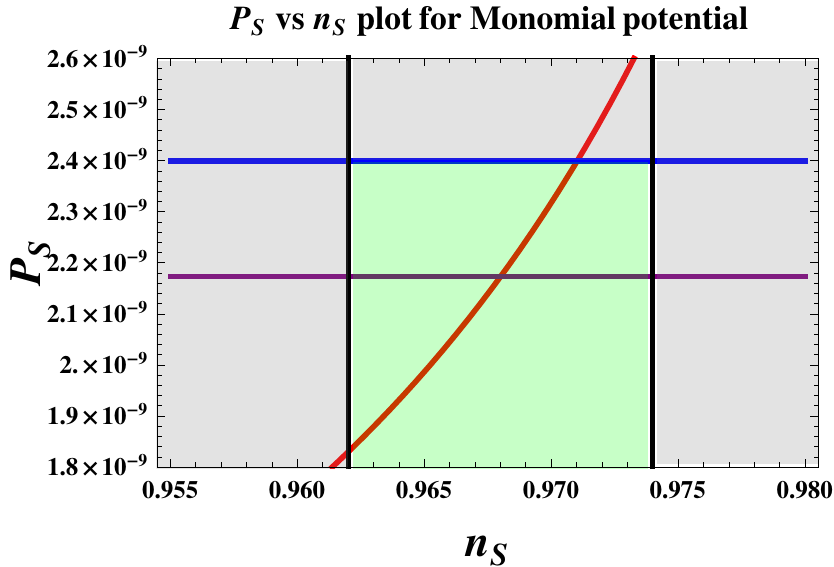}
		\label{fig3ax}
	}
	\subfigure[$P_{S}$ vs $\beta$.]{
		\includegraphics[width=7.2cm,height=7cm] {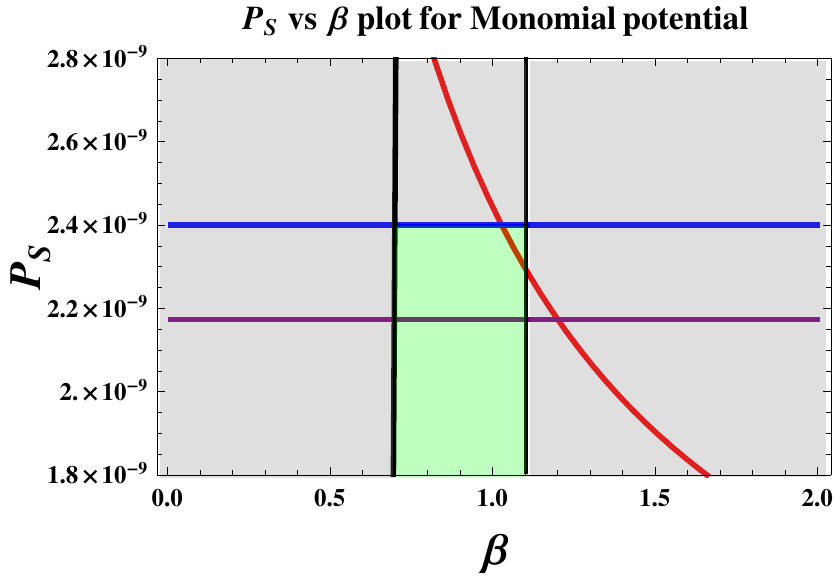}
		\label{fig3bx}
	}
	\subfigure[$n$ vs $\beta$.]{
		\includegraphics[width=9.2cm,height=8cm] {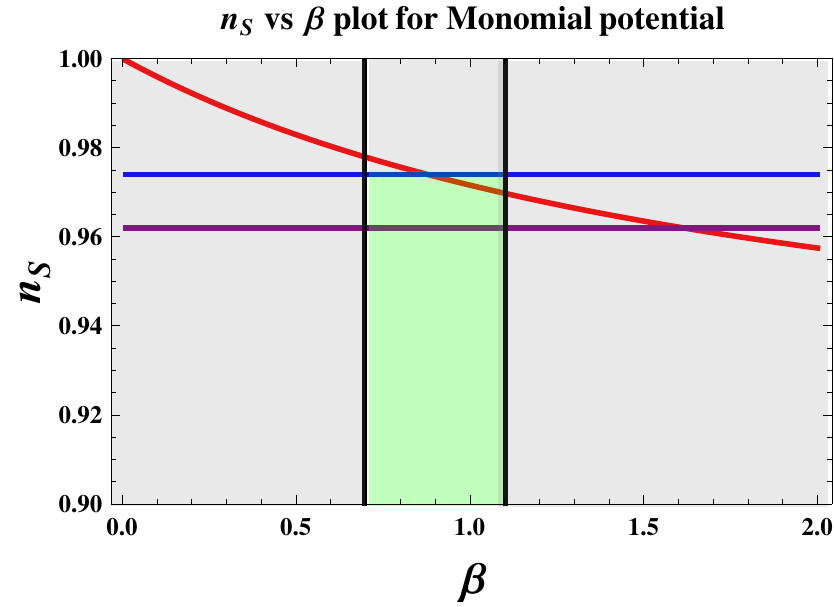}
		\label{fig3cx}
	}
	\caption[Optional caption for list of figures]{Variation of the \ref{fig3ax} scalar power spectrum $P_{S}$ vs scalar spectral index $n_{S}$, \ref{fig3bx} scalar power spectrum $P_{S}$ vs index $\beta$
		and \ref{fig3cx} scalar power spectrum $n_{S}$ vs index $\beta$. The purple and blue coloured line represent the upper and lower bound allowed by WMAP+Planck 2015 data respectively. 
		The green dotted region bounded by two vertical black coloured lines represent the Planck $2\sigma$ allowed region and the rest of the light gray shaded region is disfavoured by the Planck+WMAP constraint.} 
	\label{fig3x}
\end{figure*}
From \ref{fig3a} it 
is observed that, within $50<\Delta{\cal N}_{b}<70$ the monomial potential is favoured only for the characteristic index \be 0.7<\beta<1.1,\ee by Planck 2015 data and Planck+ BICEP2+Keck Array joint analysis.
In \ref{fig3b} I have explicitly shown that the in $r-\beta$ plane the observationally favoured window for the monomial index is $0.7<\beta<1.1$. 
Additionally it is important to note that, for monomial potentials embedded in the high energy regime of RSII braneworld, 
the consistency relation between tensor-to-scalar ratio $r$ and the scalar spectral $n_{S}$ is given by, \be r\approx 4(1-n_{S}).\ee On the other 
hand in the low energy regime of RSII braneworld or equivalently in the GR limiting situation, the consistency relation between  
tensor-to-scalar ratio $r$ and the scalar spectral $n_{S}$ is modified as, \be r\approx \frac{8}{3}(1-n_{S}).\ee This also clearly suggests that the 
estimated numerical value of the tensor-to-scalar ratio from the GR limit is different compared to its value in the high density regime of the 
RSII braneworld. To justify the validity of this statement, let me discuss a very simplest situation, where the scalar spectral index is constrained 
within \be 0.969<n_{S}<0.970,\ee as appearing in this paper. Now in such a case using the consistency relation in GR limit one can easily compute that the tensor-to-scalar is 
constrained within the window, \be 0.080<r<0.083,\ee which is pretty consistent with Planck 2015 result. 

\begin{figure*}[t]
\centering
\subfigure[$V^{1/4}_{0}$ vs $\beta$ for $\sigma\sim 10^{-9}~M^{4}_{p}$.]{
    \includegraphics[width=10.2cm,height=6cm] {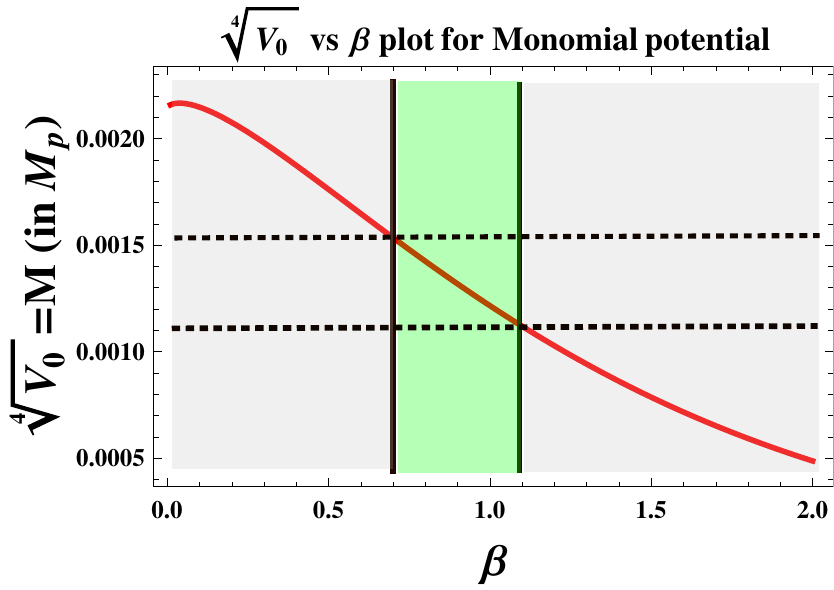}
    \label{fig4a}
}
\subfigure[$V^{1/4}_{0}$ vs $\beta$ for $\sigma\sim 5\times 10^{-16}~M^{4}_{p}$.]{
    \includegraphics[width=10.2cm,height=6cm] {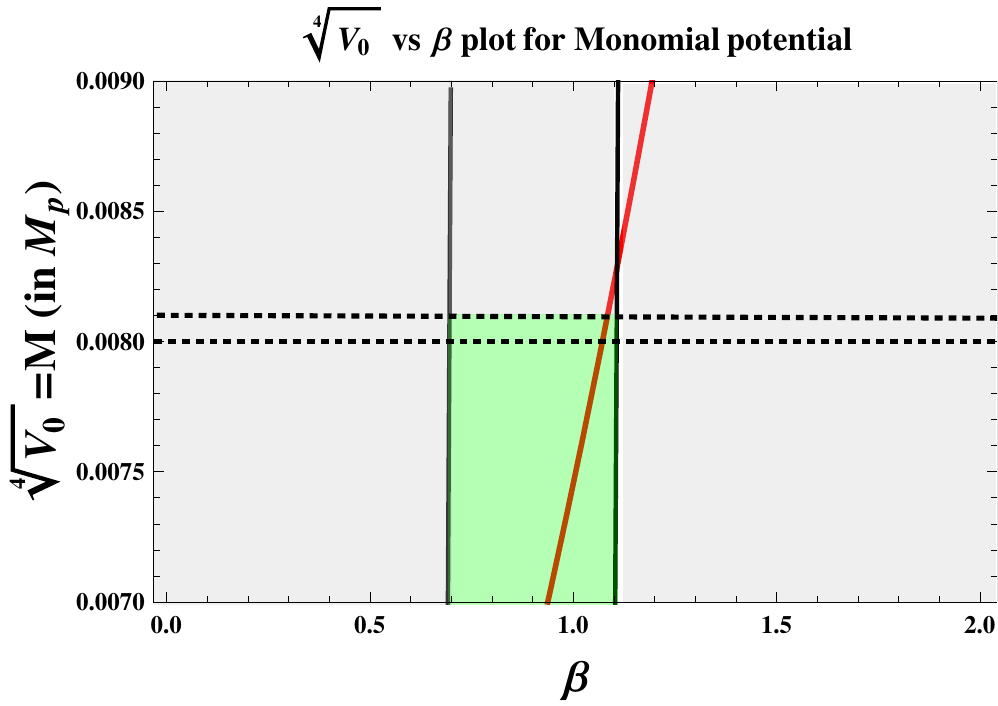}
    \label{fig4b}
}
\caption[Optional caption for list of figures]{Variation of the energy scale of the monomial potential with respect to the characteristic index $\beta$.
The green shaded region bounded by two vertical black coloured lines  and two black coloured horizontal line represent the Planck $2\sigma$ allowed region and the rest of the light gray shaded
region is disfavoured by the Planck data and Planck+ BICEP2+Keck Array joint constraint. In \ref{fig4a} and \ref{fig4b} I have fixed the value of the brane tension at $\sigma\sim 10^{-9}~M^{4}_{p}$ and 
$\sigma\sim 10^{-15}~M^{4}_{p}$ respectively. This analysis explicitly shows that 
 the $2\sigma$ allowed
window for the parameter
$\beta$ within $0.7<\beta<1.1$ constraints the scale of inflation within $ 1.1\times 10^{-3}~M_{p} <\sqrt[4]{V_{0}}<1.5\times 10^{-3}~M_{p}$
for $\sigma\sim 10^{-9}~M^{4}_{p}$ and $ 8.08\times 10^{-3}~M_{p} <\sqrt[4]{V_{0}}<8.13\times 10^{-3}~M_{p}$ for $\sigma\sim 5\times 10^{-16}~M^{4}_{p}$. For the first case the tensor-to-scalar ratio and scalar spectral
tilt are constrained within the window, $4.15\times 10^{-5}<r<1.44\times 10^{-4}$ and $n_{S}\sim 0.99$. Here for $\sigma\sim 10^{-9}~M^{4}_{p}$ the value of $r$ is consistent with the upper bound on tensor-to-scalar ratio, but the value of scalar spectral 
tilt is outside the $2\sigma$ CL. On the other hand, for the second case, the tensor-to-scalar ratio and scalar spectral
tilt are constrained within the window, $0.121<r<0.124$ and $0.969<n_{S}<0.970$, which is consistent with $2\sigma$ CL constraints.} 
\label{fig4}
\end{figure*}

Variation of the \ref{fig3ax} scalar power spectrum $P_{S}$ vs scalar spectral index $n_{S}$, \ref{fig3bx} scalar power spectrum $P_{S}$ vs index $\beta$
and \ref{fig3cx} scalar power spectrum $n_{S}$ vs index $\beta$. The purple and blue coloured line represent the upper and lower bound allowed by WMAP+Planck 2015 data respectively. 
The green dotted region bounded by two vertical black coloured lines represent the Planck $2\sigma$ allowed region and the rest of the light gray shaded region is disfavoured by the Planck+WMAP constraint.
From the fig.~\ref{fig3ax}-fig.~\ref{fig3cx} it is clearly observed that the monomial index of the the inflationary potential is constrained within the window $0.7<\beta<1.1$  for the amplitude of the scalar power spectrum, 
\be 2.3794\times 10^{-9}<P_{S}<2.3798\times 10^{-9}\ee and scalar spectral tilt, \be 0.969<n_{S}<0.970.\ee Now using Eq~(\ref{m55cv}), Eq~(\ref{m55}) and Eq~(\ref{m55v1}) one can 
write another consistency relation among the amplitude of the scalar power spectrum $P_{S}$, tensor-to-scalar ratio $r$ and scalar spectral
index $n_{S}$ for monomial potentials embedded in the high density regime of RSII braneworld 
as: \be P_{S}=\frac{V^{\frac{2}{\beta+2}}_{0}\left(2\sigma \beta^{2}\right)^{\frac{\beta}{\beta+2}}}{36\pi^2}
\left[\frac{6}{1-n_{S}}\right]^{\frac{2(\beta+1)}{\beta+2}}=\frac{V^{\frac{2}{\beta+2}}_{0}\left(2\sigma \beta^{2}\right)^{\frac{\beta}{\beta+2}}}{36\pi^2}
\left[\frac{24}{r}\right]^{\frac{2(\beta+1)}{\beta+2}}.\ee
Further using Eq~(\ref{iopk}), I get the following stringent constraint on the tunable energy scale of the monomial models of inflation:
\bea\label{m56}
V_{0}=M^{4}
&<& \frac{\left(2.12\times 10^{-11}~M^{4}_p\right)^{1+\frac{\beta}{2}}}{\left(2\sigma\beta^{2}\right)^{\frac{\beta}{2}}}.
\eea
The variation of the energy scale of the monomial potential with respect to the characteristic index $\beta$ is shown in fig.~\ref{fig4a} and fig.~\ref{fig4b},
for the fixed the value of the brane tension at $\sigma\sim 10^{-9}~M^{4}_{p}$ and 
$\sigma\sim 10^{-15}~M^{4}_{p}$ respectively. This analysis explicitly shows that for $\sigma\sim 5\times 10^{-16}~M^{4}_{p}$  the tensor-to-scalar ratio and scalar spectral
tilt are constrained within the window,\bea 0.121&<r<&0.124,\\ 0.969&<n_{S}<&0.970,\eea which is consistent with $2\sigma$ CL constraints.

\begin{figure}[htb]
\centering
\includegraphics[width=9cm,height=7cm]{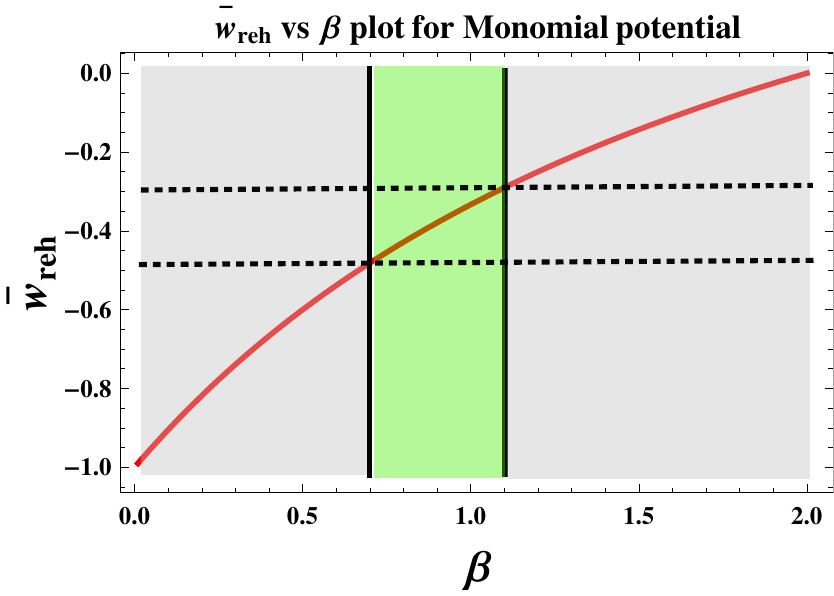}
\caption{Variation of the mean equation of state parameter with respect to the monomial index $\beta$. The green shaded region bounded by two vertical black coloured lines and two black coloured horizontal line represent the Planck $2\sigma$ allowed region and the rest of the light gray shaded
region is disfavoured by the Planck data and Planck+ BICEP2+Keck Array joint constraint. It is also observed from the plot that, if I fix the number of e-foldings within the window, $\Delta{\cal N}_{b}\approx {\cal O}(50-70)$, then the 
mean equation of state parameter $\bar{w}_{reh}$ is constrained as, $-0.48<\bar{w}_{reh}<-0.29$.
}
\label{fig5}
\end{figure}
Also using Eq~(\ref{m50}) the mean equation of state parameter during reheating can be computed as~\footnote{To satisfy the CMB constraints, if I fix $\Delta{\cal N}_{b}\approx {\cal O}(50-70)$, then the 
mean equation of state parameter $\bar{w}_{reh}$ is constrained as, \be -0.48<\bar{w}_{reh}<-0.29.\ee}:
\be\label{m57}
\bar{w}_{reh}=\frac{\beta-2}{\beta+2}.
\ee 
Variation of the mean equation of state parameter with respect to the monomial index $\beta$ is explicitly shown in fig.~\ref{fig5}. The green shaded region bounded by two vertical black coloured lines and two black coloured horizontal line represent the Planck $2\sigma$ allowed region and the rest of the light gray shaded
region is disfavoured by the Planck data and Planck+ BICEP2+Keck Array joint constraint.

Hence using Eq~(\ref{m16v2}), I get the following stringent constraint on the upper bound on the monomial index $\beta$ of the inflationary potential in terms of the magnetic field at the present epoch as:
\be \label{m58}
\beta <\frac{4\left(1+\frac{1}{\Delta\bar{\cal N}_{b}}\ln\left(\frac{\sqrt{B_0}}{(2\rho_{\gamma})^{1/4}}\right)\right)}{\left(1-\frac{2}{\Delta\bar{\cal N}_{b}}\ln\left(\frac{\sqrt{B_0}}{(2\rho_{\gamma})^{1/4}}\right)\right)}
 \ee
 where
 \be \label{m59}
 \Delta\bar{\cal N}_{b}={\cal N}_{reh;b}-{\cal N}_{cmb;b}+\Delta{\cal N}_{b}.
 \ee
 Here using Eq~(\ref{m55v2}) in Eq~ one can derive the following constraint on $\Delta\bar{\cal N}_{b}$ as:
 \be\label{m59v1}
 \Delta\bar{\cal N}_{b}<\frac{398}{\Delta{\cal N}_{b}}\ln\left(\frac{\sqrt{B_0}}{(2\rho_{\gamma})^{1/4}}\right).
 \ee
\begin{figure*}[htb]
\centering
\subfigure[$B_{0}$ vs $\beta$.]{
    \includegraphics[width=7.2cm,height=7.1cm] {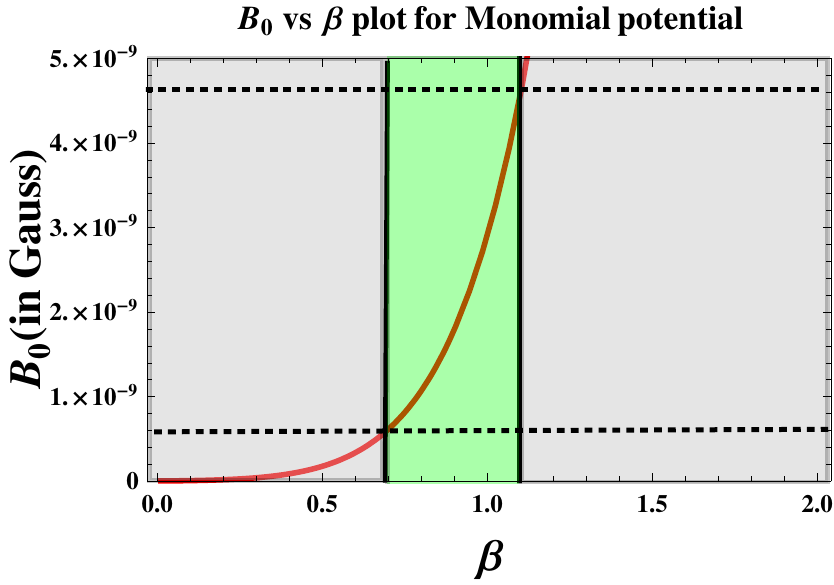}
    \label{fig6a}
}
\subfigure[$\rho_{reh}$ vs $\beta$.]{
    \includegraphics[width=7.2cm,height=7.1cm] {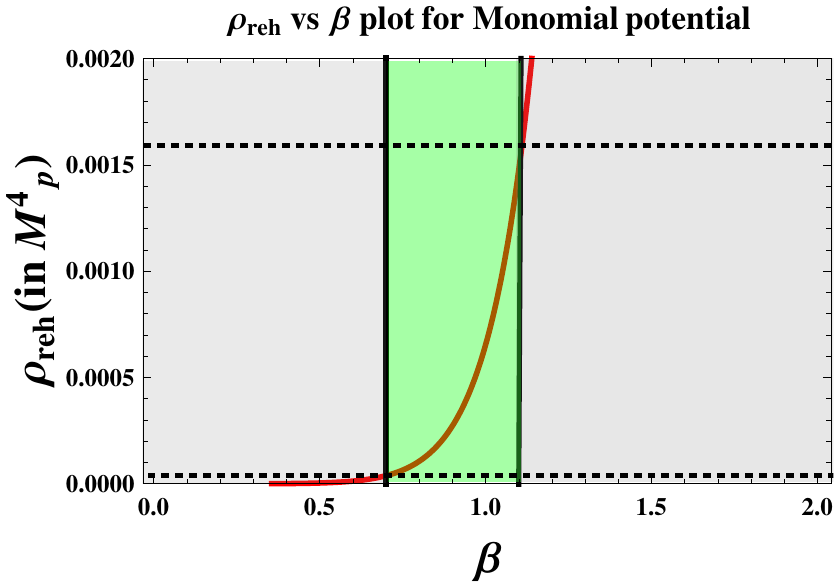}
    \label{fig6b}
}
\subfigure[$\ln(R_{sc})$ vs $\beta$.]{
    \includegraphics[width=9.2cm,height=8cm] {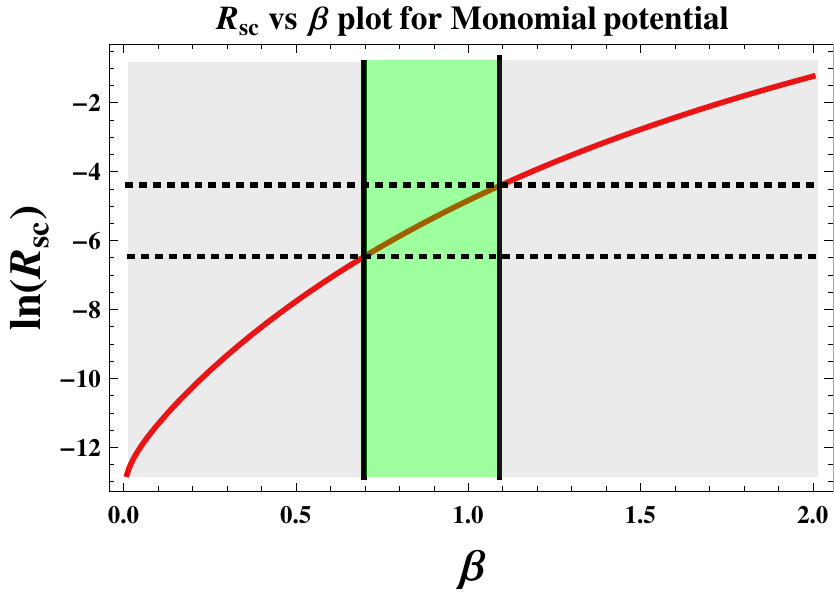}
    \label{fig6c}
}
\caption[Optional caption for list of figures]{Variation of \ref{fig6a} the magnetic field at the present epoch $B_{0}$, \ref{fig6b} reheating energy density and \ref{fig6c} logarithm
of reheating characteristic parameter with respect to the characteristic index $\beta$ of
the hilltop potential for $\Delta{\cal N}_{b}=50$, $|\Delta\bar{\cal N}_{b}|=7$ and $\sigma\sim 5\times 10^{-16}~M^{4}_{p}$..
The green shaded region bounded by two vertical black coloured lines and two black coloured horizontal line represent the Planck $2\sigma$ allowed region and the rest of the light gray shaded
region is disfavoured by the Planck data and Planck+ BICEP2+Keck Array joint constraint. In \ref{fig6a}-\ref{fig6c} the black horizontal dotted line correspond to the $2\sigma$ CL constrained value of the 
magnetic field at the present epoch, reheating energy density and $\ln(R_{sc})$.} 
\label{fig6}
\end{figure*}
\begin{figure*}[htb]
\centering
\subfigure[$\rho_{reh}$ vs $B_{0}/\sqrt{2\rho_{\gamma}}$.]{
    \includegraphics[width=12.2cm,height=8.6cm] {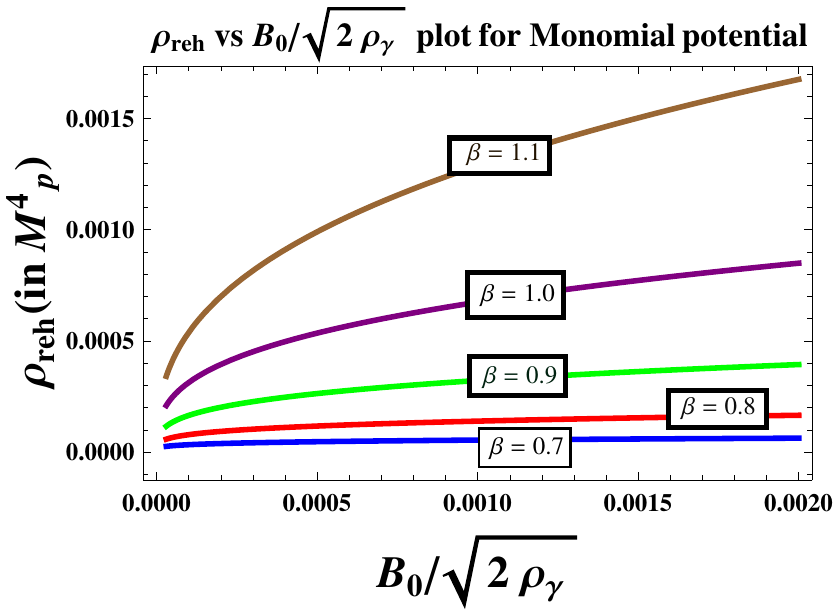}
    \label{fig7a}
}
\subfigure[$\ln(R_{sc})$ vs $B_{0}/\sqrt{2\rho_{\gamma}}$.]{
    \includegraphics[width=12.2cm,height=8.6cm] {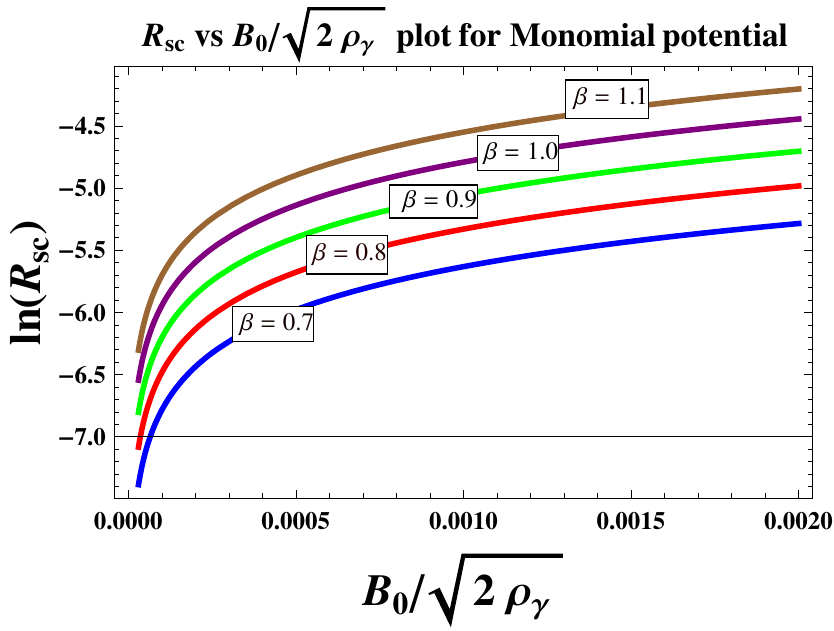}
    \label{fig7b}
}
\caption[Optional caption for list of figures]{Variation of \ref{fig7a} the reheating energy density and \ref{fig7b} logarithm of reheating characteristic parameter with respect to the scaled magnetic field at
the present epoch $\frac{B_{0}}{\sqrt{2\rho_{\gamma}}}$ for
the characteristic index $\beta=0.7(\textcolor{blue}{blue}),0.8(\textcolor{red}{red}),0.9(\textcolor{green}{green}),1.0(\textcolor{purple}{purple}),1.1(\textcolor{brown}{brown})$. Here I fix 
$\Delta{\cal N}_{b}=50$, $\Delta\bar{\cal N}_{b}=7$ and $\sigma\sim 5\times 10^{-16}~M^{4}_{p}$.} 
\label{fig7}
\end{figure*}
 Further using Eq~(\ref{m34}) the reheating energy density can be computed as:
 \bea \label{m60}
\rho_{reh}=\frac{\left(8.46\times 10^{-7}~M^{4}_p\right)}{\left[1+\left(1+\frac{2}{\beta}\right)
\Delta{\cal N}_{b}\right]}\times\exp\left[\frac{6\beta\Delta\bar{\cal N}_{b}}{\beta+2}\right]
\times\left(\frac{B_0}{\sqrt{2\rho_{\gamma}}}\right)^{-\frac{\beta}{(\beta-4)}}~~~~~~~
\eea
and also using the numerical constraint on the energy density at the end of inflation, as stated in Eq~(\ref{o2}),
I get following stringent constraint on the magnetic field measured at the present epoch in terms of model parameter $\beta$ for 
instantaneous reheating~\footnote{For instantaneous reheating the energy density of inflaton at the end of inflation is instantaneously converted to the 
reheating energy density or radiation and the instantaneous transition occurs at $\rho_{end}=\rho_{reh}$. 
Commonly this physical situation is known as instantaneous entropy generation scenario \cite{Planck:2013jfk}.}
as:
\bea\label{m60v2}
\frac{\left(8.68\times 10^{-6}\right)^{\frac{4-\beta}{\beta}}\left[1+\left(1+\frac{2}{\beta}\right)
\Delta{\cal N}_{b}\right]^{\frac{4-\beta}{\beta}}}{\exp\left[\frac{6(4-\beta)\Delta\bar{\cal N}_{b}}{\beta+2}\right]}<
\frac{B_0}{\sqrt{2\rho_{\gamma}}}<\frac{\left(7.02\times 10^{-4}\right)^{\frac{4-\beta}{\beta}}\left[1+\left(1+\frac{2}{\beta}\right)
\Delta{\cal N}_{b}\right]^{\frac{4-\beta}{\beta}}}{\exp\left[\frac{6(4-\beta)\Delta\bar{\cal N}_{b}}{\beta+2}\right]}.~~~~~~~
\eea
Next using Eq~(\ref{m47}), I get the following constraint on the dimensionless magnetic density parameter:
\be\label{m61}
 \Omega_{B_{end}}=\frac{B^{4}_{0}M^{6}_{p}}{24\sigma H^{2}_{0}\rho^2_{\gamma}\left[1+\left(1+\frac{2}{\beta}\right)\Delta{\cal N}_{b}\right]^2}\times(7.16\times 10^{-13})\times\exp\left[8\Delta\bar{\cal N}_{b}\right].
 \ee
 Finally the rescaled reheating parameter can be expressed in terms of the model parameters of the monomial  models of inflationary potential as:
 \be\label{m62}
 R_{sc}=\frac{3.03\times 10^{-2}}{\left[1+\left(1+\frac{2}{\beta}\right)\Delta{\cal N}_{b}\right]^{1/4}}\times\exp\left[\frac{3\beta\Delta\bar{\cal N}_{b}}{2(\beta+2)}
 \right]
 \times\left(\frac{B_0}{\sqrt{2\rho_{\gamma}}}\right)^{1/2}~~~~~
 \ee
 and using the numerical constraint on the rescaled reheating parameter as stated in Eq~(\ref{o1}) I get the lower bound on the present value of the magnetic field for the monomial potentials as:
 \bea\label{m63}
 \frac{B_0}{\sqrt{2\rho_{\gamma}}}>\frac{8.6\times 10^{-23}\times\left[1+\left(1+\frac{2}{\beta}\right)\Delta{\cal N}_{b}\right]^{1/2}}{\exp\left[\frac{3\beta\Delta\bar{\cal N}_{b}}{\beta+2}
 \right]}
 \eea
 In fig.~\ref{fig6a}, fog.~\ref{fig6b} and in fig.~\ref{fig6c} I have explicitly shown the variation of the magnetic field at the present epoch $B_{0}$, reheating energy density $\rho_{reh}$ 
 and logarithm of reheating characteristic parameter $\ln(R_{sc})$ with respect to the characteristic index $\beta$ of the monomial potential for the number of e-foldings $\Delta{\cal N}_{b}=50$.
The green shaded region bounded by two vertical black coloured lines and two black coloured horizontal line represent the Planck $2\sigma$ allowed region and the rest of the light gray shaded
region is disfavoured by the Planck data and Planck+ BICEP2+Keck Array joint constraint. Also in fig.~\ref{fig7a} and in fig.~\ref{fig7b} I have depicted the behaviour 
of the reheating energy density $\rho_{reh}$ and logarithm of reheating characteristic parameter $\ln(R_{sc})$, with respect to the scaled magnetic field at the present epoch $B_{0}/\sqrt{2\rho_{\gamma}}$ for
the characteristic index $0.7\leq \beta\leq 1.1$. 

Further 
using Eq~(\ref{m55}) in Eq~(\ref{eq16}) and Eq~(\ref{zaw3}) finally I get the following constraints on the regulating factor 
  within RSII setup 
 as~\footnote{After fixing $\Delta{\cal N}_{b}\approx {\cal O}(50-70)$, 
the regulating factor within RSII can be constrained as, \be 3.06\times 10^{-19}<\Sigma_{b}(k_{L}=k_0,k_{*})\times\left(\frac{M^4_p}
{\sigma}\right)^{2/5}<3.13\times 10^{-19},\ee which is consistent with
the upper bound mentioned in Eq~(\ref{tune}).}:
\bea\label{eq16sd}
     \displaystyle \Sigma_{b}(k_{L}=k_0,k_{*})\times\left(\frac{M^4_p}{\sigma}\right)^{2/5}\approx
{\cal O}(1.58\times10^{-21})\times\left[1+\left(1+\frac{2}{\beta}\right)\Delta{\cal N}_{b}\right]
    \eea
    which is compatible with the observed/measured bound on CP asymmetry and baryon asymmetry parameter.
    
From fig.~\ref{fig6a}, fig.~\ref{fig6b} and fig.~\ref{fig6c} I get the following $2\sigma$ constraints on
magneto reheating cosmological parameters computed from the monomial inflationary model:
\bea\label{f1f}
5.969\times 10^{-10}~{\rm Gauss}<B_{0}=&\sqrt{\frac{I_{\xi}(k_{L}=k_{0},k_{\Lambda})}{2\pi^2}A_{\bf B}}<&4.638\times 10^{-9}~{\rm Gauss},\\
\label{f1x1}
1.940\times 10^{-132}~M^{4}_{p}&<\rho_{B_{0}}=B^{2}_{0}/2<&1.171\times 10^{-130}~M^{4}_{p},\\
\label{f2f} 4.061\times10^{-5}~M^{4}_{p}&<\rho_{reh}<&1.591\times10^{-3}~M^{4}_{p},\\
\label{f2fx} 6.227\times10^{-4}\times g^{-1/4}_*~M_{p}&<T_{reh}<&4.836\times10^{-3}\times g^{-1/4}_*~M_{p},\\
\label{f2fxxx}\Gamma_{total}&\sim&0.24~M_{p},\\
\label{f3f}  1.55\times 10^{-3}&<R_{sc}<&1.24\times 10^{-2},\\
\label{f6f} \epsilon_{\bf CP}& \sim& {\cal O}(10^{-6}),\\
\label{f7f} \eta_{B}&\sim& {\cal O}(10^{-9}),\\
\label{f8f} 0.121&<r<&0.124,\\
\label{f9f} 0.969&<n_{S}<&0.970,\\
\label{f9x}2.3794\times 10^{-9}&<P_{S}<&2.3798\times 10^{-9},\\
\label{f10f} 8.08\times 10^{-3}~M_{p} &<\sqrt[4]{V_{0}}<&8.13\times 10^{-3}~M_{p},
\eea
for the number of e-foldings $\Delta{\cal N}_{b}=50$, $|\Delta\bar{\cal N}_{b}|=7$, mean equation of state parameter $-0.48<\bar{w}_{reh}<-0.29$ and $\Omega_{rad}h^2\sim2.5\times 10^{-5}$, along with the following restricted model parameter space:
\bea \label{f11f} 0.7&<\beta<&1.1,\\
\label{f12f} \sigma&\sim& 5\times 10^{-16}~M^{4}_{p},\\
\label{f13f} M_{5}&\sim& \left(1.042\times 10^{-32}\times\frac{M^{8}_{p}}{|\tilde{\Lambda}_{5}|}\right)^{1/3}.
\eea
It is important to note that, if I choose different parameter space by allowing fine tuning in--(1) the energy scale of monomial
potential $V_{0}=M^4$, 
 (2) the brane tension $\sigma$ and (3) the characteristic index of the monomial potential $\beta$ then the overall analysis and the obtained
results suggests that-
\begin{itemize}
 \item  For $\beta<0.7$, the amplitude of the scalar power spectrum $P_{S}$ match the Planck 2015 data and 
 also consistent with the joint constraint obtained from Planck +BICEP2 +Keck Array. But the allowed range for 
 scalar spectral tilt $n_{S}$ is outside the observational window. Also in this regime the value of tensor-to-scalar ratio $r$ is lower compared to
 the upper bound 
 i.e. $r<<0.12$. On the other hand, for very low $\beta$ the estimated value of the magnetic field at the present epoch $B_{0}$
 from the monomial model is very very small and can reach up to the lower bound \be B_{0}>10^{-15}~{\rm Gauss}.\ee Similarly for low $\beta$, the reheating 
 energy density $\rho_{reh}$ or equivalently the reheating temperature $T_{reh}$ falls down and also the rescaled reheating parameter $R_{sc}$ decrease.
 
 \item For $1.1<\beta<1.2$, both the amplitude of the scalar power spectrum $P_{S}$ and the scalar spectral tilt $n_{S}$ are perfectly consistent with
 the Planck 2015 data and also consistent with the joint constraint obtained from Planck+BICEP2+Keck Array. But for $\beta>1.2$ the estimated 
 value of the amplitude of the scalar power spectrum falls down from its predicted bound from observation. Also for $\beta>1.2$ region
 the value of tensor-to-scalar ratio $r$ is very very large compared to its the upper bound 
 i.e. $r>>0.12$. As $\beta$ increases the estimated value of the magnetic field at the present epoch $B_{0}$ exceeds the upper
 bound i.e. \be B_{0}>>10^{-9}~{\rm Gauss}\ee as obtained from Faraday rotation.
 Additionally in the large $\beta$ regime the reheating 
 energy density $\rho_{reh}$ or equivalently the reheating temperature $T_{reh}$ and the rescaled reheating parameter $R_{sc}$ are not consistent 
 with the observational constraints.
\end{itemize}

\subsection{Hilltop Models}
\begin{figure*}[htb]
	\centering
	\subfigure[For~$\mu=10~M_{p}$.]{
		\includegraphics[width=10.2cm,height=5.9cm] {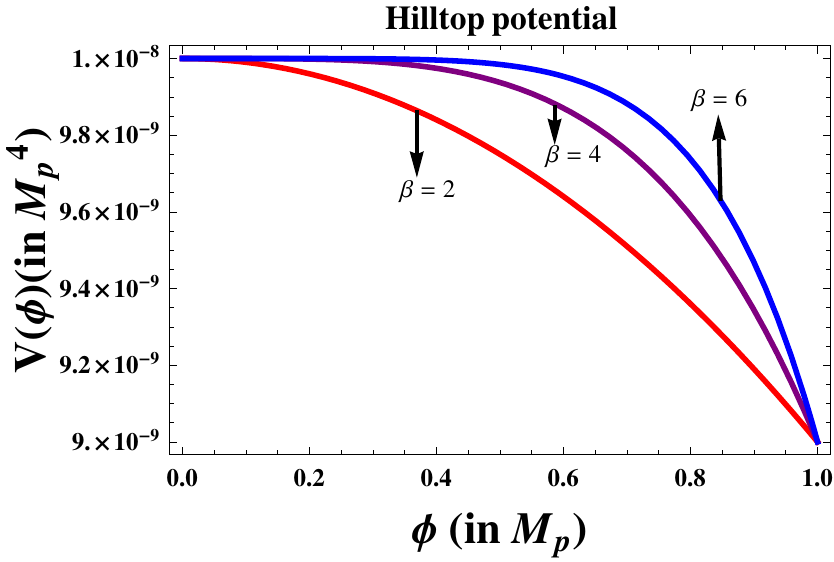}
		\label{fig2a}
	}
	\subfigure[For~$\mu=1~M_{p}$.]{
		\includegraphics[width=10.2cm,height=5.9cm] {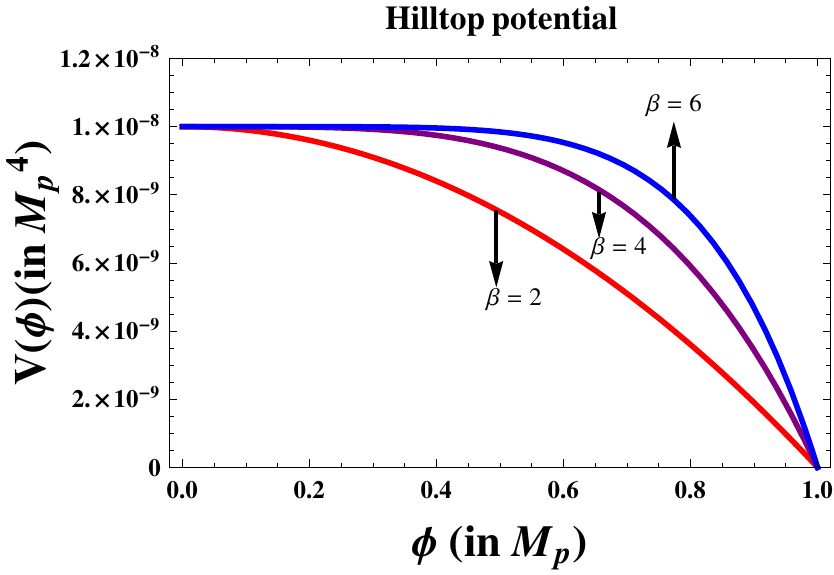}
		\label{fig2b}
	}
	\subfigure[For~$\mu=0.1~M_{p}$.]{
		\includegraphics[width=10.2cm,height=5.9cm] {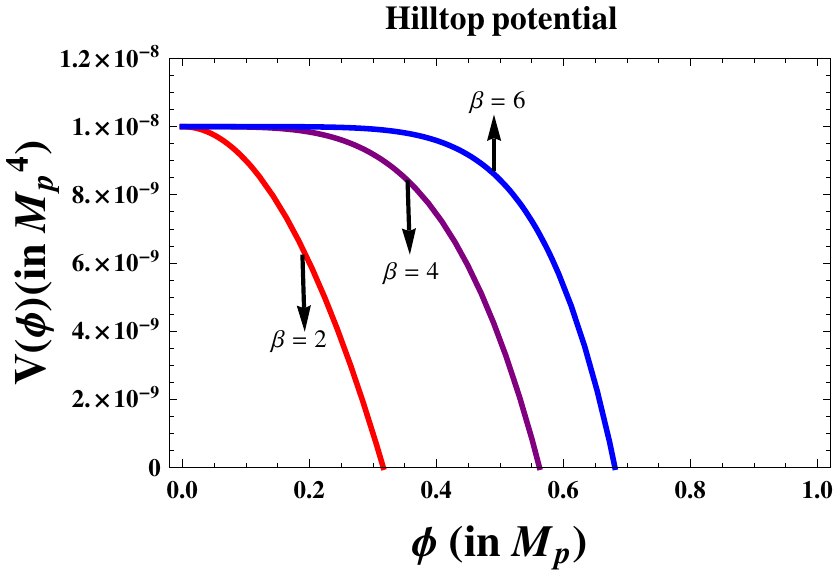}
		\label{fig2c}
	}
	\caption[Optional caption for list of figures]{Variation of the hilltop potential for the index $\beta=2, 4, 6$. Here I fix the tunable scale at $\sqrt[4]{V_{0}}=4.12\times 10^{-3}~M_{p}=10^{16}~GeV$.} 
	\label{fig2}
\end{figure*}

\begin{figure*}[htb]
	\centering
	\subfigure[$r$ vs $n_{S}$.]{
		\includegraphics[width=9.2cm,height=6.3cm] {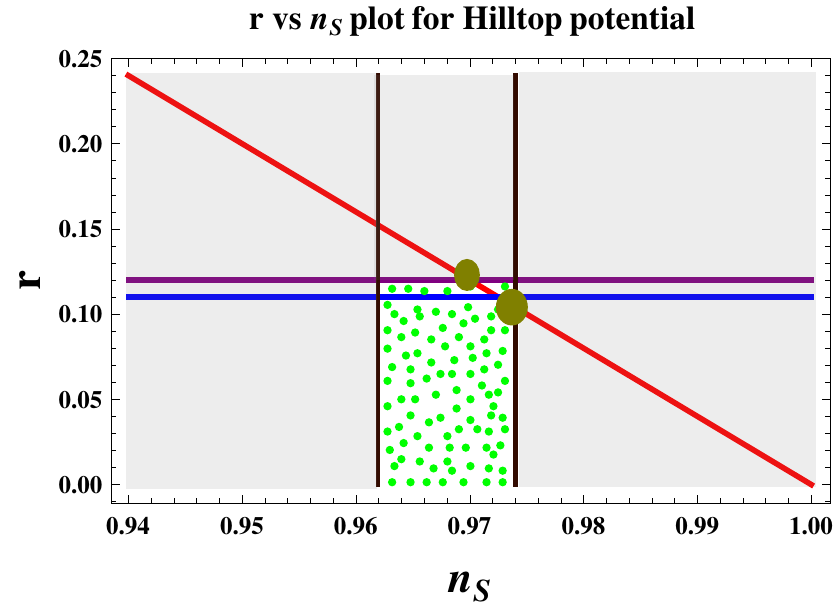}
		\label{fig8a}
	}
	\subfigure[$r$ vs $\beta$.]{
		\includegraphics[width=9.2cm,height=6.3cm] {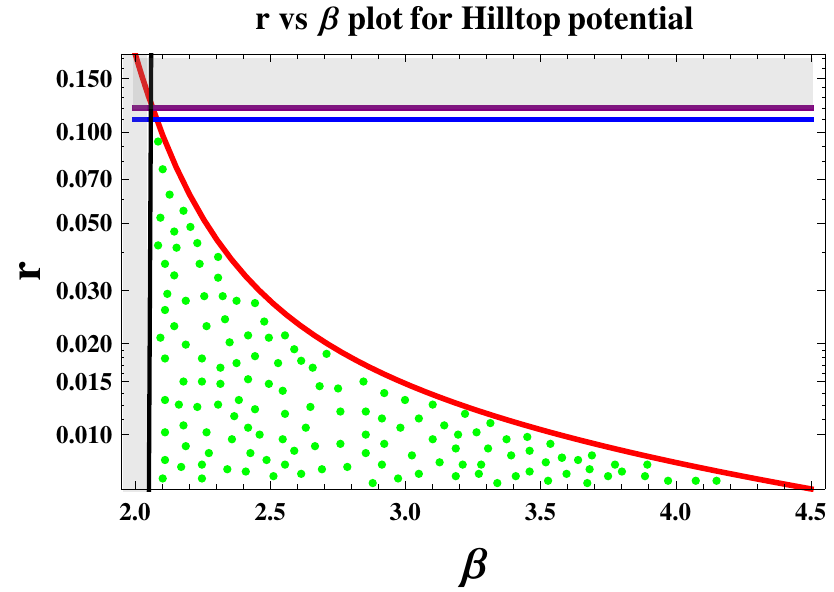}
		\label{fig8b}
	}
	\caption[Optional caption for list of figures]{Behaviour of the tensor-to-scalar ratio $r$ with respect to \ref{fig8a} the scalar
		spectral index $n_{S}$ and \ref{fig8b} the characteristic parameter of the hilltop potential $\beta$
		for the brane tension $\sigma\sim 10^{-9}~M^{4}_{p}$ and the mass scale parameter $\mu=5.17~M_p$. The purple and blue coloured line represent the upper bound of tenor-to-scalar ratio allowed by Planck+ BICEP2+Keck Array joint constraint and only Planck 2015 data respectively. The small and the big bubbles
		represent two consecutive points in $r-n_{S}$ plane, where for the small bubble $\Delta{\cal N}_{b}=50, r=0.124, n_{S}=0.969$ and for the big bubble $\Delta{\cal N}_{b}=70, r=0.121, n_{S}=0.970$ respectively.
		In \ref{fig8a} and \ref{fig8b} the green dotted region signifies the Planck $2\sigma$ allowed region and the rest of the light grey shaded region is excluded by the Planck data and Planck+ BICEP2+Keck Array
		joint constraint. From \ref{fig8a} it 
		is observed that, within $50<\Delta{\cal N}_{b}<70$ the hilltop potential is favoured for the characteristic index $\beta>2.04$, by Planck 2015 data and Planck+ BICEP2+Keck Array joint analysis.
		In \ref{fig8b} I have explicitly shown that the in $r-\beta$ plane the observationally favoured lower bound for the characteristic index of the hilltop potential is $\beta>2.04$.} 
	\label{fig8}
\end{figure*}

In case of hilltop models the inflationary potential can be represented by the following functional form:
\be\label{m64}
V(\phi)=V_{0}\left[1-\left(\frac{\phi}{\mu}\right)^{\beta}\right]
\ee
where $V_{0}=M^4$ is the tunable energy scale, which is necessarily required to fix the amplitude of the CMB anisotropies and
$\beta$ is the characteristic index which characterizes the feature of the potential.
In the present context $V_{0}$ mimics the role of vacuum energy and the scale of inflation is fixed by this correction term.
The variation of the hilltop potential for the index $\beta=2, 4, 6$, mass parameter $\mu=0.1~M_p ,1~M_p ,10~M_p$ and 
the tunable scale \be \sqrt[4]{V_{0}}=4.12\times 10^{-3}~M_{p}=10^{16}~{\rm GeV}\ee
is shown in fig.~\ref{fig2a}, fig.~\ref{fig2b} and fig.~\ref{fig2c} respectively. 
To analyze the detailed features of the hilltop potential here I start with the definition of number of e-foldings
$\Delta{\cal N}_{b}(\phi)$ in the high energy regime of RSII setup (see Appendix \ref{a7} for details), using which I get: 
\be\label{m65}
\Delta{\cal N}_{b}(\phi)\approx\frac{V_{0}\mu^{p}}{2\sigma \beta\left(\beta-2\right)M^{2}_{p}}\left(\phi^{2-\beta}-\phi^{2-\beta}_{end}\right).
\ee
Further setting $\phi=\phi_{cmb}$ in Eq~(\ref{m65}), the field value at the horizon crossing can be computed as:
\be\label{m66}
\phi_{cmb}\approx\phi_{end}\left[1+\frac{2\sigma\beta\left(\beta-2\right)M^{2}_{p}\Delta{\cal N}_{b}}{\phi^{2-\beta}_{end}\mu^{\beta}V_0}\right]^{\frac{1}{2-\beta}}
\ee
where $\phi_{end}$ represents the field value of inflaton at the end of inflation. Within RSII setup from the violation of the slow-roll conditions one can compute:
\be\label{m67}
\phi_{end}\approx\left(\frac{V_{0}}{2\sigma \beta^{2}}\right)^{\frac{1}{2(\beta-1)}}\left(\frac{\mu}{M_p}\right)^{\frac{\beta}{\beta-1}}~M_p.
\ee
From hilltop models of inflation the scale of the potential at the horizon crossing and at the end of inflation can be computed as:
\bea\label{m68}
\rho_{cmb}&\approx& V(\phi_{cmb})=V_{0}\left[1-\left(\frac{\phi_{cmb}}{\mu}\right)^{\beta}\right]\nonumber\\
&=&V_{0}\left[1-
\left(\frac{V_{0}}{2\sigma \beta^{2}}\right)^{\frac{\beta}{2(\beta-1)}}
\left(\frac{\mu}{M_p}\right)^{\frac{\beta}{\beta-1}}\left\{1+\frac{2\sigma\beta\left(\beta-2\right)M^{\beta}_{p}
\Delta{\cal N}_{b}}{\left(\frac{V_{0}}{2\sigma \beta^{2}}\right)^{\frac{2-\beta}{2(\beta-1)}}\left(\frac{\mu}{M_p}
\right)^{\frac{\beta(2-\beta)}{\beta-1}}\mu^{\beta}V_0}\right\}^{\frac{\beta}{2-\beta}}\right],~~~~~~~~\\
\rho_{end}&\approx& V(\phi_{end})=V_{0}\left[1-\left(\frac{\phi_{end}}{\mu}\right)^{\beta}\right]
=V_{0}\left[1-\left(\frac{V_{0}}{2\sigma \beta^{2}}\right)^{\frac{\beta}{2(\beta-1)}}
\left(\frac{\mu}{M_p}\right)^{\frac{\beta}{\beta-1}}\right].
\eea
Further using the consistency condition in the high energy regime of RSII braneworld, as stated in Eq~(\ref{cv1}) of the Appendix \ref{a7}, one can derive the following expressions for the tensor to scalar ratio and scalar spectral tilt as:
\bea\label{m69cv} P_{S}(k_*)&=&
\frac{V_{0}\left[1-
\left(\frac{V_{0}}{2\sigma \beta^{2}}\right)^{\frac{\beta}{2(\beta-1)}}
\left(\frac{\mu}{M_p}\right)^{\frac{\beta}{\beta-1}}\left\{1+\frac{2\sigma\beta\left(\beta-2\right)M^{\beta}_{p}
\Delta{\cal N}_{b}}{\left(\frac{V_{0}}{2\sigma \beta^{2}}\right)^{\frac{2-\beta}{2(\beta-1)}}\left(\frac{\mu}{M_p}
\right)^{\frac{\beta(2-\beta)}{\beta-1}}\mu^{\beta}V_0}\right\}^{\frac{\beta}{2-\beta}}\right]}{36\pi^2\left\{1+\frac{2\sigma\beta\left(\beta-2\right)
\Delta{\cal N}_{b}}{\left(\frac{V_{0}}{2\sigma \beta^{2}}\right)^{\frac{2-\beta}{2(\beta-1)}}\left(\frac{\mu}{M_p}
\right)^{\frac{\beta}{\beta-1}}V_0}\right\}^{\frac{2(\beta-1)}{2-\beta}}}
,~~~~~~~~\\
\label{m69} r(k_*)&=&24\left\{1+\frac{2\sigma\beta\left(\beta-2\right)
\Delta{\cal N}_{b}}{\left(\frac{V_{0}}{2\sigma \beta^{2}}\right)^{\frac{2-\beta}{2(\beta-1)}}\left(\frac{\mu}{M_p}
\right)^{\frac{\beta}{\beta-1}}V_0}\right\}^{\frac{2(\beta-1)}{2-\beta}},\\
\label{m70} n_{S}(k_*)-1&\approx&-6\left\{1+\frac{2\sigma\beta\left(\beta-2\right)
\Delta{\cal N}_{b}}{\left(\frac{V_{0}}{2\sigma \beta^{2}}\right)^{\frac{2-\beta}{2(\beta-1)}}\left(\frac{\mu}{M_p}
\right)^{\frac{\beta}{\beta-1}}V_0}\right\}^{\frac{2(\beta-1)}{2-\beta}}.\eea
and to satisfy the joint constraint on the scalar spectral tilt and upper bound of tensor-to-scalar ratio as observed by Planck (2013 and
2015) and Planck+BICEP2+Keck Array, one need the following constraint on the parameters of the inflationary potential:
\be\label{m71}
\frac{2\sigma
\Delta{\cal N}_{b}}{V_0}<\frac{\exp\left[\frac{2.65(\beta-2)}{(\beta-1)}\right]-1}{\beta(\beta-2)}\left(\frac{V_{0}}{2\sigma \beta^{2}}\right)^{\frac{2-\beta}{2(\beta-1)}}\left(\frac{\mu}{M_p}
\right)^{\frac{\beta}{\beta-1}}.
\ee 
The behaviour of the tensor-to-scalar ratio $r$ with respect to the scalar spectral index $n_{S}$ and the characteristic parameter of
the hilltop potential $\beta$ are plotted in fig.~\ref{fig8a} and fig.~\ref{fig8b} respectively. 

\begin{figure*}[htb]
	\centering
	\subfigure[$P_{S}$ vs $n_{S}$.]{
		\includegraphics[width=7.2cm,height=7.5cm] {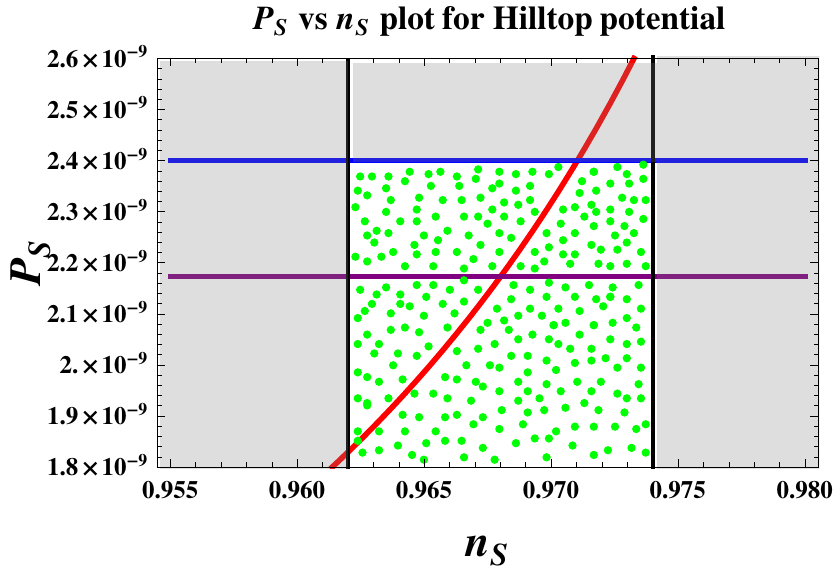}
		\label{fig3axx}
	}
	\subfigure[$P_{S}$ vs $\beta$.]{
		\includegraphics[width=7.2cm,height=7.5cm] {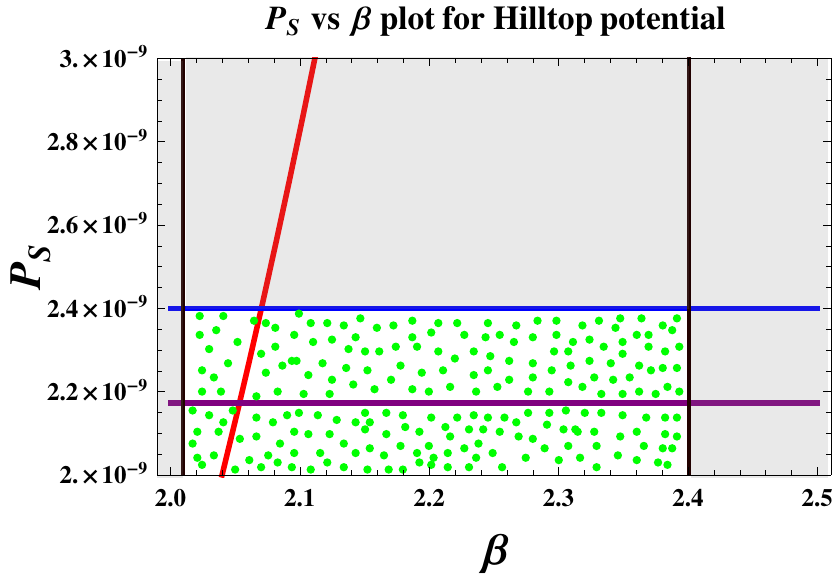}
		\label{fig3bxx}
	}
	\subfigure[$n$ vs $\beta$.]{
		\includegraphics[width=9.8cm,height=8.5cm] {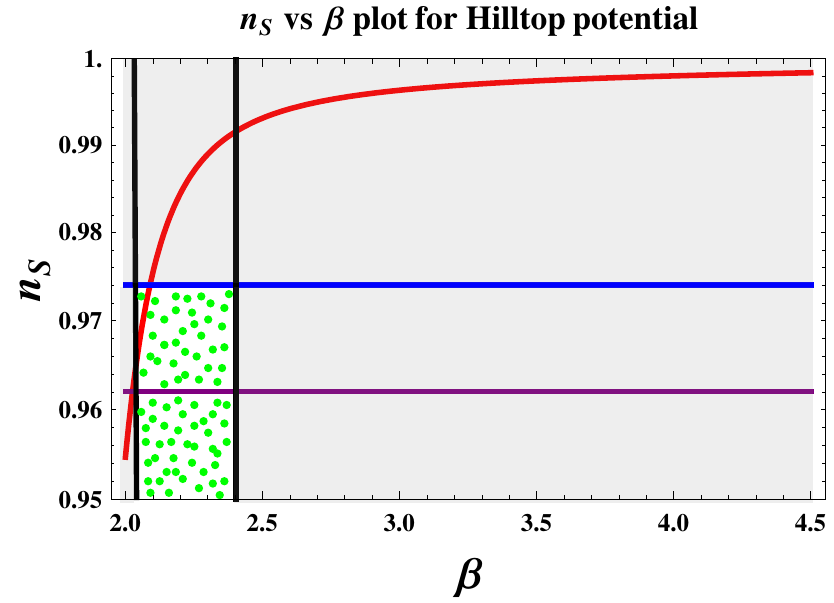}
		\label{fig3cxx}
	}
	\caption[Optional caption for list of figures]{Variation of the \ref{fig3axx} scalar power spectrum $P_{S}$ vs scalar
		spectral index $n_{S}$, \ref{fig3bxx} scalar power spectrum $P_{S}$ vs index $\beta$
		and \ref{fig3cxx} scalar power spectrum $n_{S}$ vs index $\beta$. The purple and blue coloured line represent the upper and lower bound allowed by WMAP+Planck 2015 data respectively. 
		The green dotted region bounded by two vertical black coloured lines represent the Planck $2\sigma$ allowed region and the rest of the light gray shaded region is disfavoured by the Planck+WMAP constraint.} 
	\label{fig3xx}
\end{figure*}
From \ref{fig8a} it 
is observed that, within $50<\Delta{\cal N}_{b}<70$ the hilltop potential is favoured for the characteristic index
\be\beta>2.04,\ee by Planck 2015 data and Planck+ BICEP2+Keck Array joint analysis.
In \ref{fig8b} I have explicitly shown that the in $r-\beta$ plane the observationally favoured window for the characteristic index is $\beta>2.04$.
Additionally it is important to note that, for hilltop potentials embedded in the high energy regime of RSII braneworld, 
the consistency relation between tensor-to-scalar ratio $r$ and the scalar spectral $n_{S}$ is given by, \be r\approx 4(1-n_{S}).\ee On the other 
hand in the low energy regime of RSII braneworld or equivalently in the GR limiting situation, the consistency relation between  
tensor-to-scalar ratio $r$ and the scalar spectral $n_{S}$ is modified as,\be r\approx \frac{8}{3}(1-n_{S}).\ee This also clearly suggests that the 
estimated numerical value of the tensor-to-scalar ratio from the GR limit is different compared to its value in the high density regime of the 
RSII braneworld. To justify the validity of this statement, let me discuss a very simplest situation, where the scalar spectral index is constrained 
within \be 0.969<n_{S}<0.970,\ee as appearing in this paper. Now in such a case using the consistency relation in GR limit one can easily compute that the tensor-to-scalar is 
constrained within the window, \be 0.080<r<0.083,\ee which is pretty consistent with Planck 2015 result.

\begin{figure}[t]
\centering
\includegraphics[width=12cm,height=7cm]{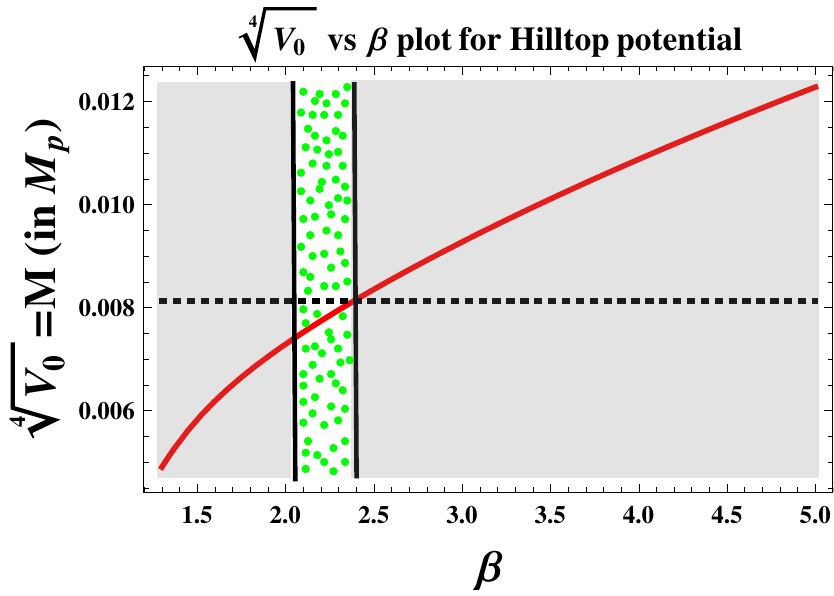}
\caption{Variation of the energy scale of the hilltop potential with respect to the characteristic index $\beta$ for the brane tension $\sigma\sim 10^{-9}~M^{4}_{p}$ and the mass scale parameter $\mu=~M_p$.
The green dotted region bounded by two vertical black coloured lines  and one black coloured horizontal line represent the Planck $2\sigma$ allowed region and the rest of the light gray shaded
region is disfavoured by the Planck data and Planck+ BICEP2+Keck Array joint constraint. This analysis explicitly shows that 
 the $2\sigma$ allowed
window for the parameter
$\beta$ within $2.04<\beta<2.4$ constraints the scale of inflation within $8.08\times 10^{-3}~M_{p} <\sqrt[4]{V_{0}}<8.13\times 10^{-3}~M_{p}$.
Here for $\sigma\sim 10^{-9}~M^{4}_{p}$ the tensor-to-scalar ratio and scalar spectral
tilt are constrained within the window, $0.121<r<0.124$ and $0.969<n_{S}<0.970$, which is consistent with $2\sigma$ CL constraints.
}
\label{fig9}
\end{figure}

Variation of the \ref{fig3axx} scalar power spectrum $P_{S}$ vs scalar spectral index $n_{S}$, \ref{fig3bxx} scalar power
spectrum $P_{S}$ vs index $\beta$
and \ref{fig3cxx} scalar power spectrum $n_{S}$ vs index $\beta$. The purple and blue coloured line
represent the upper and lower bound allowed by WMAP+Planck 2015 data respectively. 
The green dotted region bounded by two vertical black coloured lines represent the Planck $2\sigma$ allowed
region and the rest of the light gray shaded region is disfavoured by the Planck+WMAP constraint.
From the fig.~\ref{fig3axx}-fig.~\ref{fig3cxx} it is clearly observed that the characteristic index of the the
inflationary potential is constrained within the window \be 2.04<\beta<2.4\ee  for the amplitude of the scalar power spectrum, 
\be 2.3794\times 10^{-9}<P_{S}<2.3798\times 10^{-9}\ee and scalar spectral tilt, \be 0.969<n_{S}<0.970.\ee Now using Eq~(\ref{m69cv}), Eq~(\ref{m69}) and Eq~(\ref{m70}) one can 
write another consistency relation among the amplitude of the scalar power spectrum $P_{S}$, tensor-to-scalar ratio $r$ and scalar spectral
index $n_{S}$ for hilltop potentials embedded in the high density regime of RSII braneworld 
as: \bea P_{S}&=&\frac{V_{0}\left[1-
\left(\frac{V_{0}}{2\sigma \beta^{2}}\right)^{\frac{\beta}{2(\beta-1)}}
\left(\frac{\mu}{M_p}\right)^{\frac{\beta}{\beta-1}}\left(\frac{1-n_{S}}{6}\right)^{\frac{\beta}{2(\beta-1)}}\right]}{6\pi^2(1-n_{S})}\nonumber\\
&=&\frac{2V_{0}\left[1-
\left(\frac{V_{0}}{2\sigma \beta^{2}}\right)^{\frac{\beta}{2(\beta-1)}}
\left(\frac{\mu}{M_p}\right)^{\frac{\beta}{\beta-1}}\left(\frac{r}{24}\right)^{\frac{\beta}{2(\beta-1)}}\right]}{3\pi^2 r}. \eea
Further using Eq~(\ref{iopk}), I get the following stringent constraint on the tunable energy scale of the hilltop models of inflation:
\bea\label{m72}
V_{0}=M^{4}
&<& 5.98\times 10^{-8}~M^{4}_p.
\eea
The variation of the energy scale of the hilltop potential with respect to the characteristic index $\beta$ for the brane tension $\sigma\sim 10^{-9}~M^{4}_{p}$ and the mass scale parameter $\mu=~M_p$
is shown in fig.~\ref{fig9}. This analysis explicitly shows that for $\sigma\sim 10^{-9}~M^{4}_{p}$  the tensor-to-scalar ratio and scalar spectral
tilt are constrained within the window, $0.121<r<0.124$ and $0.969<n_{S}<0.970$, which is consistent with $2\sigma$ CL constraints.
\begin{figure*}[htb]
\centering
\subfigure[$B_{0}$ vs $\beta$.]{
    \includegraphics[width=7.2cm,height=7.5cm] {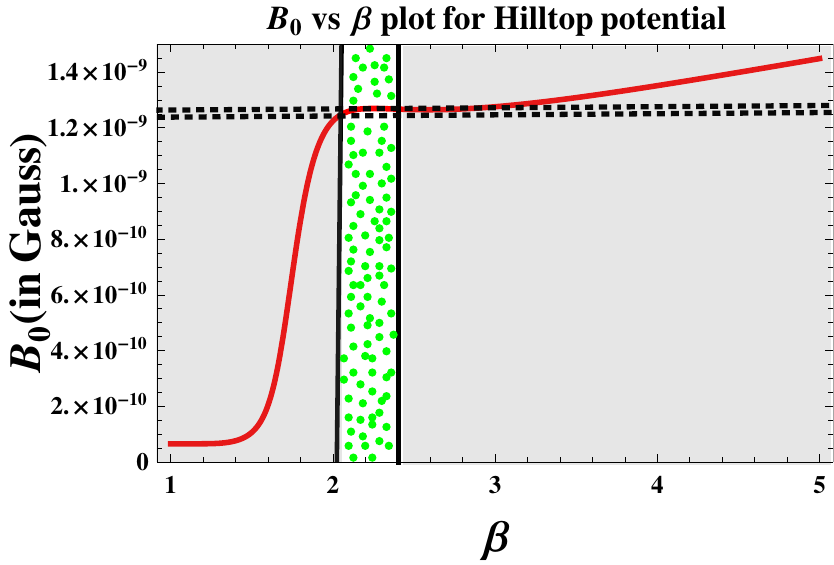}
    \label{fig10a}
}
\subfigure[$\rho_{reh}$ vs $\beta$.]{
    \includegraphics[width=7.2cm,height=7.5cm] {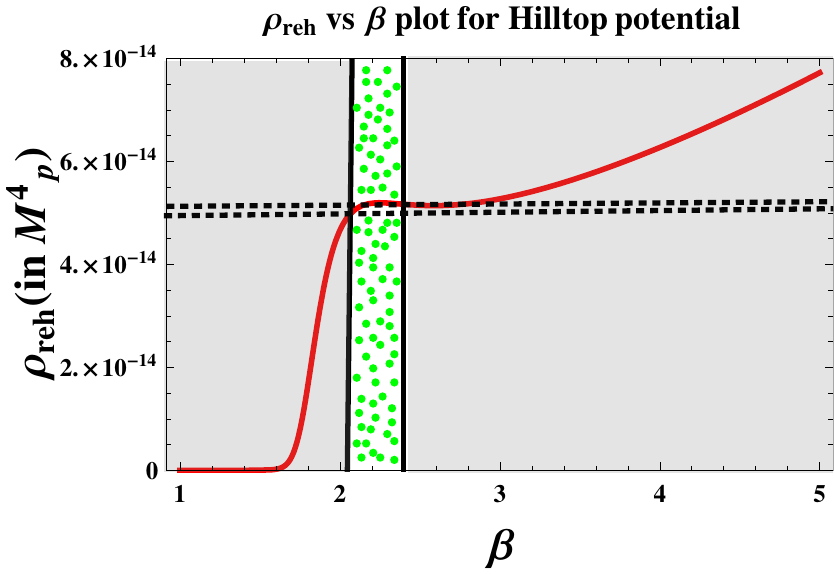}
    \label{fig10b}
}
\subfigure[$\ln(R_{sc})$ vs $\beta$.]{
    \includegraphics[width=9.2cm,height=7.5cm] {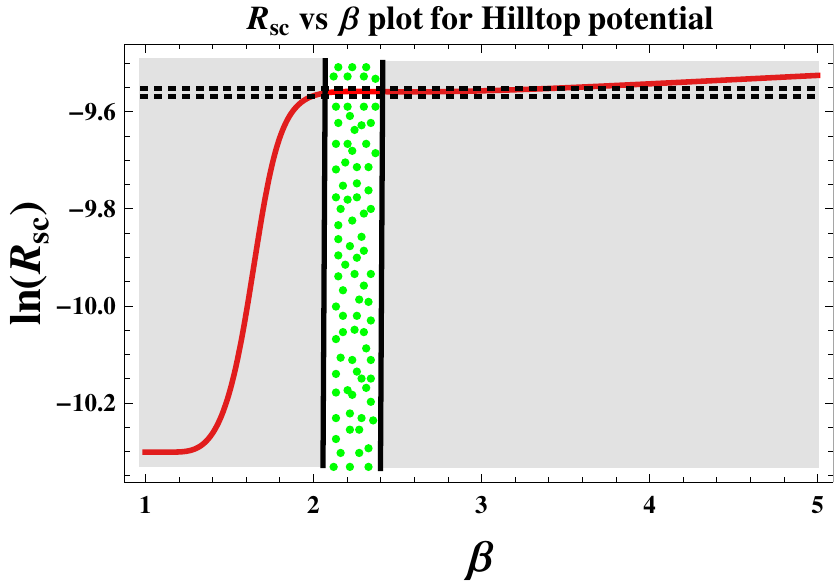}
    \label{fig10c}
}
\caption[Optional caption for list of figures]{Variation of \ref{fig10a} the magnetic field at the present epoch $B_{0}$, \ref{fig10b} reheating energy density and \ref{fig10c} logarithm
of reheating characteristic parameter with respect to the characteristic index $\beta$ of the hilltop potential for $\Delta{\cal N}_{b}=50$,
$|\Delta\bar{\cal N}_{b}|=11.5$, $\sigma\sim 10^{-9}~M^{4}_{p}$, $\mu=1~M_p$ and $\bar{w}_{reh}=0$.
The green dotted region bounded by two vertical black coloured lines and two black coloured horizontal line represent the Planck $2\sigma$ allowed region and the rest of the light grey shaded
region is disfavoured by the Planck data and Planck+ BICEP2+Keck Array joint constraint. In \ref{fig10a}-\ref{fig10c} the black horizontal dotted line correspond to the $2\sigma$ CL constrained value of the 
magnetic field at the present epoch, reheating energy density and $\ln(R_{sc})$.} 
\label{fig10}
\end{figure*}
\begin{figure*}[htb]
\centering
\subfigure[$\rho_{reh}$ vs $B_{0}/\sqrt{2\rho_{\gamma}}$.]{
    \includegraphics[width=12.2cm,height=9cm] {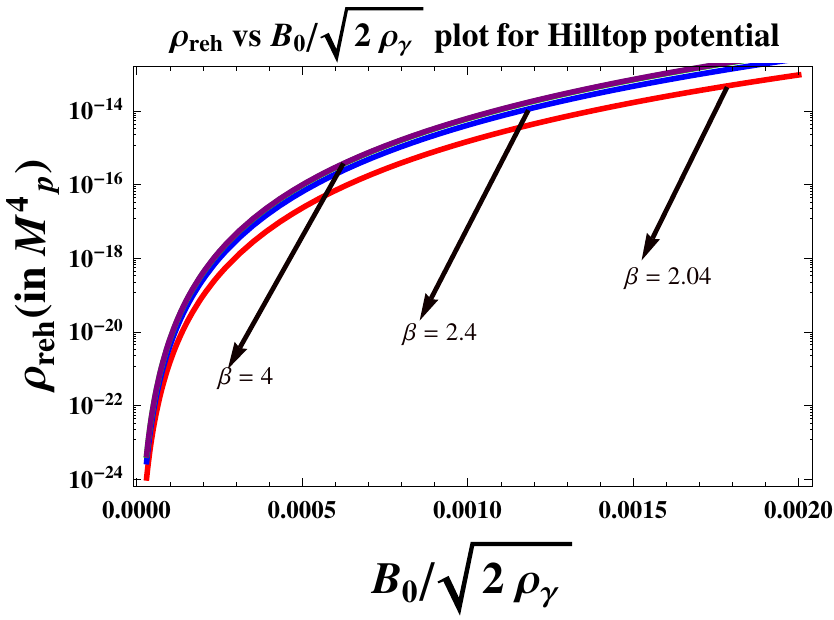}
    \label{fig11a}
}
\subfigure[$\ln(R_{sc})$ vs $B_{0}/\sqrt{2\rho_{\gamma}}$.]{
    \includegraphics[width=12.2cm,height=9cm] {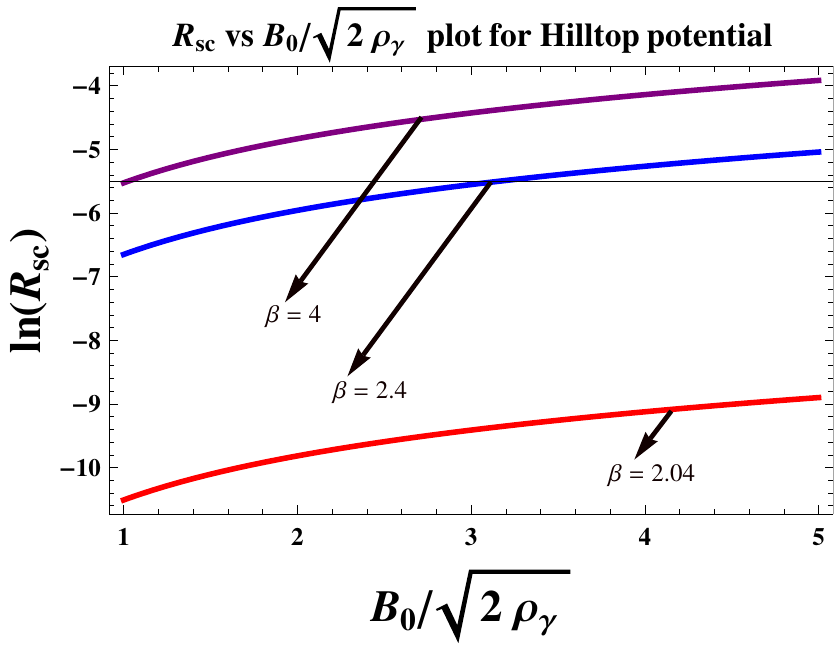}
    \label{fig11b}
}
\caption[Optional caption for list of figures]{Variation of \ref{fig11a} the reheating energy
density and \ref{fig11b} logarithm of reheating characteristic parameter with respect to the
scaled magnetic field at the present epoch $\frac{B_{0}}{\sqrt{2\rho_{\gamma}}}$ for
the characteristic index $\beta=2.04(\textcolor{red}{red}),2.4(\textcolor{blue}{blue}),
4(\textcolor{purple}{purple})$.} 
\label{fig11}
\end{figure*}
 Further using Eq~(\ref{m34}) and Eq~(\ref{m35}) the reheating energy density can be computed as:
 \be\begin{array}{lll}\label{m73}
 \displaystyle\rho_{reh}\geq(8.46\times 10^{-7}~M^4_p)\times\left\{1+\frac{2\sigma\beta\left(\beta-2\right)
\Delta{\cal N}_{b}}{\left(\frac{V_{0}}{2\sigma \beta^{2}}\right)^{\frac{2-\beta}{2(\beta-1)}}\left(\frac{\mu}{M_p}
\right)^{\frac{\beta}{\beta-1}}V_0}\right\}^{\frac{2(\beta-1)}{2-\beta}}\\~~~~~~~~~~~~~~~~~~~~~~~~~~~~~~~~~~\times\left\{\begin{array}{ll}
                    \displaystyle  \exp\left[3\Delta\bar{\cal N}_{b}\right]\times\left(\frac{B_0}{\sqrt{2\rho_{\gamma}}}\right)^{6} &
 \mbox{\small {\bf for \underline{$\bar{w}_{reh}=0$}}}  \\ 
         \displaystyle \exp\left[4\Delta\bar{\cal N}_{b}\right]\left(\frac{B_0}{\sqrt{2\rho_{\gamma}}}
\right)^{-\frac{4{\Delta\bar{\cal N}_{b}}}{\ln\left(\frac{B_0}{\sqrt{2\rho_{\gamma})}}\right)}} & \mbox{\small {\bf for \underline{$\bar{w}_{reh}\neq 0$}}}.
          \end{array}
\right.
\end{array}\ee
Next using Eq~(\ref{m47}), I get the following constraint on the dimensionless magnetic density parameter:
\be\label{m74}
 \Omega_{B_{end}}=\frac{B^{4}_{0}M^{6}_{p}}{24\sigma H^{2}_{0}\rho^2_{\gamma}}\times(7.16\times 10^{-13})\times\left\{1+\frac{2\sigma\beta\left(\beta-2\right)
\Delta{\cal N}_{b}}{\left(\frac{V_{0}}{2\sigma \beta^{2}}\right)^{\frac{2-\beta}{2(\beta-1)}}\left(\frac{\mu}{M_p}
\right)^{\frac{\beta}{\beta-1}}V_0}\right\}^{\frac{4(\beta-1)}{2-\beta}}\times\exp\left[8\Delta\bar{\cal N}_{b}\right].
 \ee
 Finally the rescaled reheating parameter can be expressed in terms of the model parameters of the hilltop models
 of inflationary potential as~\footnote{The CMB constraint on the lower bound of the rescaled reheating parameter for hilltop models 
 within $2\sigma$ CL is given by \cite{Martin:2010kz}:
 \be\label{cmb} R_{sc}>9.29\times 10^{-11}.\ee }:
 \bea\label{m75}
 R_{sc}&=&3.03\times 10^{-2}\times\left\{1+\frac{2\sigma\beta\left(\beta-2\right)
\Delta{\cal N}_{b}}{\left(\frac{V_{0}}{2\sigma \beta^{2}}\right)^{\frac{2-\beta}{2(\beta-1)}}\left(\frac{\mu}{M_p}
\right)^{\frac{\beta}{\beta-1}}V_0}\right\}^{\frac{\beta-1}{2(2-\beta)}}\exp\left[\Delta\bar{\cal N}_{b}
 \right]
 \times\left(\frac{B_0}{\sqrt{2\rho_{\gamma}}}\right).~~~~~~~~~~~
 \eea
 and using the numerical constraint on the rescaled reheating parameter as stated in Eq~(\ref{cmb}) I get the lower bound on the present value of the magnetic field as:
 \bea\label{m76}
 \frac{B_0}{\sqrt{2\rho_{\gamma}}}>\frac{4.45\times 10^{-12}\times\left\{1+\frac{2\sigma\beta\left(\beta-2\right)
\Delta{\cal N}_{b}}{\left(\frac{V_{0}}{2\sigma \beta^{2}}\right)^{\frac{2-\beta}{2(\beta-1)}}\left(\frac{\mu}{M_p}
\right)^{\frac{\beta}{\beta-1}}V_0}\right\}^{\frac{2(\beta-1)}{3(\beta-2)}}}{\exp\left[\frac{4}{3}\Delta\bar{\cal N}_{b}
 \right]}.
 \eea
 In fig.~\ref{fig10a}, fig.~\ref{fig10b} and fig.~\ref{fig10c} I have explicitly shown the variation of the magnetic field at the present epoch $B_{0}$, reheating energy density $\rho_{reh}$ 
 and logarithm of reheating characteristic parameter $\ln(R_{sc})$ with respect to the characteristic index $\beta$ of the hilltop potential for the number of e-foldings $\Delta{\cal N}_{b}=50$,
$|\Delta\bar{\cal N}_{b}|=11.5$, brane tension $\sigma\sim 10^{-9}~M^{4}_{p}$, mass scale parameter $\mu=1~M_p$ and mean equation of state parameter $\bar{w}_{reh}=0$.
The green dotted region bounded by two vertical black coloured lines and two black coloured horizontal lines represent the Planck $2\sigma$ allowed region and the rest of the light grey shaded
region is disfavoured by the Planck data and Planck+ BICEP2+Keck Array joint constraint. Also in fig.~\ref{fig11a} and in fig.~\ref{fig11b} I have depicted the behaviour 
of the reheating energy density $\rho_{reh}$ and logarithm of reheating characteristic parameter $\ln(R_{sc})$, with respect to the scaled magnetic field at the present epoch $B_{0}/\sqrt{2\rho_{\gamma}}$ for
the characteristic index $2.04\leq \beta\leq 2.4$. 

Using Eq~(\ref{m69}) in Eq~(\ref{eq16}) and Eq~(\ref{zaw3}), finally I get the following constraints on the magnetic regulating factor 
  within RSII setup 
 as~\footnote{After fixing $\Delta{\cal N}_{b}\approx {\cal O}(50-70)$, 
the regulating factor within RSII can be constrained as, \be \Sigma_{b}(k_{L}=k_0,k_{*})\times\left(\frac{M^4_p}{\sigma}\right)^{2/5}<3.98\times 10^{-19},\ee which is consistent with
the upper bound mentioned in Eq~(\ref{tune}).}:
\bea\label{eq16sdx}
     \displaystyle \Sigma_{b}(k_{L}=k_0,k_{*})\times\left(\frac{M^4_p}{\sigma}\right)^{2/5}\approx
{\cal O}(1.99\times10^{-21})\times\left\{1+\frac{2\sigma\beta\left(\beta-2\right)
\Delta{\cal N}_{b}}{\left(\frac{V_{0}}{2\sigma \beta^{2}}\right)^{\frac{2-\beta}{2(\beta-1)}}\left(\frac{\mu}{M_p}
\right)^{\frac{\beta}{\beta-1}}V_0}\right\}^{\frac{2(\beta-1)}{\beta-2}}
    \eea
    which is compatible with the observed/measured bound on CP asymmetry and baryon asymmetry parameter.

From fig.~\ref{fig8a}-fig.~\ref{fig10c}, I get the following $2\sigma$ constraints on
cosmological parameters computed from the hilltop inflationary model:
\bea\label{w1}
1.238\times 10^{-9}~{\rm Gauss}<B_{0}=&\sqrt{\frac{I_{\xi}(k_{L}=k_{0},k_{\Lambda})}{2\pi^2}A_{\bf B}}<&1.263\times 10^{-9}~{\rm Gauss},\\
\label{w1x1}
8.345\times 10^{-132}~M^{4}_{p}&<\rho_{B_{0}}=B^{2}_{0}/2<&8.685\times 10^{-132}~M^{4}_{p},\\
\label{w2} 4.945\times10^{-14}~M^{4}_{p}&<\rho_{reh}<&5.128\times10^{-14}~M^{4}_{p},\\
\label{w2x} 6.227\times10^{-4}\times g^{-1/4}_*~M_{p}&<T_{reh}<&6.283\times10^{-4}\times g^{-1/4}_*~M_{p},\\
\label{w2xxx}\Gamma_{total}&\sim&1.7\times10^{-4}~M_{p},\\
\label{w3}  7\times 10^{-5}&<R_{sc}<&7.11\times 10^{-5},\\
\label{w6} \epsilon_{\bf CP}& \sim& {\cal O}(10^{-6}),\\
\label{w7} \eta_{B}&\sim& {\cal O}(10^{-9}),\\
\label{w8} 0.121&<r<&0.124,\\
\label{w9} 0.969&<n_{S}<&0.970,\\
\label{w9x}2.3794\times 10^{-9}&<P_{S}<&2.3798\times 10^{-9},\\
\label{w10} 8.08\times 10^{-3}~M_{p} &<\sqrt[4]{V_{0}}<&8.13\times 10^{-3}~M_{p},
\eea
for the number of e-foldings $\Delta{\cal N}_{b}=50$, $|\Delta\bar{\cal N}_{b}|=11.5$, mean equation of state parameter $\bar{w}_{reh}=0$ and $\Omega_{rad}h^2\sim2.5\times 10^{-5}$, along with the following restricted model parameter space:
\bea \label{w11} 2.04&<\beta<&2.4,\\
\label{w12} \mu&\sim& {\cal O}(M_{p}),\\
\label{w13} \sigma&\sim& 10^{-9}~M^{4}_{p},\\
\label{w14} M_{5}&\sim& \left(4.170\times 10^{-20}\times\frac{M^{8}_{p}}{|\tilde{\Lambda}_{5}|}\right)^{1/3}.
\eea
It is important to mention here that, if I choose different parameter space by allowing fine tuning in-(1) the energy scale of hilltop potential $V_{0}=M^4$, (2) 
the mass scale parameter $\mu$ of the
hilltop model, (3) the brane tension $\sigma$ and (4) the characteristic index of the hilltop potential $\beta$ then the overall analysis and the obtained
results suggests that-
\begin{itemize}
 \item  For $\beta<2.04$, the amplitude of the scalar power spectrum $P_{S}$ and the scalar spectral tilt $n_{S}$ match the Planck 2015 data and 
 also consistent with the joint constraint obtained from Planck+BICEP2+Keck Array. But in this regime the value of tensor-to-scalar ratio $r$ exceeds the upper bound 
 i.e. $r>0.12$. On the other hand, for very low $\beta$ the estimated value of the magnetic field at the present epoch $B_{0}$
 from the hilltop model is very very small and can be able to reach up to the lower bound \be B_{0}>10^{-15}~{\rm Gauss}.\ee Similarly for low $\beta$, the reheating 
 energy density $\rho_{reh}$ or equivalently the reheating temperature $T_{reh}$ falls down and also the rescaled reheating parameter $R_{sc}$ decrease.
 
 \item For $\beta>2.4$, the amplitude of the scalar power spectrum $P_{S}$ and the scalar spectral tilt $n_{S}$ are perfectly consistent with
 the Planck 2015 data and also consistent with the joint constraint obtained from Planck+BICEP2+Keck Array. But in this case
 the value of tensor-to-scalar ratio $r$ is very very small compared to its the upper bound 
 i.e. $r<<0.12$. As $\beta$ increases the estimated value of the magnetic field at the present epoch $B_{0}$ exceeds the upper
 bound i.e. \be B_{0}>>10^{-9}~{\rm Gauss}\ee as obtained from Faraday rotation.
 Additionally in the large $\beta$ regime the reheating 
 energy density $\rho_{reh}$ or equivalently the reheating temperature $T_{reh}$ and the rescaled reheating parameter $R_{sc}$ are not consistent 
 with the observational constraints.
\end{itemize}
\section{Summary}
 \label{a6}
To summarize, in the present article, I have addressed the following points:
\begin{itemize}
 \item I have established
a theoretical constraint relationship on inflationary
magnetic field in the framework of Randall-Sundrun braneworld gravity (RSII) from: (1) tensor-to-scalar ratio ($r$), (2) reheating,
(3) leptogenesis and (4) baryogenesis for
a generic large and small field model of inflation with a flat potential,
where inflation is driven by slow-roll.

\item For such a class of model it is also possible to predict
amount of magnetic field at the present epoch ($B_{0}$) by measuring
non-vanishing CP asymmetry ($\epsilon_{CP}$) in leptogenesis and baryon asymmetry ($\eta_{B}$) in baryogenesis or the tensor-to-scalar
ratio in the inflationary setup.

\item Most significantly, once the signature of primordial
gravity waves will be predicted by in any near future observational probes, it will be possible to comment
on the associated CP asymmetry and baryon asymmetry and vice versa.

\item  In this paper I
have used important cosmological and particle physics constraints arising from Planck 2015 and Planck+BICEP2/Keck Array joint data on the 
amplitude of scalar power spectrum, scalar spectral tilt, the upper bound on tenor to scalar ratio,
lower bound on rescaled characteristic reheating parameter
and the bound on the reheating energy density
within $1.5\sigma-2\sigma$ statistical CL.

\item Further assuming the conformal invariance to be restored after inflation in the framework of Randall-Sundrum single
braneworld gravity (RSII), I have explicitly shown that the requirement of the 
sub-dominant feature of large scale magnetic field after inflation gives two fold
non-trivial characteristic constraints- on equation of state parameter ($w$) and the corresponding 
energy scale during reheating ($\rho^{1/4}_{rh}$) epoch.

\item  Hence avoiding the contribution of back-reaction from the magnetic field I have established
a bound on the reheating characteristic
parameter ($R_{rh}$) and its rescaled version ($R_{sc}$), to achieve large scale magnetic field within the prescribed setup
and apply the Cosmic Microwave Background (CMB) constraints as obtained from recent Planck 2015 data and Planck+BICEP2/Keck Array joint data.

\item  To this end I have explicitly 
shown the cosmological consequences from two specific models of 
brane inflation- monomial (large field) and hilltop (small field) after applying all the constraints obtained from inflationary magnetic field. 
For monomial models I get, $5.969\times 10^{-10}~{\rm Gauss}<B_{0}<4.638\times 10^{-9}~{\rm Gauss},$
$ 4.061\times10^{-5}~M^{4}_{p}<\rho_{reh}<1.591\times10^{-3}~M^{4}_{p},$ $1.940\times 10^{-132}~M^{4}_{p}<\rho_{B_{0}}<1.171\times 10^{-130}~M^{4}_{p},$ $6.227\times10^{-4}\times g^{-1/4}_*~M_{p}<T_{reh}<4.836\times10^{-3}\times g^{-1/4}_*~M_{p},$
$\Gamma_{total}\sim0.24~M_{p},$
$ 1.55\times 10^{-3}<R_{sc}<1.24\times 10^{-2}$, $\epsilon_{\bf CP}\sim {\cal O}(10^{-6}),$
$\eta_{B}\sim {\cal O}(10^{-9}),$  $0.121<r<0.124$, $0.969<n_{S}<0.970$, $2.3794\times 10^{-9}<P_{S}<2.3798\times 10^{-9}$, $ 8.08\times 10^{-3}~M_{p} <\sqrt[4]{V_{0}}<8.13\times 10^{-3}~M_{p}$
for $0.7<\beta<1.1$, $-0.48<\bar{w}_{reh}<-0.29$, $\Delta{\cal N}_{b}=50$, $\Delta\bar{\cal N}_{b}=7$ and $\sigma\sim 5\times 10^{-16}~M^{4}_{p}$.
Similarly for hilltop models I get, $1.238\times 10^{-9}~{\rm Gauss}<B_{0}<1.263\times 10^{-9}~{\rm Gauss},$
$ 4.945\times10^{-14}~M^{4}_{p}<\rho_{reh}<5.128\times10^{-14}~M^{4}_{p},$ $8.345\times 10^{-132}~M^{4}_{p}<\rho_{B_{0}}<8.685\times 10^{-132}~M^{4}_{p},$ $6.227\times10^{-4}\times g^{-1/4}_*~M_{p}<T_{reh}<6.283\times10^{-4}\times g^{-1/4}_*~M_{p},$
$\Gamma_{total}\sim 1.7\times10^{-4}~M_{p},$
$  7\times 10^{-5}<R_{sc}<7.11\times 10^{-5}$, $\epsilon_{\bf CP}\sim {\cal O}(10^{-6}),$
$\eta_{B}\sim {\cal O}(10^{-9}),$  $0.121<r<0.124$, $0.969<n_{S}<0.970$, $2.3794\times 10^{-9}<P_{S}<2.3798\times 10^{-9}$, $ 8.08\times 10^{-3}~M_{p} <\sqrt[4]{V_{0}}<8.13\times 10^{-3}~M_{p}$
for $2.04<\beta<2.4$, $\bar{w}_{reh}=0$, $\Delta{\cal N}_{b}=50$, $\Delta\bar{\cal N}_{b}=11.5$,
$\sigma\sim 10^{-9}~M^{4}_{p}$ and $\mu=1~M_p$.

\item The prescribed analysis performed in this paper also shows that the estimated cosmological parameters for both of the models confronts well with the Planck 2015 data  and Planck+BICEP2+Keck Array joint constraint
within $2\sigma$ CL for restricted choice of the parameter space of the model parameters within the framework of Randall-Sundrum single braneworld. Also it is important mention here that by doing parameter 
estimation from both of these simple class of models, it is clearly observed that the magneto-reheating constraints serve the purpose of breaking the degeneracy between the inflationary observables estimated
from both of these inflationary models.

\end{itemize}
Further my
aim is to carry forward this work in a more broader sense,
where I will apply all the derived results to the rest of the inflationary models within RSII setup. The other promising future prospects of this work are-
\begin{enumerate}
\item One can follow the prescribed methodology to derive the cosmological constraints 
in the context of various modified gravity framework i.e. Dvali-Gabadadze-Porrati (DGP) braneworld \cite{Dvali:2000hr}, Einstein-Hilbert-Gauss-Bonnet (EHGB) gravity \cite{Choudhury:2013eoa,Maitra:2013cta},
Einstein-Gauss-Bonnet-Dilaton (EGBD) gravity \cite{Choudhury:2013yg,Choudhury:2013aqa,Choudhury:2014hna,Choudhury:2015wfa} and $f(R)$ theory of gravity \cite{DeFelice:2010aj,Sotiriou:2008rp} etc. 
\item Hence using the derived constraints one can constrain various classes of large and small field inflationary
models \cite{Mazumdar:2010sa,Martin:2014vha,Martin:2014rqa,Martin:2014rqa,Martin:2013nzq,Linde:2014nna,Lyth:1998xn,Lyth:2007qh} within the framework of  other
modified theories of gravity.
\item One can explore various hidden cosmological features of CMB E-mode and B-mode polarization
spectra from the various modified gravity frameworks, which can be treated as a significant probe to put further stringent constraint on various classes of large and
small field inflationary models.
\item One can study the model independent prescription of describing the origin of primordial magnetic field by reconstructing inflationary potential \cite{Choudhury:2014kma,Choudhury:2015pqa} from various cosmological constraints from the observed data.
\item One can also implement the methodology for the alternative theories of inflation i.e. bouncing frameworks and related ideas. For an example one can investigate for the cosmological implications of cosmic hysteresis scenario \cite{Choudhury:2015baa} in the 
generation of primordial magnetic field.
\item Explaining the origin of primordial magnetic field in presence of non-standard/ non-canonical kinetic term, using non-minimal inflaton coupling to gravity sector, multi-field sector and also exploring the highly non-linear regime of field theory are serious of open issues in this literature. String theory originated DBI and tachyonic inflationary frameworks are the two prominent and well known examples of non-standard field theoretic setup through which one can explore various open questions in this area.
\end{enumerate}
\section*{Acknowledgments}
SC would like to thank Department of Theoretical Physics, Tata Institute of Fundamental
Research, Mumbai for providing me Visiting (Post-Doctoral) Research Fellowship.  SC take this opportunity to thank sincerely
to Prof. Sandip P. Trivedi, Prof. Shiraz Minwalla, Prof. Varun Sahni, Prof. Soumitra SenGupta, Prof. Sudhakar Panda,
Prof. Sayan Kar, Prof. Subhabrata Majumdar and Dr. Supratik Pal
for their constant support and inspiration. SC take this opportunity to thank all the active members and the
regular participants of weekly student discussion meet ``COSMOMEET'' from Department of Theoretical Physics and Department of Astronomy and Astrophysics, Tata Institute of Fundamental
Research for their strong support. SC additionally take this opportunity to thank the organizers of STRINGS, 2015, International Centre for Theoretical Science, Tata Institute of Fundamental Research (ICTS,TIFR), Indian Institute of Science (IISC) and specially Prof. Shiraz Minwalla for giving me the opportunity to participate in STRINGS, 2015 and also providing the local hospitality during the work. SC also thanks The Inter-University Centre for Astronomy and Astrophysics (IUCAA), Pune, India and specially Prof. Varun Sahni for providing the academic visit during the work. Last but not the least, I would all like to acknowledge our debt to the people of
India for their generous and steady support for research in natural sciences, especially for various areas in
theoretical high energy physics i.e. cosmology, string theory and particle physics.

\section{Appendix}
\subsection{ Inflationary consistency relations in RSII}
\label{a7}
In the context of RSII the spectral tilts $(n_S, n_T)$, running of the tilts $(\alpha_S, \alpha_T)$ and running of the running of tilts $(\kappa_T,\kappa_S)$ at the momentum pivot scale $k_*$ 
can be expressed as:
\begin{eqnarray}
 n_S(k_*) -1&=& 2\eta_{b}(\phi_*)-6\epsilon_{b}(k_*),\\
\label{cv1}n_T(k_*) &=& -3\epsilon_{b}(k_*)=-\frac{r(k_*)}{8},\\
\alpha_S(k_*) &=& 16\eta_{b}(k_*)\epsilon_{b}(k_*)-18\epsilon^{2}_{b}(k_*)-2\xi^{2}_{b}(k_*),\\
\alpha_T(k_*) &=& 6\eta_{b}(k_*)\epsilon_{b}(k_*)-9\epsilon^{2}_{b}(k_*),\\
\kappa_S(k_*)&=&152\eta_{b}(k_*)\epsilon^{2}_{b}(k_*)-32\epsilon_{b}(k_*)\eta^{2}_{b}(k_*)-108\epsilon^{3}_{b}(k_*)\nonumber
\\ &&~~~~~-24\xi^{2}_{b}(k_*)\epsilon_{b}(k_*)+2\eta_{b}(k_*)\xi^{2}_{b}(k_*)+2\sigma^{3}_{b}(k_*),\\
\kappa_T(k_*)&=&66\eta_{b}(k_*)\epsilon^{2}_{b}(k_*)-12\epsilon_{b}(k_*)\eta^{2}_{b}(k_*)
-54\epsilon^{3}_{b}(k_*)-6\epsilon_{b}(k_*)\xi^{2}_{b}(k_*).
\end{eqnarray}
 In terms of slow-roll parameters in RSII setup one can also write the following sets of consistency conditions for brane inflation:
\begin{eqnarray}
 \label{wqc1}n_T(k_*)-n_S(k_*)+1&=&\left(\frac{d\ln r(k)}{d\ln k}\right)_*=\left[\frac{r(k_*)}{8}-2\eta_{b}(k_*)\right],\\
\label{wqc2}\alpha_T(k_*)-\alpha_S(k_*)&=&\left(\frac{d^2\ln r(k)}{d\ln k^2}\right)_*=\left[\left(\frac{r(k_*)}{8}\right)^2-\frac{20}{3}\left(\frac{r(k_*)}{8}\right)+2\xi^2_{b}(k_*)\right],\\
\label{wqc3}\kappa_T(k_*)-\kappa_S(k_*)&=&\left(\frac{d^3\ln r(k)}{d\ln k^3}\right)_* \nonumber\\&=&
\left[2\left(\frac{r(k_*)}{8}\right)^3-\frac{86}{9}\left(\frac{r(k_*)}{8}\right)^2
\right.\\&&\left.~~+\frac{4}{3}\left(6\xi^2_{b}(k_*)+5\eta^{2}_{b}(k_*)\right)\left(\frac{r(k_*)}{8}\right)
+2\eta_{b}(k_*)\xi^2_{b}(k_*)+2\sigma^{3}_{b}(k_*)\right].\nonumber
\end{eqnarray}
Here Eq~(\ref{wqc1}-\ref{wqc3})) represent the running, running of the running and running of the double running of tensor-to-scalar ratio in RSII brane inflationary setup. 
Within high energy limit $\rho>>\sigma$ the slow-roll parameters in the visible brane can be expressed as:
\begin{eqnarray}
 \epsilon_{b}(\phi)&\approx& \frac{2M^{2}_{p}\sigma (V^{'}(\phi))^{2}}{V^{3}(\phi)},\\
\eta_{b}(\phi)&\approx& \frac{2M^{2}_{p}\sigma V^{''}(\phi)}{V^{2}(\phi)},\\
\xi^{2}_{b}(\phi)&\approx& \frac{4M^{4}_{p}\sigma^{2} V^{'}(\phi)V^{'''}(\phi)}{V^{4}(\phi)},\\
\sigma^{3}_{b}(\phi)&\approx& \frac{8M^{6}_{p}\sigma^{3} (V^{'}(\phi))^{2}V^{''''}(\phi)}{V^{6}(\phi)}.
\end{eqnarray}
and consequently the number of e-foldings can be written as:
\begin{eqnarray}\label{efolda}
\Delta {\cal N}_{b}={\cal N}_{b}(\phi_{cmb})-{\cal N}_{b}(\phi_{e})&\approx&  \frac{1}{2\sigma M^2_p}\int^{\phi_{cmb}}_{\phi_{e}} d\phi\frac{V^{2}(\phi)}{V^{'}(\phi)}
\end{eqnarray}
where $\phi_{e}$ corresponds to the field value at the end of inflation, which can be obtained from 
the following constraint equation:
\begin{eqnarray}
 \max_{\phi=\phi_{e}}\left[\epsilon_{b},|\eta_{b}|,|\xi^{2}_{b}|,|\sigma^{3}_{b}|\right]&=&1.
\end{eqnarray}
\subsection{ Evaluation of $I_{\xi}(k_{L},k_{\Lambda})$ integral kernel}
\label{a8}
 \be\begin{array}{lll}\label{eqintk1}
 I_{\xi}(k_{L};k_{\Lambda})=
\left\{\begin{array}{ll}
                    \displaystyle \frac{\sqrt{\pi}}{2\xi k_{*}}\left[{\rm erf(\xi k_{\Lambda})}-{\rm erf(\xi k_{L})}\right]~~~~~~~&
 \mbox{\small {\bf for \underline{Case I}}}  
\\ 
         \displaystyle \frac{1}{2\left(\xi k_{*}\right)^{n_B+3}} \left[\Gamma \left(\frac{\left(n_B+3\right)}{2} ,\xi ^2 k^2_{L}\right)
         -\Gamma \left(\frac{\left(n_B+3\right)}{2} ,\xi ^2 k^2_{\Lambda}\right)\right]~~~ & \mbox{\small {\bf for \underline{Case II}}}
         \\ 
         \displaystyle \left[\frac{\sqrt{\pi}~{\rm erf}\left(\xi k\right)}{2\xi k_{*}}\left\{1+{\cal Q}\ln\left(\frac{k}{k_{*}}\right)
+{\cal P}\ln^{2}\left(\frac{k}{k_{*}}\right)\right\}\right.\\ \left. 
\displaystyle~~~~~~+\left(\frac{k}{k_{*}}\right)
\left\{2{\cal P}~_PF_Q\left[\left\{\frac{1}{2},\frac{1}{2},\frac{1}{2}\right\}~;\left\{\frac{3}{2},\frac{3}{2},\frac{3}{2}\right\}~;
-\xi^{2}k^{2}\right]
\right.\right.\\ \left.\left.\displaystyle 
~~~~~~
\displaystyle -\left({\cal Q}+2{\cal P}\ln\left(\frac{k}{k_{*}}\right)\right)
~_PF_Q\left[\left\{\frac{1}{2},\frac{1}{2}\right\}~;\left\{\frac{3}{2},\frac{3}{2}\right\}~;
-\xi^{2}k^{2}\right]\right\}\right]^{k=k_{\Lambda}}_{k=k_{L}}  & \mbox{\small {\bf for \underline{Case III}}}
\\ 
         \displaystyle  \left[\frac{\sqrt{\pi}~{\rm erf}\left(\xi k\right)}{2\xi k_{*}}\left\{1+{\cal Q}\ln\left(\frac{k}{k_{*}}\right)
+{\cal P}\ln^{2}\left(\frac{k}{k_{*}}\right)+{\cal F}\ln^{3}\left(\frac{k}{k_{*}}\right)\right\}\right.\\ \left. 
\displaystyle~~~~~~+\left(\frac{k}{k_{*}}\right)
\left\{-6{\cal F}~_PF_Q\left[\left\{\frac{1}{2},\frac{1}{2},\frac{1}{2},\frac{1}{2}\right\}~;\left\{\frac{3}{2},\frac{3}{2},\frac{3}{2},\frac{3}{2}\right\}~;
-\xi^{2}k^{2}\right]\right.\right.\\ \left.\left.\displaystyle 
~~~~~~~+2\left({\cal P}+3{\cal F}\ln\left(\frac{k}{k_{*}}\right)\right)
~_PF_Q\left[\left\{\frac{1}{2},\frac{1}{2},\frac{1}{2}\right\}~;\left\{\frac{3}{2},\frac{3}{2},\frac{3}{2}\right\}~;
-\xi^{2}k^{2}\right]\right.\right.\\ \left.\left.\displaystyle 
~~~~~~
\displaystyle -\left({\cal Q}+2{\cal P}\ln\left(\frac{k}{k_{*}}\right)+6{\cal F}\ln^{2}\left(\frac{k}{k_{*}}\right)\right)\right.\right.\\ \left.\left.\displaystyle
~~~~~~\times~_PF_Q\left[\left\{\frac{1}{2},\frac{1}{2}\right\}~;\left\{\frac{3}{2},\frac{3}{2}\right\}~;
-\xi^{2}k^{2}\right]\right\}\right]^{k=k_{\Lambda}}_{k=k_{L}} & \mbox{\small {\bf for \underline{Case IV}}}
          \end{array}
\right.
   \end{array}\ee
where ${\cal Q}=n_{ B}+2$, ${\cal P}=\alpha_{ B}/2$ and ${\cal F}=\kappa_{ B}/6$.
\subsection{ Evaluation of $J(k_{L},k_{\Lambda})$ integral kernel}
\label{a9}
 \be\begin{array}{lll}\label{eqintk11}
 J(k_{L};k_{\Lambda})=
\left\{\begin{array}{ll}
                    \displaystyle  \frac{1}{3} \left[\left(\frac{k_{\Lambda}}{k_{*}}\right)^{3}-\left(\frac{k_{L}}{k_{*}}\right)^{3}\right]~~~~~~~&
 \mbox{\small {\bf for \underline{Case I}}}  
\\ 
         \displaystyle \frac{1}{(n_B+3)} \left[\left(\frac{k_{\Lambda}}{k_{*}}\right)^{n_{B}+3}-\left(\frac{k_{L}}{k_{*}}\right)^{n_{B}+3}\right]~~~ & \mbox{\small {\bf for \underline{Case II}}}
         \\ 
         \displaystyle \left\{\left(\frac{k}{k_{*}}\right)\left[(1+2{\cal P}-{\cal Q})+({\cal Q}-2{\cal P})\ln\left(\frac{k}{k_{*}}\right)
+{\cal P}\ln^{2}\left(\frac{k}{k_{*}}\right)\right]\right\}^{k=k_{\Lambda}}_{k=k_{L}}  & \mbox{\small {\bf for \underline{Case III}}}
\\ 
         \displaystyle  \left\{\left(\frac{k}{k_{*}}\right)\left[(1-6{\cal F}+2{\cal P}-{\cal Q})+(6{\cal F}-2{\cal P}+{\cal Q})\ln\left(\frac{k}{k_{*}}\right)
\right.\right.\\ \left.\left.\displaystyle~~~~~~-(3{\cal F}-{\cal P})\ln^{2}\left(\frac{k}{k_{*}}\right)+{\cal F}\ln^{3}\left(\frac{k}{k_{*}}\right)\right]\right\}^{k=k_{\Lambda}}_{k=k_{L}} & \mbox{\small {\bf for \underline{Case IV}}}
          \end{array}
\right.
   \end{array}\ee
where ${\cal Q}=n_{ B}+2$, ${\cal P}=\alpha_{ B}/2$ and ${\cal F}=\kappa_{ B}/6$.

\end{document}